\documentclass[submitting]{nst}

\usepackage{subfigure,dcolumn}
\usepackage{epstopdf}
\usepackage{mhchem}
\usepackage{upgreek}
\usepackage{float}
\usepackage{graphicx}
\usepackage{hyperref}
\usepackage{color}
\usepackage{xcolor}
\usepackage{ulem}
\usepackage{ragged2e}
\usepackage{wrapfig}
\usepackage{amsmath}

\usepackage{tikz,xcolor,hyperref}

\definecolor{lime}{HTML}{A6CE39}
\DeclareRobustCommand{\orcidicon}{
	\begin{tikzpicture}
	\draw[lime, fill=lime] (0,0) 
	circle [radius=0.16] 
	node[white] {{\fontfamily{qag}\selectfont \tiny ID}};
	\draw[white, fill=white] (-0.0625,0.095) 
	circle [radius=0.007];
	\end{tikzpicture}
	\hspace{-2mm}
}
\foreach \x in {A, ..., Z}{%
	\expandafter\xdef\csname orcid\x\endcsname{\noexpand\href{https://orcid.org/\csname orcidauthor\x\endcsname}{\noexpand\orcidicon}}
}

\begin{document}

\title{Properties of the QCD Matter \\
An Experimental Review of Selected Results from RHIC BES Program}\thanks{This review is dedicated to Professor Wenqing Shen on the occasion to celebrate his leadership of the Chinese STAR Collaboration, the development and production of the STAR MRPC TOF detector in China and many physics analyses. All authors contributed equally to this work.}

\author{Jinhui Chen}
\affiliation{Key Laboratory of Nuclear Physics and Ion-beam Application (MOE), Institute of Modern Physics, Fudan University, Shanghai 200433, China}
\author{Xin Dong}
\affiliation{Nuclear Science Division, Lawrence Berkeley National Laboratory, Berkeley, CA 94720, USA}
\author{Xionghong He}
\affiliation{QMRC, Institute of Modern Physics, Chinese Academy of Sciences, Lanzhou, China}
\author{Huanzhong Huang}
\affiliation{Department of Physics and Astrophysics, University of California, Los Angeles, CA 90095, USA}
\author{Feng Liu}
\affiliation{Key Laboratory of Quark \& Lepton Physics (MOE) and Institute of Particle Physics, Central China Normal University, Wuhan 430079, China}
\author{Xiaofeng Luo}
\affiliation{Key Laboratory of Quark \& Lepton Physics (MOE) and Institute of Particle Physics, Central China Normal University, Wuhan 430079, China}
\author{Yugang Ma}
\affiliation{Key Laboratory of Nuclear Physics and Ion-beam Application (MOE), Institute of Modern Physics, Fudan University, Shanghai 200433, China}
\author{Lijuan Ruan}
\affiliation{Department of Physics, Brookhaven National Laboratory, Upton, NY 11973, USA}
\author{Ming Shao}
\affiliation{Department of Modern Physics, University of Science and Technology of China, Hefei, China}
\author{Shusu Shi}
\affiliation{Key Laboratory of Quark \& Lepton Physics (MOE) and Institute of Particle Physics, Central China Normal University, Wuhan 430079, China}
\author{Xu Sun}
\affiliation{QMRC, Institute of Modern Physics, Chinese Academy of Sciences, Lanzhou, China}
\author{Aihong Tang}
\affiliation{Department of Physics, Brookhaven National Laboratory, Upton, NY 11973, USA}
\author{Zebo Tang}
\affiliation{Department of Modern Physics, University of Science and Technology of China, Hefei, China}
\author{Fuqiang Wang}
\affiliation{Department of Physics, Purdue University, West Lafayette, IN 47907, USA}
\author{Hai Wang}
\affiliation{Key Laboratory of Nuclear Physics and Ion-beam Application (MOE), Institute of Modern Physics, Fudan University, Shanghai 200433, China}
\author{Yi Wang}
\affiliation{Department of Engineering Physics, Tsinghua University, Beijing, China}
\author{Zhigang Xiao}
\affiliation{Department of Physics, Tsinghua University, Beijing, China}
\author{Guannan Xie}
\affiliation{School of Nuclear Science and Technology, University of Chinese Academy of Sciences, Beijing 100049, China}
\author{Nu Xu} 
\affiliation{Nuclear Science Division, Lawrence Berkeley National Laboratory, Berkeley, CA 94720, USA}
\author{Qinghua Xu}
\affiliation{Key Laboratory of Particle Physics and Particle Irradiation (MOE), Institute of Frontier and Interdisciplinary Science, Shandong University, Qingdao 266237, China}
\author{Zhangbu Xu}
\affiliation{Department of Physics, Kent State University, Kent, OH 44242, USA}
\author{Chi Yang}
\affiliation{Key Laboratory of Particle Physics and Particle Irradiation (MOE), Institute of Frontier and Interdisciplinary Science, Shandong University, Qingdao 266237, China}
\author{Shuai Yang}
\affiliation{Institute of Quantum Matter, South China Normal University, Guangzhou 510006, China}
\author{Wangmei Zha}
\affiliation{Department of Modern Physics, University of Science and Technology of China, Hefei, China}
\author{Yapeng Zhang}
\affiliation{QMRC, Institute of Modern Physics, Chinese Academy of Sciences, Lanzhou, China}
\author{Yifei Zhang}
\affiliation{Department of Modern Physics, University of Science and Technology of China, Hefei, China}
\author{Jie Zhao}
\affiliation{Key Laboratory of Nuclear Physics and Ion-beam Application (MOE), Institute of Modern Physics, Fudan University, Shanghai 200433, China}
\author{Xianglei Zhu}
\affiliation{Department of Engineering Physics, Tsinghua University, Beijing, China}

%---================================================

\begin{abstract}
%This mini review is written on the occasion of Professor Wenqing Shen's symposium. 
In the paper, we discuss the development of the multi-gap resistive plate chamber Time-of-Flight (TOF) technology and the production of the STAR TOF detector in China at the beginning of the 21st century. Then 
we review recent experimental results from the first beam energy scan program (BES-I) at the Relativistic Heavy Ion Collider (RHIC). Topics cover measurements of collectivity, chirality, criticality, global polarization, strangeness, heavy-flavor, di-lepton and light nuclei productions.
\end{abstract}

%\begin{keyword} keyword 1 \sep keyword 2 \sep keyword 3 \sep keyword 4
%\end{keyword}

\maketitle
\nolinenumbers

\tableofcontents

%%%%%%%%%%%%%%%%%%  Main article  %%%%%%%%%%%%%%%%%%
\section{Introduction }\label{introduction}

Quantum Chromodynamics (QCD) is the theory of the strong force and is the cornerstone for understanding the fundamental nature of matter under the most extreme conditions \cite{Gross:1973id,Politzer:1973fx}. Of the myriad phenomena it encompasses, perhaps one of the most fascinating is the behavior of QCD matter at extreme temperatures and densities, where quarks and gluons, the fundamental constituents of matter, undergo phase transitions to become hadronic matter through hadronization. The experiments at Relativistic Heavy Ion Collider (RHIC) have provided unique experimental evidences for the transition (e.g., \cite{STAR:2005gfr}), yet it is far from clear at exactly what temperature and/or densities this phase transition occurs and what the nature of the phase transition is. Thus, it can be said that we are still mystified about the true nature of QCD, especially at extreme temperatures and densities. In the fiery furnaces of the early universe or in the cores of neutron stars, matter undergoes epic transformations, transitioning between different phases dictated by the intricate dynamics of QCD. It is in these extreme environments that the search for the properties of QCD matter has faced its greatest challenges and most profound revelations \cite{Bzdak:2019pkr,Luo:2022mtp}.

RHIC at Brookhaven National Laboratory stands as a beacon in the quest to unravel the mysteries of QCD matter. Through its Beam Energy Scan (BES) program \cite{STAR:2010vob}, RHIC has probed the properties of QCD matter across a wide range of collision energies in various aspects, providing a comprehensive experimental landscape to explore the phases and transitions of this extreme form of matter.

In this review, we start with a brief description of the early development and production of STAR MRPC TOF detector in China which marked first significant detector contribution to an international experiment from Chinese nuclear physics community. We then embark on a journey through the rich tapestry of experimental results gleaned from the RHIC BES program. We delve into the intricate interplay of phenomena such as the quark-gluon plasma (QGP), hadronization, and the evolution of collective behavior in heavy ion collisions. We highlight key findings from selected topics which have reshaped our understanding of the QCD matter and its manifestations in the laboratory. Topics cover basic observables including collectivity, chirality, criticality, global polarization, strangeness, heavy-flavor, di-lepton as well as light nuclei etc. 

From the onset of RHIC's operation to its latest experimental endeavors, this review attempts to encapsulate the progress made in deciphering the properties of QCD matter. Through precision measurements and innovative analysis techniques, RHIC has made strides to unravel the phase diagram of QCD matter, revealing its intricate structure and elucidating the fundamental forces that govern the Universe.

This review article is arranged as follows. Section~\ref{Sec:tof} describes the development of Time-of-Flight detector. Section~\ref{Sec:data} presents selective STAR measurements of identified particles which are enabled with the TOF detector. A brief summary and outlook is presented in Section~\ref{Sec:summary}.

%--==========================================================
%---===============================================
\section{Development and Production of the STAR MRPC TOF Detector}\label{Sec:tof}

The multi-gap resistive plate chamber (MRPC) technology was first realized in the mid-1990s by the ALICE Time-of-Flight (TOF) group~\cite{Crispin1996}. The MRPC technology enabled the construction of a cost-effective TOF detector to identify charged particles copiously produced in relativistic heavy ion collisions. The basic structure of MRPC features a stack of parallel resistive plates, usually with gaps of $\sim 0.2-0.3 mm$. High voltages are applied to the stack through the outermost plates through resistive conductive graphite, while the inner plates are electrically floating. When a charged particle passes through the MRPC, primary electrons are produced by ionization in the gaps (filled with Freon-rich gas mixture), then triggers gas avalanche amplification in the strong electric field (usually $\sim 100 kV/cm$ or more). Fast signals are induced on the outer readout strips. Usually differential signals are used to input preamplifier to reduce noise.  Multiple narrow gaps are beneficial to reduce the time fluctuation of avalanche, thus improving the timing performance. The inner electrically-floating plates take right potentials due to the gain-feedback in different gaps, and guarantee the gain uniformity. This striking feature greatly simplifies the manufacture and operation of MRPC. In short, MRPC is a new type of cost effective gas detector with an excellent timing performance.

\begin{figure}[htb]       
\centering
\includegraphics[width=0.45\textwidth]{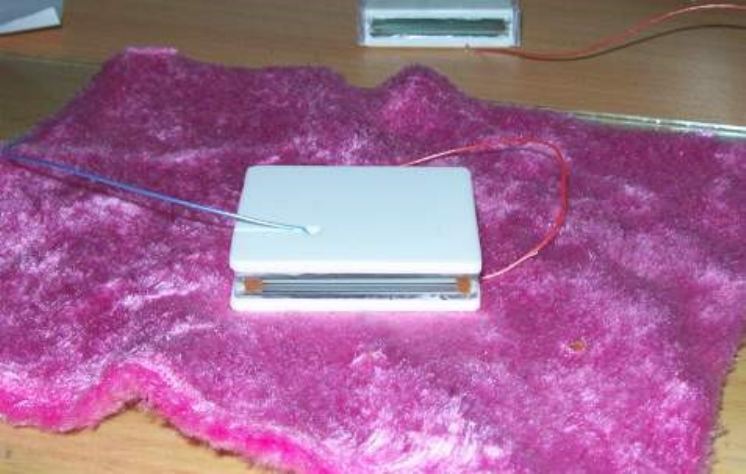}\vspace{-.5cm}
\caption{The first MRPC prototype produced in USTC, with an active area of $4\times4~cm^2$ and single-channel readout.}
\label{fig_first_mrpc}
\end{figure}

By developing the MRPC-based barrel TOF for the Solenoidal Tracker at RHIC (STAR) experiment, the China-US cooperation in heavy ion physics started in 2000. The first MRPC prototype in China was soon developed by University of Science and Technology of China (USTC)~\cite{mrpc2000}, as illustrated in Fig. \ref{fig_first_mrpc}.
In May 2001 the Chinese STAR team was officially established, led by Prof. Wenqing Shen, and decided to build a TOF tray (TOFr) demonstrator with 28 MRPCs. One month later, the STAR collaboration accepted all 6 institutions of the Chinese team, including Shanghai Institute of Applied Physics, Chinese Academy of Sciences (SINAP-CAS), Institute of High Energy Physics (IHEP-CAS), Institute of Modern Physics (IMP-CAS), Central China Normal University (CCNU), Tsinghua University (THU) and USTC, to be institutional members of the collaboration.

In 2002, the TOFr demonstrator was successfully developed jointly by Chinese and American teams. The Chinese side developed 24 MRPCs, and the US side developed 4 MRPCs and all electronics. Through this effort, Chinese researchers had gained deep understanding of the MRPC technology, in both detector physics and module production. TOFr had all features suitable for installation and operation in STAR, and joined physics run of STAR in 2003. The physics and experimental results from TOFr were so fruitful \cite{STAR:2003oii,STAR:2004ocv,STAR:2005ryu, tofr2004, tofr2005, pid2006} that Dr. Hallman, the spokesperson of STAR, wrote a letter specially to Prof. Wenqing Shen to express his congratulations. The major technical progress from the Chinese STAR team ultimately led STAR to decide to produce all the MRPC modules for the barrel TOF in China. %\cite{startof2004}.

In 2006, the project “Research of relativistic nuclear collision physics at STAR and development of time of flight detector” was jointly funded by NSFC, CAS and the Ministry of Science and Technology (MOST) of China. The Chinese STAR team cooperated to develop the STAR-TOF and RHIC physics research. By 2009, all 4000 MRPC modules were produced by THU and USTC. Due to the understanding of MRPC technology and strict quality control, the final yield is up to $95\%$, with very good stability and consistency \cite{qc_ustc, qc_thu}. Since the TOFr demonstrator, STAR TOF has maintained a systematic time resolution of $\sim 80 ps$ (MRPC intrinsic resolution $\sim 60 ps$) \cite{tof2012}, which was highly evaluated by experts of the US Department of Energy (DOE) and the STAR collaboration.

\begin{figure}[htb]       
\centering
\includegraphics[width=0.47\textwidth]{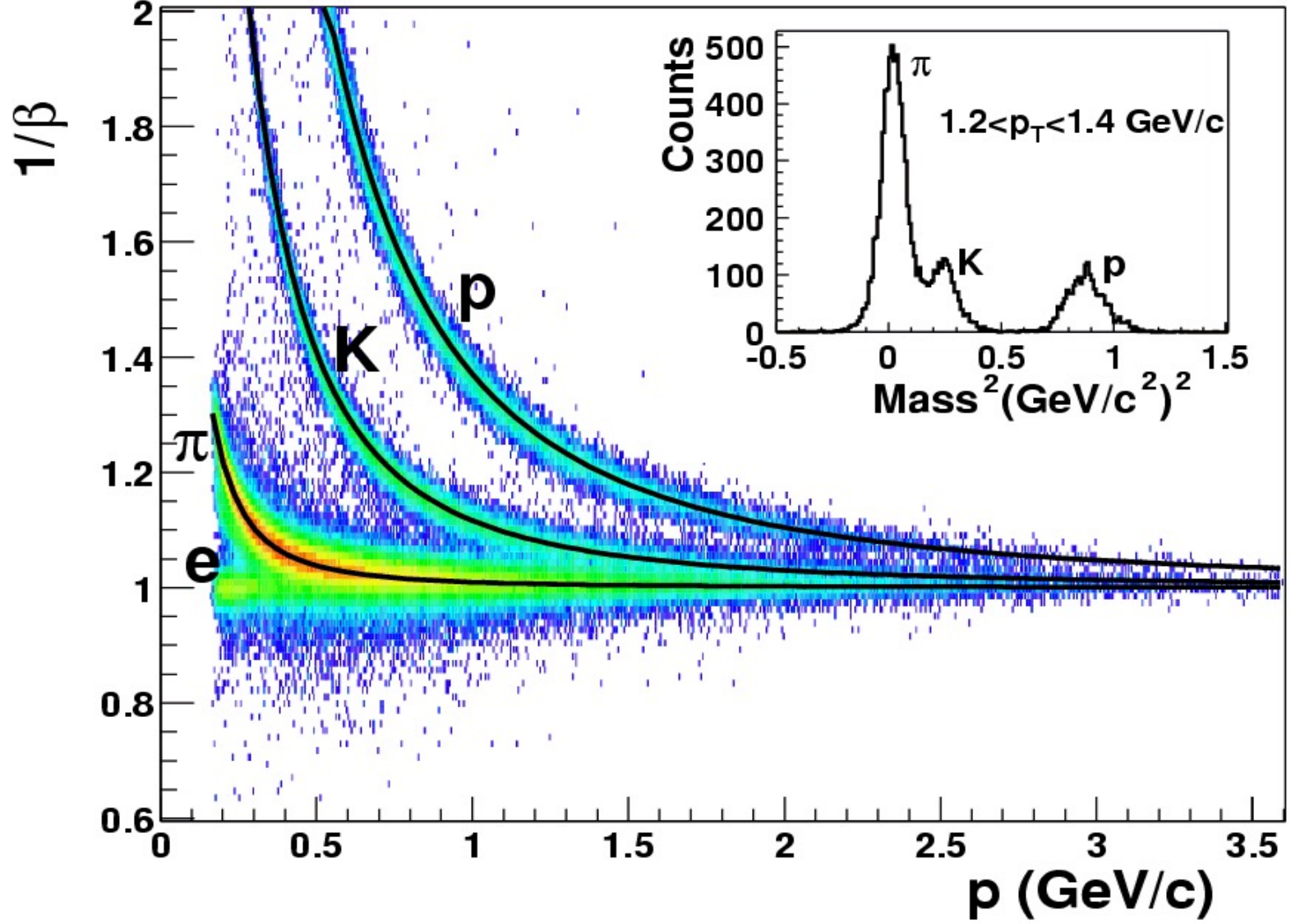}\vspace{-.5cm}
\caption{The particle velocity, $1/\beta$, as a function of particle momentum.}
\label{fig_TOFPID}
\end{figure}

\begin{figure}[htb]       
\centering
\includegraphics[width=0.47\textwidth]{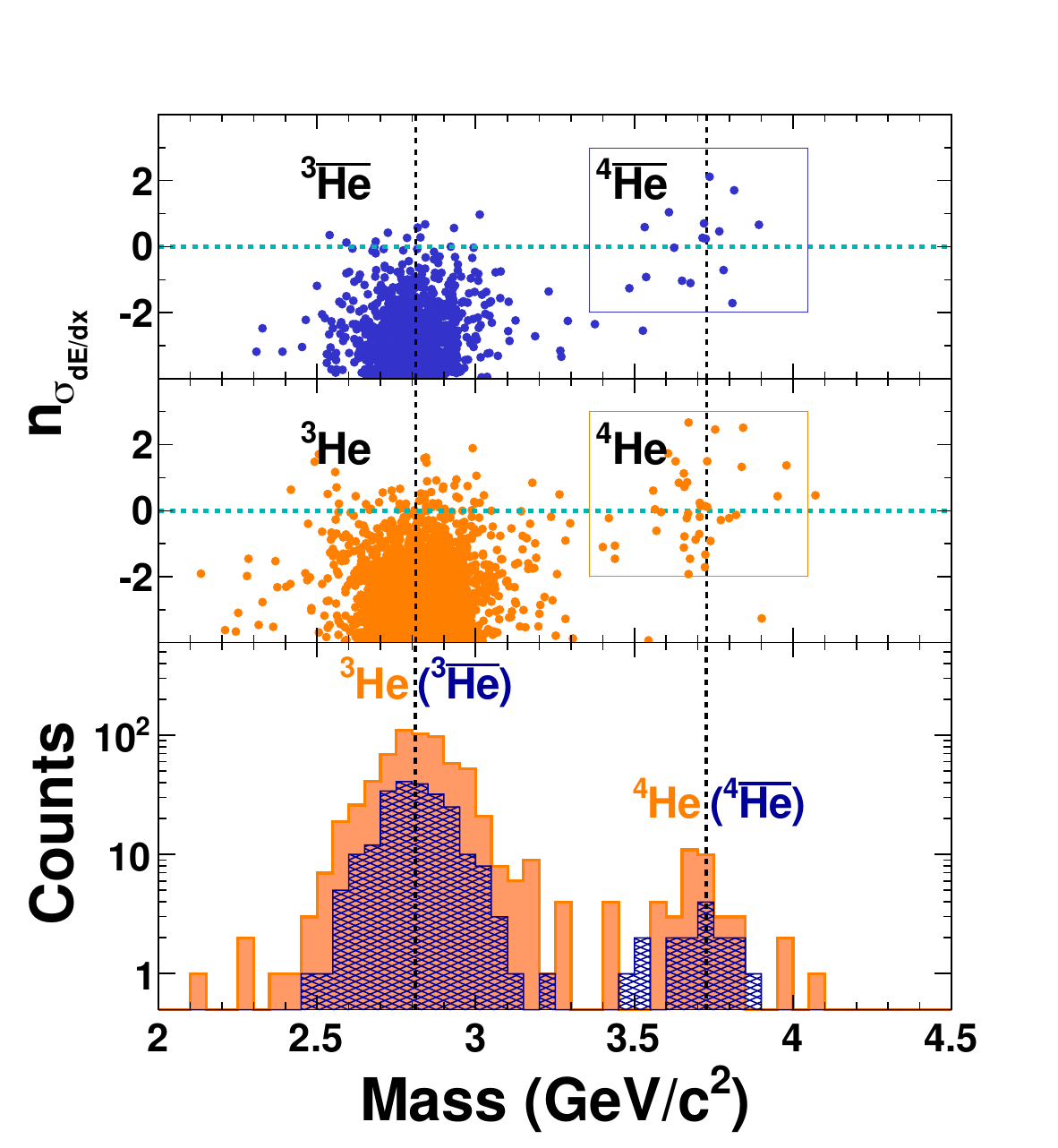}\vspace{-.5cm}

\caption{Top two panels show the $\langle dE/dx \rangle$ in units of multiples of $\sigma_{\rm dE/dx}$, $n\sigma_{\rm dE/dx}$, of negatively charged particles (first panel) and positively charged particles (second panel) as a function of mass measured by the TOF system. Rectangular boxes highlight areas for ${}^4\overline{\rm He}$ (${}^4$He) selections.
Bottom panel shows a projection of entries in the upper two panels onto the mass axis for particles in the selected window. A total of 16 candidates of $^4\overline{\rm He}$ are identified using the combined measurements of energy loss and time of flight. See Ref.~\cite{STAR:2011eej} for more details.}
\label{fig_III_H_antiHe4_PID}
\end{figure}

The TOF detector significantly extended the STAR particle identification capabilities. As seen in Fig.~\ref{fig_TOFPID}, with 2$\sigma$ separation, protons/(pions + kaons) and kaons/pions were identified up to 3 and 1.6 GeV/$c$, respectively. Without TOF, these two groups could only be identified up to 1.0 and 0.7 GeV/$c$, respectively. The successful construction and smooth operation of the ToF system also contributes to the observation of the heaviest antimatter helium-4 nucleus~\cite{STAR:2011eej}. By measuring the mean energy loss per unit track length in the Time Projection Chamber (TPC)~\cite{Anderson:2003ur}, which helps distinguish particles with different masses or charges, and the time of flight of particles arriving at the TOF surrounding the TPC, anti-helium nuclei can be identified unambiguously, see Fig.~\ref{fig_III_H_antiHe4_PID}.

The successful development and operation of STAR TOF and its significant promotion to STAR physics research have greatly boosted the application of MRPC technology. In 2008, the long-strip (length: $87 cm$) MRPC (MRPC) was developed in USTC \cite{lmrpc2008}. With the strong support of NSFC, the Chinese STAR team completed the development and construction of LMRPC-based MTD \cite{mtd2009}, whose successful performance \cite{mtd2014} further improved the research of lepton physics in STAR. In China, the successful operation of STAR TOF also triggered the endcap TOF upgrade of Beijing Spectrometer Experiment (BESIII) with MRPC technology \cite{bes3etof2020}.

With the success of the RHIC beam energy scan (BES) program (phase-I), high-luminosity heavy ion collision experiments at lower center-of-mass energies have become an important frontier to search for the phase boundary and critical end-point of the quark-gluon plasma phase transition. To adapt to the high-luminosity physics run, STAR TOF is required to have a magnitude-higher counting rate capability, especially in the endcap region. The STAR and CBM collaborations had jointly carried out research and development for this purpose. USTC adopts ultra-thin float glass to increase the MRPC counting rate from a few hundred $Hz/cm^{2}$ to $kHz/cm^{2}$, while THU successfully developed the MRPC that can operate at the counting rate of tens of $kHz/cm^{2}$, based on the special low-resistivity glass plates (bulk resistivity $\sim 10^{10}  \Omega \cdot cm$) \cite{highrate2010, highrate2013}. Both MRPCs were installed into the STAR endcap TOF and met the required performance.

\begin{figure}[htb]       
\centering
\includegraphics[width=0.49\textwidth]{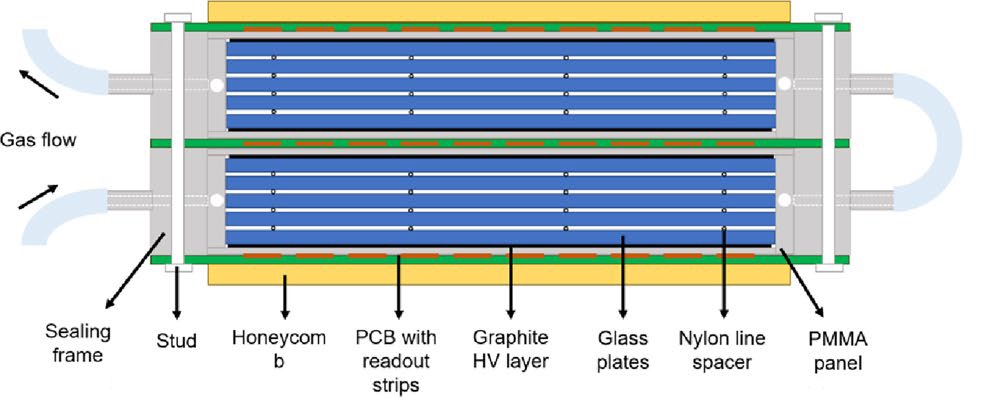}\vspace{-.5cm}
\caption{The latest design of the structure of sealed MRPC.}
\label{sealedmrpc}
\end{figure}

Another important application is the development of MRPC-TOF for CEE (The Cooler-storage-ring External-target Experiment, located at Lanzhou, is
the first spectrometer of China for heavy ion collision studies that operates in the GeV level energy regime.) It is committed to the studies on the phase structure of the nuclear matter, the nuclear equation of state, the symmetry energy, the production of hypernucleus, etc.. In order to improve the gas exchange speed and reduce gas consumption greatly, a new style sealed MRPC is developed \cite{2020etof}. The structure is shown in Fig. \ref{sealedmrpc}. The time resolution is better than 60 $ps$ and the efficiency is larger than 97 \%. In the cosmic test, this sealed MRPC can work at gas flush lower than 10 Sccm per square meter detector. It has been applied to the CEE-eTOF wall given a 70 \% reduction of the necessary gas flow rate and maintained performances and stability.

\begin{figure}[htb]       
\centering
\includegraphics[width=0.45\textwidth]{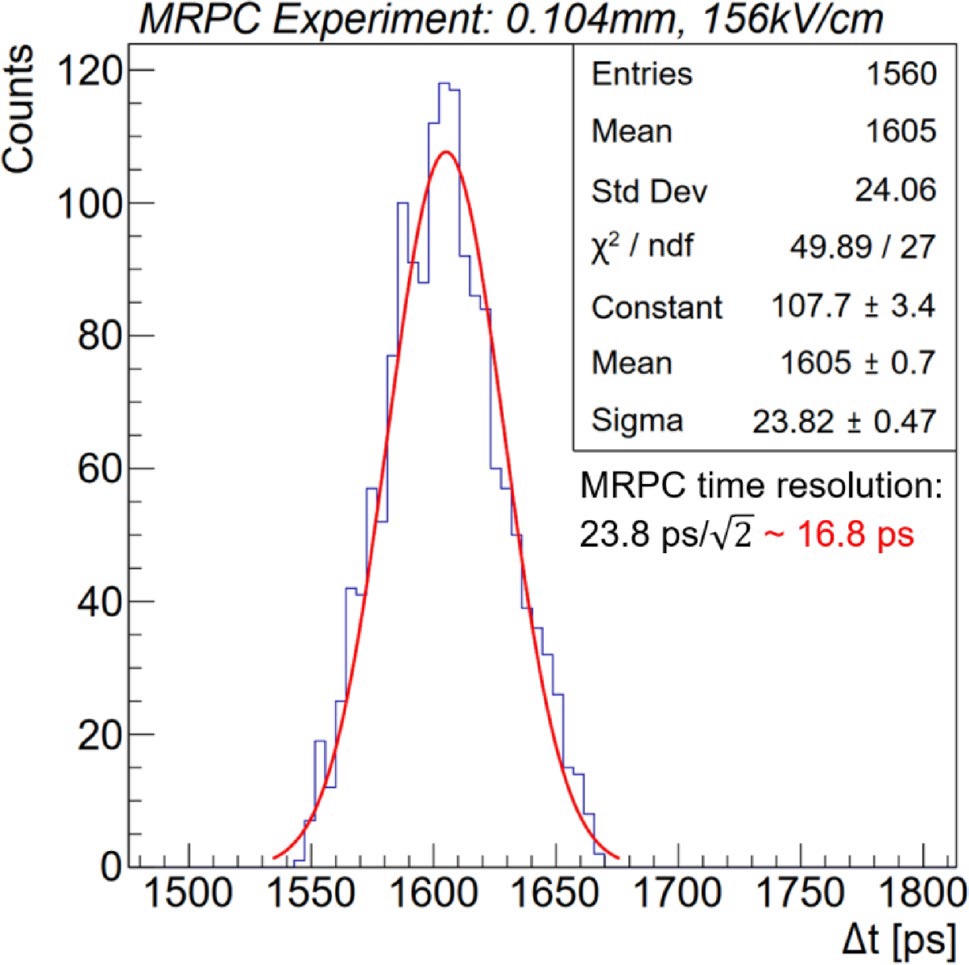}\vspace{-.5cm}
\caption{The intrinsic time resolution of MRPC has reached 16.8 $ps$, as indicated in the figure. It has 32 gas gaps and the width of gap is 0.104 mm.  }
\label{16psmrpc}
\end{figure}

The experiences of MRPC-TOF in STAR, CBM and CEE not only significantly promoted the particle detection technology, but also provided a powerful tool for many physics programs. In the next-generation nuclear and particle physics experiments based on high-luminosity accelerators, MRPC will continue to provide reliable technical options for particle identification and trigger, thanks to the new development where low-resistivity glass plates and high-speed waveform sampling technology \cite{electronic2013, electronic2019}, which enables MRPC time resolution better than 20 $ps$ (shown in Fig. \ref{16psmrpc} \cite{2019mrpc-tof} ) with high counting rate. In the mean time, modern technology such as machine learning and neural network are also studied to reconstruct the timing of MRPC \cite{2020NN}. Works never stop to improve the performance of MRPC to meet the requirement of future experiment such as new detector material, new fast electronics, new analysis method and eco-friendly working gas.  

%--==========================================================
%---===============================================
\section{Experimental Results and Discussions}
\label{Sec:data}
%--==========================================================
%---===============================================
\subsection{Charged Particle Spectra and Yields}
Relativistic heavy ion collisions experiments are designed for the search and study of the QGP. In head-on relativistic heavy ion collisions, two nuclei can be represented as two thin disks approaching each other at high speed because of the Lorentz contraction effect in the moving direction. During the initial stage of the collisions, the energy density is higher than the critical energy density from the lattice QCD calculation, thus the quarks and gluons will be deconfined from nucleons and form the QGP. The large cross section of interaction may lead to the thermalization of the QGP. In this stage, the high transverse momentum jets and heavy-flavor pair will be produced due to the large momentum transfer. After that, the QGP will expand and cool down and enter into the mixed-phase expansion. The chemical freeze-out point will be formed after the inelastic interaction stop. Here it means that the particle yields and ratios will not change. After the chemical freeze-out, the elastic interaction between hadrons will change the $p_T$ distribution of particles. The particles will freeze out finally from the system after the elastic interaction stop, which is the so-called kinetic freeze-out point. Studying the bulk properties of the system, such as the spectra, the yields ($dN/dy$), particle ratios and freeze-out properties will provide insight into the particle production mechanisms and the evolution of the QCD matter.  

\begin{figure*}[htb]
\centering 
\includegraphics[width=0.85\textwidth]{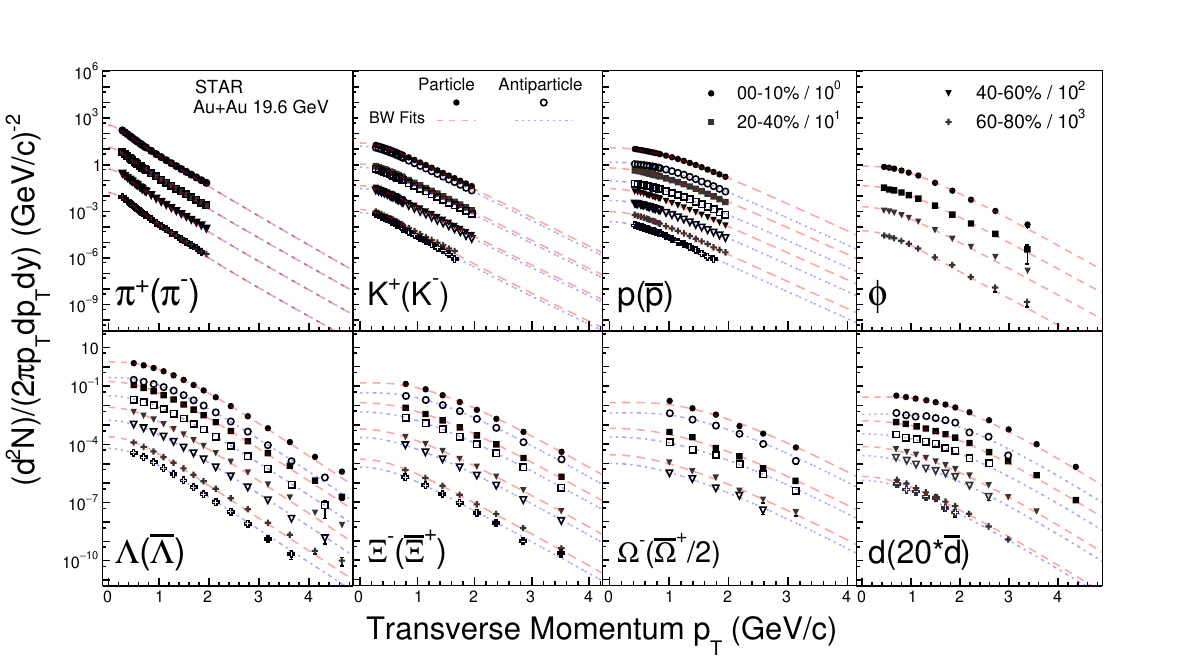}\vspace{-0.5cm}
\caption{Invariant yields of identified particles as a function of transverse momentum measured in various collision centralities in Au+Au collisions at $\sqrt{s_{NN}}=19.6$ GeV~\cite{STAR:2017sal,STAR:2019bjj,STAR:2015vvs,STAR:2019sjh}. Filled points represent particles and open points are the antiparticles, curves are fits on the data points by thermal blast wave distribution~\cite{Schnedermann:1993ws}.}
\label{spectra_fig}
\end{figure*}

For experimental observations, we first present the invariant yields of various particles as a function of transverse momentum $p_T$. 
Figure~\ref{spectra_fig} shows the invariant yields of pions ($\pi^{\pm}$), kaons ($K^{\pm}$), protons ($p$), anti-protons ($\bar{p}$), phi-mesons ($\phi$), Lambda baryons ($\Lambda$), anti-Lambda baryons ($\bar{\Lambda}$), Cascades ($\Xi^{-}$), anti-Cascades ($\overline{\Xi}^{+}$), Omegas ($\Omega^{-}$), anti-Omegas ($\overline{\Omega}^{+}$), deuterium ($d$), and anti-deuterium ($\bar{d}$). The results are shown for Au+Au collisions at $\sqrt{s_{\rm NN}}=19.6$ GeV in four collision centralities, the 0--10\%, 20--40\%, 40--60\%, and 60--80\%~\cite{STAR:2017sal,STAR:2015vvs,STAR:2019bjj,STAR:2019sjh}. The invariant yields show a decrease as a function of increasing $p_T$ and going from central to peripheral collisions. The curves represent the blast wave fits to the spectra~\cite{Schnedermann:1993ws}. The yields, $dN/dy$, are obtained by integrating these measured spectra and the fitting functions where the measurements are not available. 

\begin{figure*}[htb]
\centering  
\includegraphics[width=0.85\textwidth]{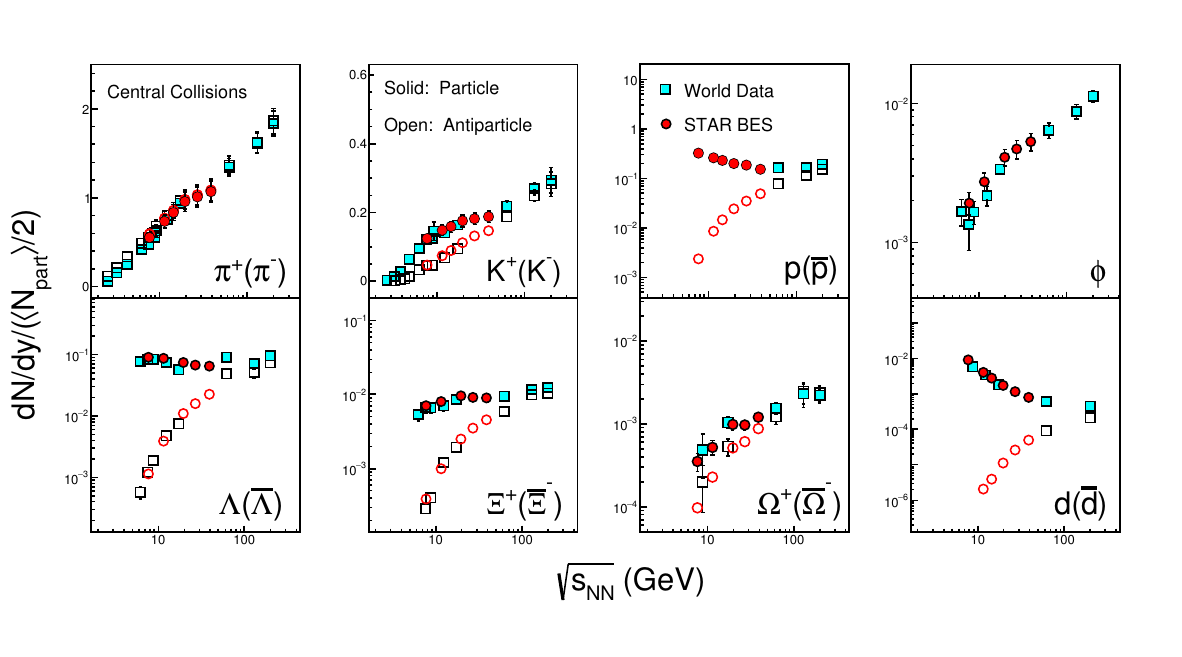}\vspace{-0.5cm}
  \caption{Energy dependence of particle yields $dN/dy$ in central collisions~\cite{STAR:2010vob}. Solid points represent particle and open points are the antiparticles. The data are normalized by the number of participants.}
\label{BES_yield}
\end{figure*}
Figure~\ref{BES_yield} shows the energy dependence of particle yields for 
$\pi^{\pm}$, $K^{\pm}$, $p$, $\bar{p}$, $\phi$, $\Lambda$, $\bar{\Lambda}$, $\Xi^{-}$, $\overline{\Xi}^{+}$, $\Omega^{-}$, $\overline{\Omega}^{+}$, $d$, and $\bar{d}$.
Results from STAR BES-I~\cite{STAR:2017sal,STAR:2015vvs,STAR:2019bjj,STAR:2019sjh} are compared with previously published STAR results at higher energies and other world experiments. We refer to the topical review for data collection~\cite{STAR:2010vob}. The yields of anti-baryons increase rapidly with the increasing collision energy, showing the increasing contribution of pair production. However the yields of baryons and $K^+$ show non-trivial energy dependence in BES energy range, indicating the interplay of baryon stopping/association production and pair production.

\begin{figure}[htb]
\centering  
\includegraphics[width=0.50\textwidth]{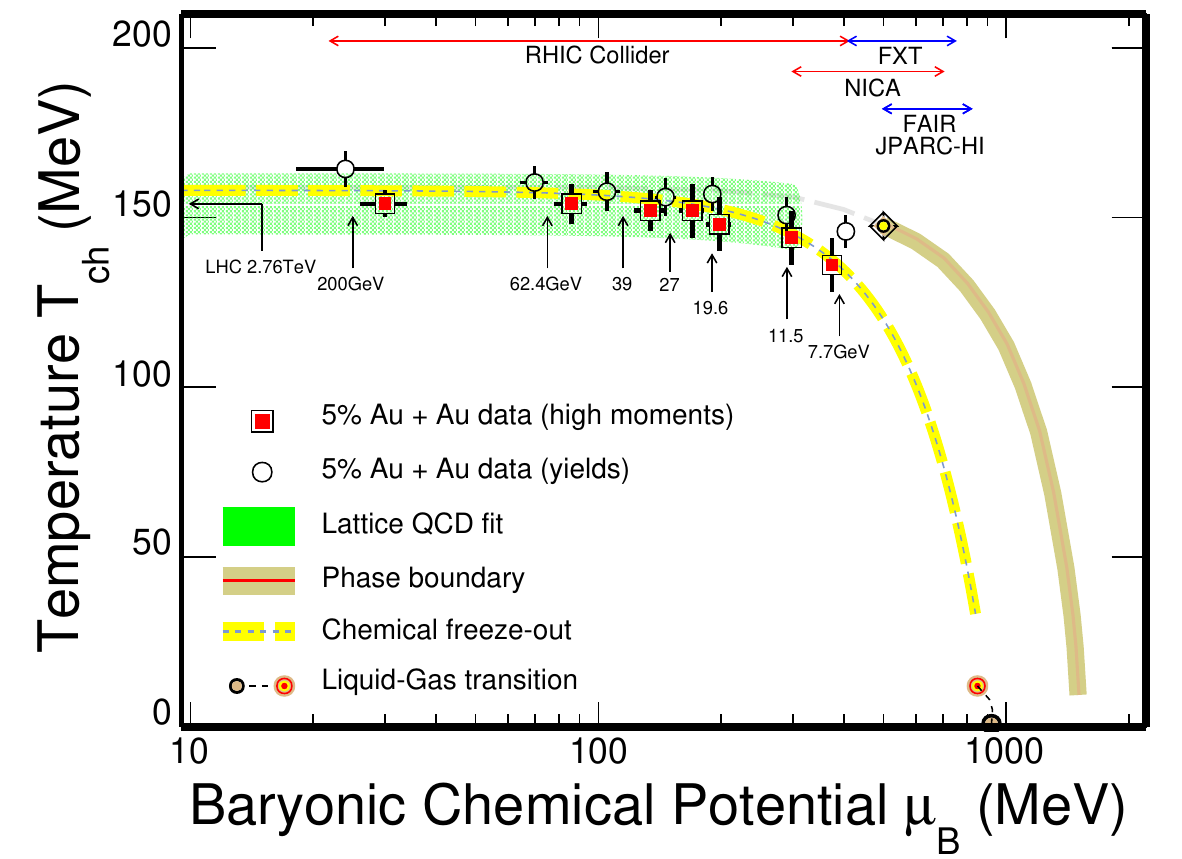}%\vspace{-0.5cm}
  \caption{
  A summary of the chemical freeze-out temperature $T_{\rm ch}(\mu_{\rm B})$ distribution~\cite{Fukushima:2020yzx}. Data points are from the 0-5\% central Au+Au collisions at STAR BES~\cite{STAR:2017sal,STAR:2020tga}. 
 }
 \label{BES_Tch}
\end{figure}
The hadron yields keep the footprint of hot and dense hadronic matter as the evolution of the collision system, presumably when the system underwent a crossover at phase transition~\cite{STAR:2017sal}. As observed experimentally, relative abundances of hadrons obey the thermal distribution at common $T$ and $\mu_{\rm B}$, so that the thermal fit can fix $T$ and $\mu_{\rm B}$~\cite{STAR:2017sal}. The temperatures of chemical freeze-out ($T_{\rm ch}$ for central Au+Au collisions at different collision energies are shown in Fig.~\ref{BES_Tch}. With the increasing energy, the $T_{\rm ch}$ increases and becomes constant at $\sim$160 MeV after $\sqrt{s_{\rm NN}}$= 11.5 GeV. The parameters extracted from net-proton higher moments~\cite{STAR:2020tga} are also presented in the figure. They are consistent with the results from hadron yield fit. The extracted parameters from BES data are consistent with results from lattice QCD calculation and thermal fit to global hadron yield data~\cite{Fukushima:2020yzx}. The coverage of the RHIC BES program, STAR fixed target program, and future experimental facilities, as shown in the figure, will yield more precise description of the QCD phase diagram.

One of the foundations of the BES program is the promise of a sweeping variation of the chemical potential across the QCD phase diagram by changing the beam energy of the heavy ion collisions. The chemical potential could be extracted empirically from the final-state particle distributions. It is an important subject itself how the baryons are shifted from target and projectile rapidity to the midrapidity. A puzzling feature of ultra-relativistic nucleus-nucleus collisions is the experimental observation of substantial baryon asymmetry in the central rapidity (mid-rapidity) region both at RHIC~\cite{Bearden:2003hx,STAR:2008med,STAR:2017sal} and at LHC energies ($\sqrt{s_{\rm NN}}=$900 GeV) ~\cite{ALICE:2010hjm, ALICEabbas2013mid}. Such a phenomenon is striking, as baryon number is strictly conserved, therefore, net-baryon number cannot be created in the system and must come from the colliding target and projectile. In a conventional picture, the valence quarks carry baryon quantum number in a nucleus. At sufficiently high energies one expects these valence quarks to pass through each other and end up far from mid-rapidity in the fragmentation regions~\cite{Kharzeev:1996sq}. RHIC BES program covers a wide range of baryon stopping of over an order of magnitude of net proton yields at midrapidity~\cite{STAR:2008med,STAR:2017sal}. 

\begin{figure}[htb]
%  \vspace{-0.5cm}
  \begin{center}
    \includegraphics[width=0.5\textwidth]{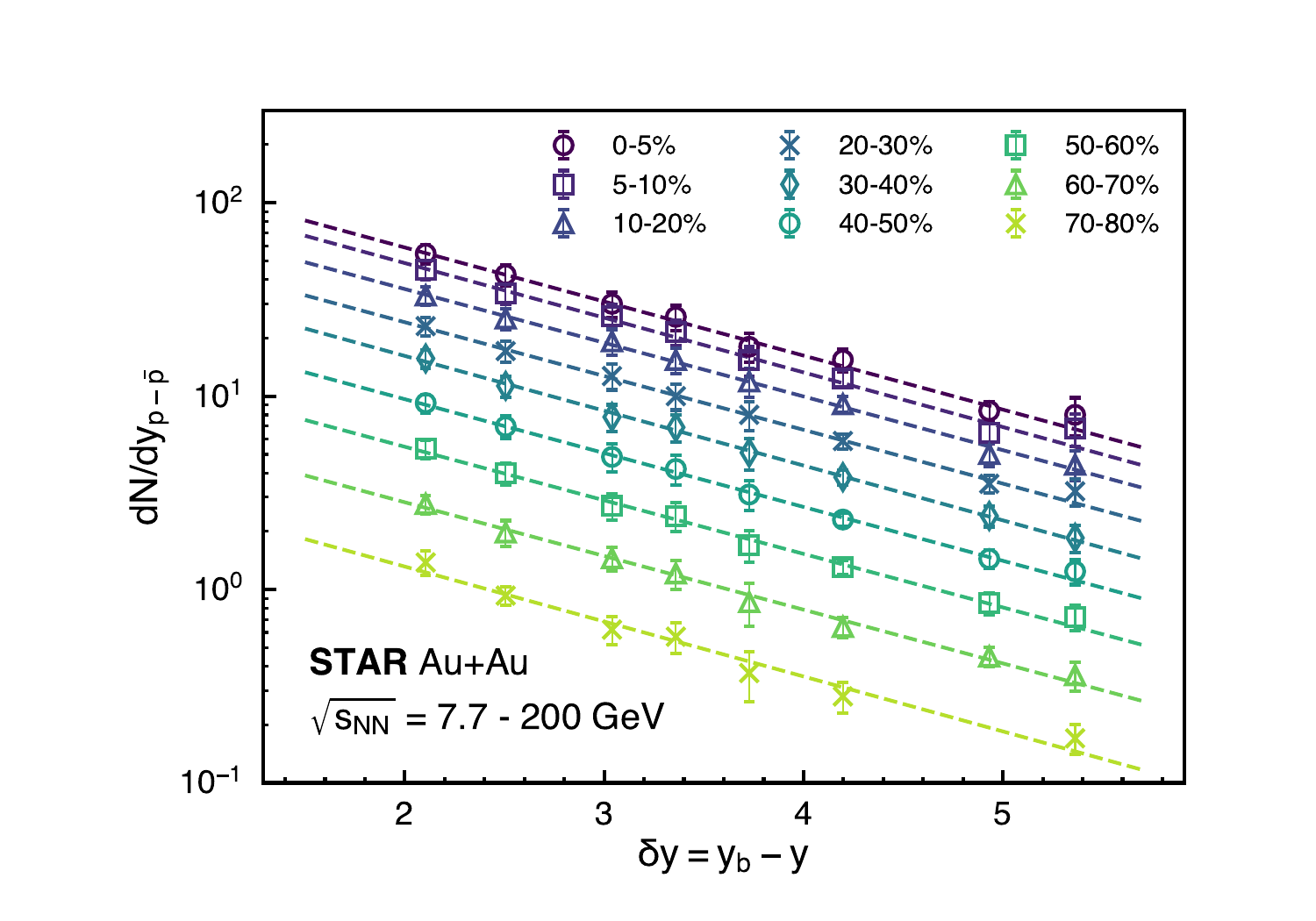}
 %   \vspace{-0.48cm}
  \end{center}
  \caption{\label{fig_baryon} Exponential dependence of midrapidity ($y\approx0$) baryon density per participant pair in heavy ion collisions with $Y_{\rm beam}$ which is equal to the rapidity difference between beam and detector midrapidity ($\delta y$)~\cite{Lewis:2022arg}. An exponential fit function of $A\times\exp{(-\alpha_B\delta y)}$ is also included. Figure from Ref.~\cite{Lewis:2022arg}.}
 %\vspace{-0.25cm}
\end{figure}

Figure~\ref{fig_baryon} presents the net-proton yields at midrapidity in Au+Au collisions at $\sqrt{s_{\rm NN}}$= 7.7 to 200 GeV~\cite{Lewis:2022arg}. We see that for all centralities in heavy ion collisions the midrapidity net-baryon density follows an exponential distribution with the variable $\delta y=Y_{\rm beam}-Y_{\rm cm}$, where $Y_{\rm beam}$ is the beam rapidity and $Y_{\rm cm}$ is the center-of-mass rapidity. This variable $\delta y$ can be referred to as the ``rapidity loss" which for midrapidity protons produced in a collider experiment is equal to beam rapidity: $\delta y=Y_{\rm beam}$ as $Y_{\rm cm}=0$. 
A single collision energy therefore gives rise to a single point on Fig.~\ref{fig_baryon}. The data points at each centrality can be fitted with the exponential function of $A\exp{(-\alpha_B\delta y)}$. 
The baryon stopping is often characterized by the average rapidity loss~\cite{BRAHMS:2009wlg}, which shows the complicated beam energy dependence and is usually skewed by the large proton yields close to beam rapidity. It was concluded~\cite{BRAHMS:2009wlg} that the ``rapidity loss" of projectile baryons at RHIC breaks the linear scaling observed at lower energies. Another way of characterizing the baryon stopping is to use the $\bar{p}/p$ ratio~\cite{STAR:2001rbj,ALICE:2010hjm, ALICEabbas2013mid}. Both pair production and baryon stopping contribute to this ratio.
Most of the dynamic models of the heavy ion collisions parametrize the baryon stopping to reproduce the experimental data while at the fundamental level, there is still a lack of understanding of how the baryons are stopped. 
A recent modeling of heavy ion collisions indicates that the inclusion of the baryon junction is essential for describing net-proton density at RHIC~\cite{Shen:2022oyg}. Clearly some of the earlier implementations of baryon junctions, which attempted to match the earlier experimental data with certain parameter tunes, do not reproduce the experimental results presented in the Fig.~\ref{fig_baryon}.

%--==========================================================
%---===============================================
\subsection{Strangeness Production}

On the physics of QCD phase boundary and searching for the onset of deconfinement, strange hadrons are excellent probes. Strangeness enhancement in heavy ion with respect to hadron collisions has long been suggested as a signature of quark-gluon plasma~\cite{Shor:1984ui,Koch:1986ud,Rafelski:1982pu}. Thus motivates it a popular measurement in many experiments at different accelerator facilities. In general, the yields of strange hadrons in nuclear collisions are close to those expected from statistical models~\cite{CE_Cleymans:2006,Andronic_2018Naure,Andronic:2005yp}. The precise measurement of these yields in the phase-I of RHIC BES experiment has lead to a better understanding of strange quark production mechanisms in nuclear collisions and a more reliable extraction of the chemical freeze-out parameters ~\cite{STAR:2017sal} as shown in Fig.~\ref{BES_Tch}. In the higher beam energies, the formation of a thermalized system is expected and the strangeness is abundant produced. However, at lower beam energies, the strangeness is less produced which requires special attention and the local treatment on the canonical ensemble is needed. This particular part will be discussed around Fig.~\ref{3GeV_phi2K}, while $\phi(1020)$ meson with zero net strangeness number (S=0) offer a unique test to scrutinize thermodynamic properties of strange quarks in the hot and dense QCD environment~\cite{Chen:2006ub}.

On the other hand, the precise measurement of the strange hadron production at different $p_{\rm T}$ ranges and centralities in heavy ion collisions are also crucial for the better understanding of the production mechanism and the medium properties created in the system. At high $p_{\rm T}$, it has been observed that the nuclear modification factor $R_{CP}$ of various particles at top RHIC energy is much less than unity~\cite{STAR:2003fka,STAR:2006egk,STAR:2007mum}, indicating a significant energy loss of the scattered partons in the dense nuclear matter, known as ``jet quenching''~\cite{STAR:2017ieb}. At intermediate $p_{\rm T}$, the baryon to meson ratios, $p/\pi$ and $\Lambda/K_{s}^{0}$ are found to be larger than unity and much higher than those observed in the peripheral A+A collisions and in the elementary collisions. This baryon to meson ratio enhancement can be explained by the recombination/coalescence models which requires constituent quarks in the partonic medium to coalesce into hadrons, or soft and hard partons to recombine into hadrons~\cite{Hwa:2006vb,Fries:2003vb,Greco:2003xt}. Thus the measurements of $R_{\rm CP}$ and baryon to meson ratios of strange hadrons are one of the corner-stone pieces of evidence for the formation of the strongly interacting QGP medium. The precise measurement of these variables in heavy ion collisions at lower beam energies can potentially reveal the medium properties at finite $\mu_{B}$, and help to locate the energy point at which the onset of the deconfinement happens.

Beside light hadrons, Fig.~\ref{BES_yield} also shows the energy dependence of strange particle yields at midrapidity for $K^{\pm}$, $\phi$, $\Lambda(\bar{\Lambda})$, $\Xi^{-} (\overline{\Xi}^{+})$ and $\Omega^{-} (\overline{\Omega}^{+})$ from central heavy ion collisions. Results from STAR BES-I~\cite{STAR:2015vvs,STAR:2019bjj} are compared with previously published STAR results at higher energies and other corresponding world data including experiments at AGS and CERN~\cite{Ahmad:1991nv,Ahmad:1998sg,E917:2001eko,Albergo:2002tn,STAR:2003jis,STAR:2002fhx,STAR:2006egk,STAR:2011fbd,NA57:2006aux,NA57:2004nxc,NA49:2008ysv}. The yields $dN/dy/\langle N_{part}/2\rangle $ of the anti-hyperons ($\bar{\Lambda}, \overline{\Xi}^{+}, \overline{\Omega}^{+}$) and $\phi$ meson increase rapidly with energy increase, while there seems to be a non-trivial energy dependence for the $\Lambda$, $\Xi^{-}$ and $\Omega^{-}$ yields. The $\Xi^{-}$ and $\Omega^{-}$ yield first increases with energy from 7.7 to 19.6 GeV, then remains almost constant up to energies around 39 GeV, then rising again toward higher energies. The $\Lambda$ yield first decreases from 7.7 to 39 GeV, then increase toward higher energies. The $\Lambda$ behavior similarly as the trend for proton in these measured energy regions~\cite{STAR:2017sal}, which reflecting a significant increase in baryon density at lower collision energy. The observed $\Lambda$ behavior also can be the interplay of the pair production of $\Lambda$-$\bar{\Lambda}$ and the associated production of $\Lambda$ along with kaons, the former process increase strongly with the increasing collision energy while the later one strongly increases with increasing net-baryon density.

\begin{figure}[htb]
\centering  
\includegraphics[width=0.45\textwidth]{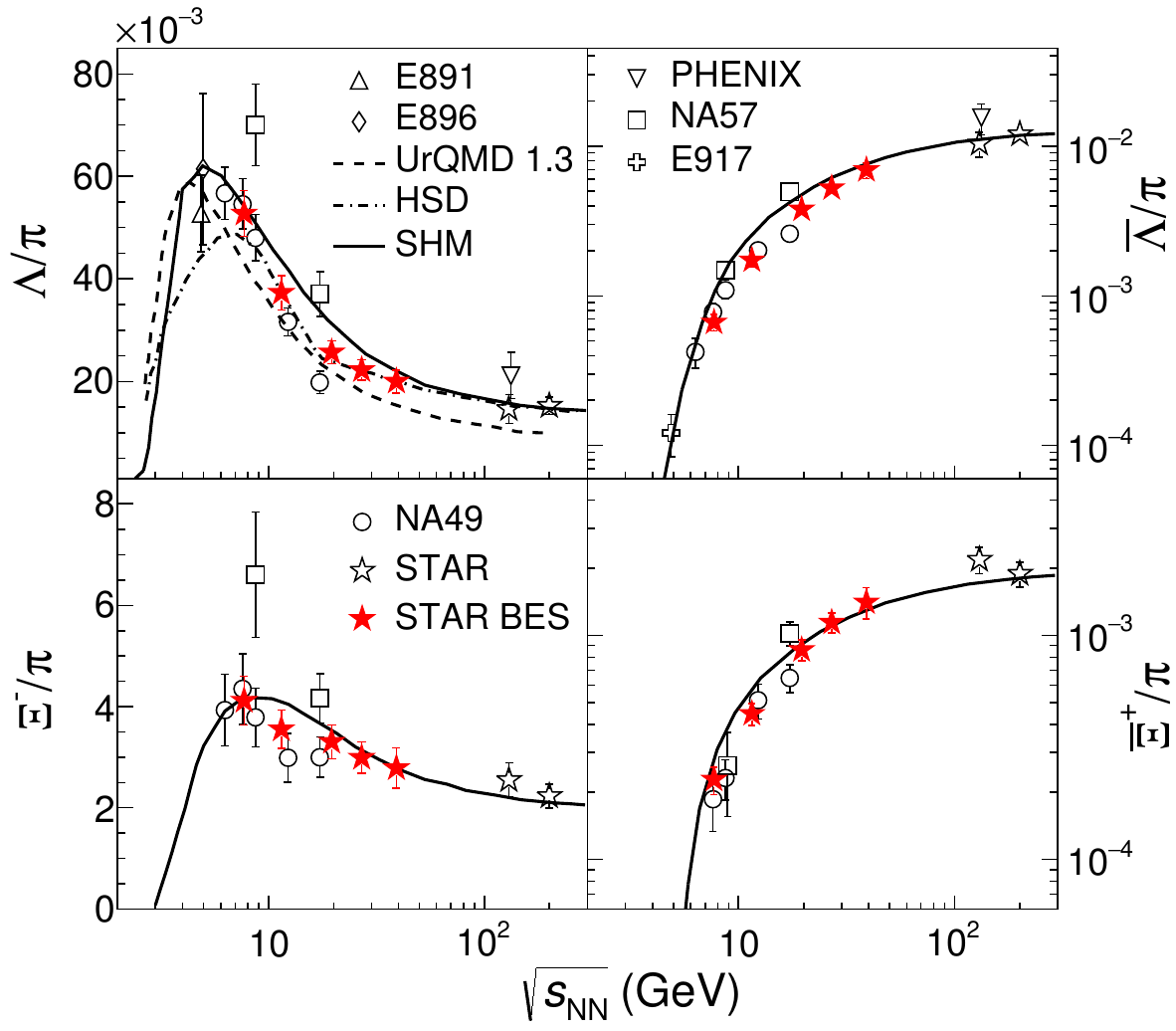}
\vspace{-0.65cm}
  \caption{Energy dependence of $\Lambda$, $\bar{\Lambda}$, $\Xi^{-}$ and $\overline{\Xi}^{+}$ midrapidity yields ratio to that of pions (1.5($\pi^{+}+\pi^{-}$)) in central Au+Au collisions from STAR BES compared to the existing data from various other experiments.}
\label{BES_strange_ratio}
\end{figure}

Figure~\ref{BES_strange_ratio} shows the energy dependence of $\Lambda(\bar{\Lambda})$ and $\Xi^{-}(\overline{\Xi}^{+})$ midrapidity yields ratio to that of pions in central Au+Au collisions from RHIC STAR BES, as well as the existing data from various experiments~\cite{STAR:2019bjj,Ahmad:1991nv,Ahmad:1998sg,E917:2001eko,Albergo:2002tn,NA49:2008ysv,NA49:2002pzu,NA49:2007stj} and the calculations from hadronic transport models (UrQMD 1.3, HSD Hadron-String Dynamics) and statistical hadron gas model (SHM)~\cite{Bass:1998ca,Bleicher:1999xi,Andronic:2005yp}. The STAR BES data are in good agreement with the trend of the existing experimental data. Though the hadronic models (UrQMD 1.3 and HSD) seem to reproduce the $\Lambda/\pi$ data, indicating that the hadronic rescatterings might play an important role in hyperon production in heavy ion collisions at this energy range, however the default UrQMD (v1.3) fails in reproducing the $\Xi/\pi$ ratio due to a smaller $\Xi$ yield in the model. On the other hand, the SHM model predictions agree well with data across the whole energy range from AGS to top RHIC energies. The SHM model used here is based on a grand-canonical ensemble and assumes chemical equilibrium. The energy dependence of the parameters $T_{ch}$ and $\mu_{B}$ in the model were obtained with a smooth parametrization of the original fitting parameters to the midrapidity particle ratios from heavy ion experiments at SPS and RHIC. Both the $\Lambda/\pi$ and $\Xi^{-}/\pi$ ratios show a maximum at $\sim$ 8 GeV, which seems to be consistent with the picture of maximum net-baryon density at freeze-out at this collision energy.

\begin{figure}[htb]
\centering  
\includegraphics[width=0.45\textwidth]{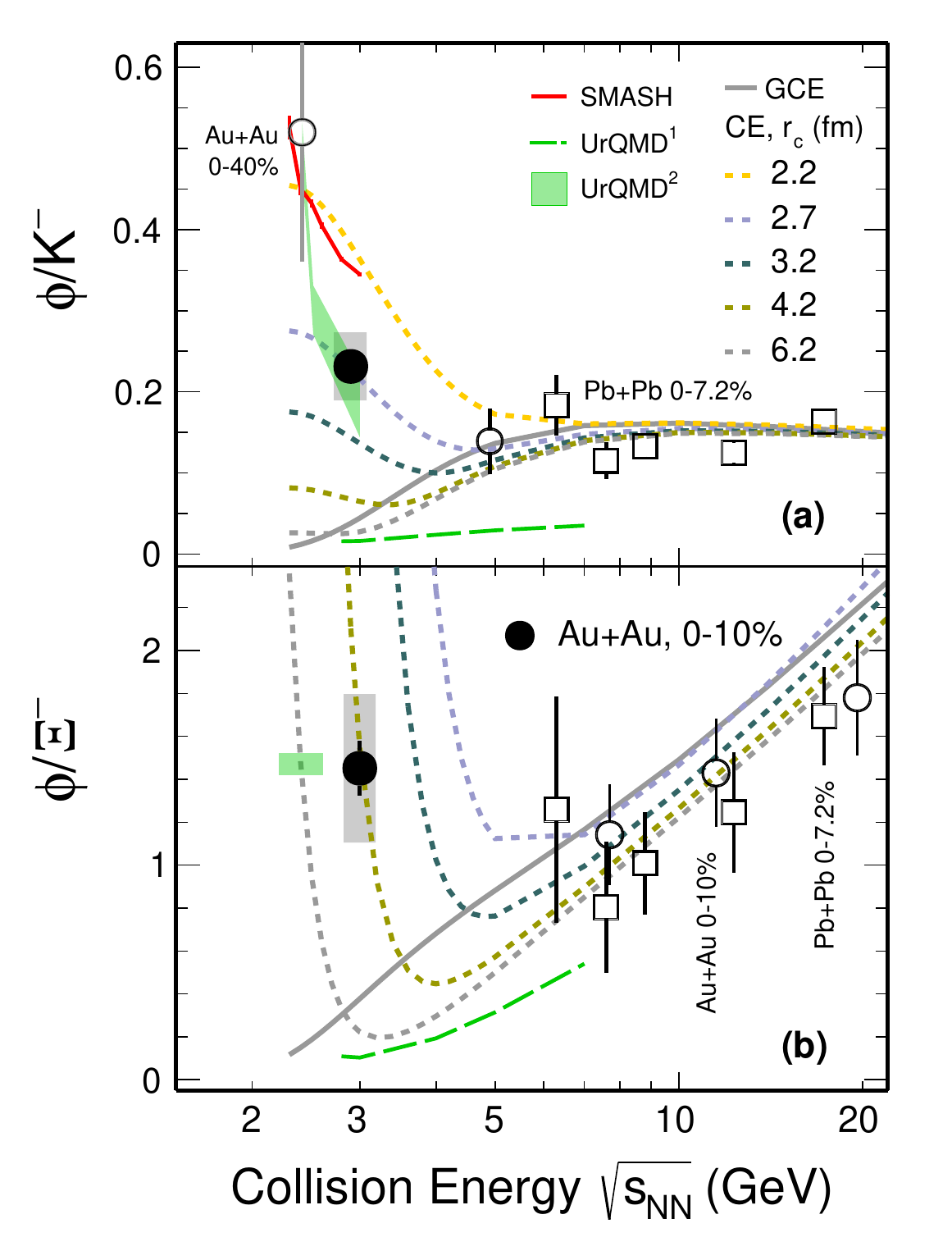}
\vspace{-0.65cm}
  \caption{$\phi/K^-$ (a) and $\phi/\Xi^-$ (b) ratio in central heavy ion collisions as a function of collision energy, $\sqrt{s_{\rm NN}}$, compared to various thermal and transport model calculations.}
\label{3GeV_phi2K}
\end{figure}

Thermodynamic properties of strange quarks play an important role in understanding the QCD matter Equation of State (EOS) at high-density regions. In the statistical thermal models, grand canonical ensemble (GCE) and canonical ensemble (CE) statistical descriptions applied differently to conserve strangeness number in order to compute the final state particle yields. It has been argued that at lower energies, strangeness number needs to be conserved locally on an event-by-event basis described by the CE, which leads to a reduction in the yields of hadrons with non-zero strangeness number (``Canonical Suppression"), but not for the $\phi(1020)$ meson with zero net strangeness number (S=0)~\cite{Redlich:2001kb}. Fig.~\ref{3GeV_phi2K} shows the measurements of $\phi/K^-$ and $\phi/\Xi^-$ ratio in the central heavy ion collisions as a function of collision energy~\cite{STAR:2021hyx,STAR:2019bjj,NA49:2008goy,HADES:2017jgz} compared to various thermal and transport model calculations~\cite{Bass:1998ca,Bleicher:1999xi,Steinheimer:2015sha}. As shown in the plot, both GCE and CE models are able to describe the measured ratios at $\sqrt{s_{\rm NN}}$ greater than 7.7\,GeV, while clearly the GCE fails when the collision energies approaching the production threshold (2.89 GeV for $\phi$ and 3.25 GeV for $\Xi^-$). The measurements favor the CE calculations with a small strangeness correlation length ($r_c$) while more detailed investigation requires more precise and differential data.

Beside thermal model, transport model calculations from modified UrQMD with high mass strange resonances can reasonably reproduce the data in Fig.~\ref{3GeV_phi2K} implying that the feed down is relevant~\cite{Steinheimer:2015sha,Shao:2019xpj}. In heavy ion collisions, the near/sub-threshold production of multi-strange hadrons can be achieved from the multiple collisions of nucleons, produced particles, and short-lived resonances. Meanwhile the particle production below its free nucleon-nucleon (NN) threshold is expected to be sensitive to the stiffness of the nuclear EoS at high density~\cite{Yong:2021npa}.

\begin{figure}[htb]
\centering  
\includegraphics[width=0.4\textwidth]{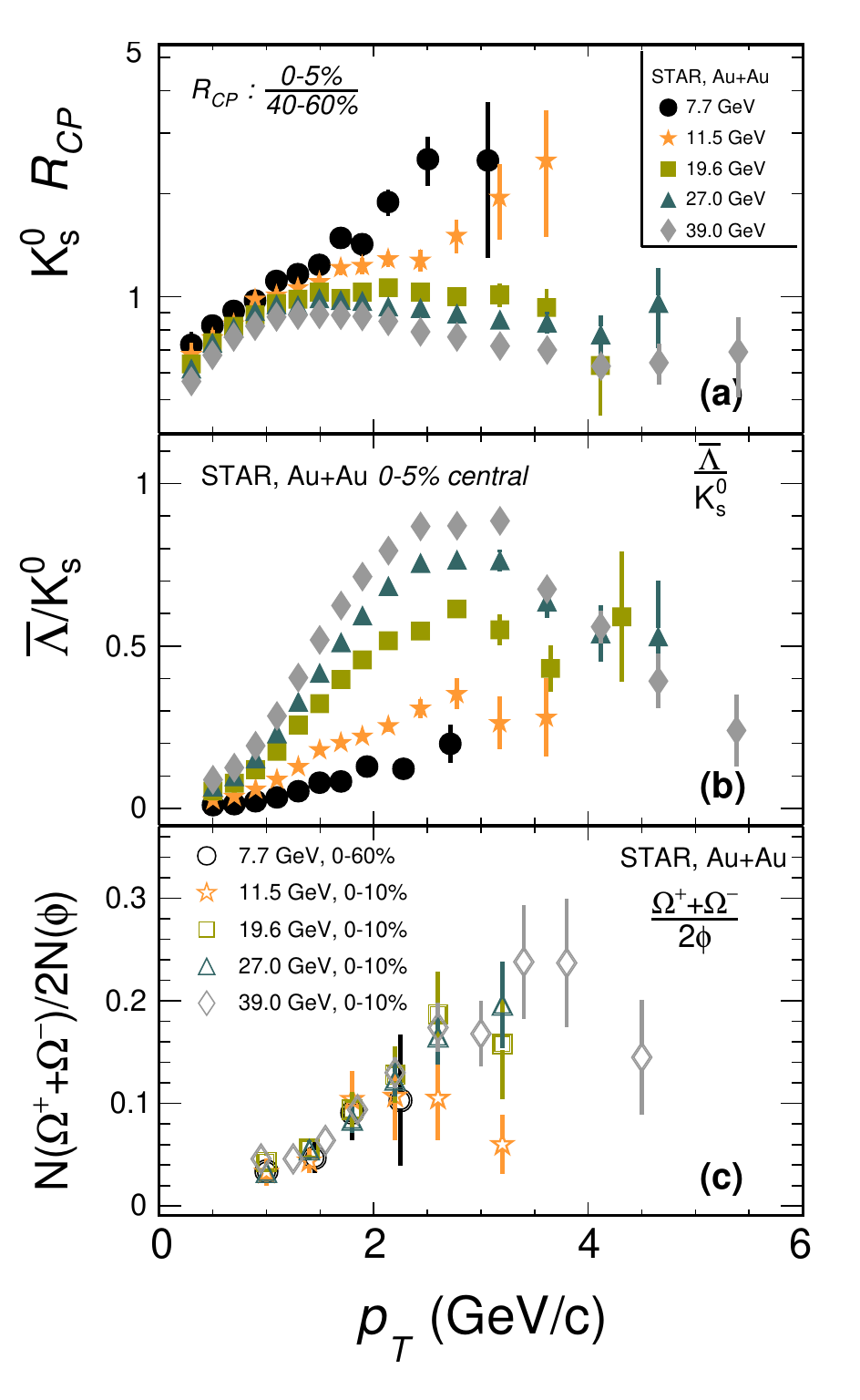}
\vspace{-0.65cm}
  \caption{ (a) $K_{S}^{0}$ nuclear modification factor, $R_{\rm CP}$, at midrapidity in Au+Au collisions at STAR BES from 7.7 to 39 GeV. (b) $\bar{\Lambda}/K_{S}^{0}$ ratios as a function of $p_{\rm T}$ in central Au+Au collisions at STAR BES. (c) baryon-to-meson ratio, $\Omega/\phi$, as a function of $p_{\rm T}$ in central Au+Au collisions from STAR BES.}
\label{BES_Ks0LambdaOmega_ratio}
\end{figure}

Figure~\ref{BES_Ks0LambdaOmega_ratio} panel (a) shows the nuclear modification factor, $R_{\rm CP}$, of $K_{S}^{0}$, in Au+Au collisions at STAR BES from 7.7 to 39 GeV~\cite{STAR:2019bjj}. For $p_{\rm T}$ $\approx$ 4 GeV/c, the $K_{S}^{0}$ $R_{\rm CP}$ is below unity at $\sqrt{s_{\rm NN}}$ = 39 GeV, which is similar to the observation at top RHIC energy though the lowest $R_{\rm CP}$ value is larger here. Then the $K_{S}^{0}$ $R_{\rm CP}$ at $p_{\rm T}$ $>$ 2 GeV/c keeps increasing with decreasing collision energies, indicating that the partonic energy loss effect becomes less important. And eventually, the $K_{S}^{0}$ $R_{\rm CP}$ shape become differently at $\sqrt{s_{\rm NN}}$ = 11.5 and 7.7 GeV although the maximum accessible $p_{\rm T}$ is smaller at these two energies. This suggest that the cold nuclear matter effect (Cronin effect) starts to take over from these energies and enhances all the hadron yields at intermediate $p_{\rm T}$ (to $\approx$ 3.5 GeV/c). Similarly to the observation for identified charged hadrons, the energy evolution of strange hadron $R_{\rm CP}$ reflects the decreasing partonic effects with decreasing beam energies~\cite{STAR:2017ieb}.

Figure~\ref{BES_Ks0LambdaOmega_ratio} panel (b) shows the $\bar{\Lambda}/K_{S}^{0}$ ratios as a function of $p_{\rm T}$ in central Au+Au collisions at STAR BES from $\sqrt{s_{\rm NN}}$ = 7.7 to 39 GeV~\cite{STAR:2019bjj}.
The $\bar{\Lambda}$ is chosen because it is newly produced baryons in the baryon-rich medium created in the lower BES energies.
The enhancement of baryon-to-meson ratios at intermediate $p_{\rm T}$ in central A+A collisions compared to peripheral A+A or p+p collisions at the same energy is observed for the energies $\sqrt{s_{\rm NN}}$ $\geq$ 19.6 GeV.
The maximum value of $\bar{\Lambda}/K_{S}^{0}$ reach a maximum value of unity at $p_T$ $\approx$ 2.5 GeV/c in the most central collisions, while in the peripheral collisions, the maximum value is significantly lower, only about 0.3 – 0.5 which is not shown in the plot. The enhancement of baryon-to-meson ratio in central collisions in these energies was interpreted as a consequence of hadron formation through parton recombination and parton collectivity. Therefore, the baryon-to-meson ratios are expected to be sensitive to the parton dynamics of the collision system. But unfortunately, for the $\sqrt{s_{\rm NN}}$ $\leq$ 11.5 GeV, the maximum $p_{\rm T}$ reach and the statistics in different centralities are limited, hence whether baryon-to-meson enhancement still persists at these low energies remains unclear with the current data.

Figure~\ref{BES_Ks0LambdaOmega_ratio} panel (c) shows the baryon-to-meson ratio, $\Omega/\phi$, as a function of $p_{\rm T}$ in central Au+Au collisions from $\sqrt{s_{\rm NN}}$ = 11.5 to 39 GeV and $\sqrt{s_{\rm NN}}$ = 7.7 GeV 0-60\% centrality~\cite{STAR:2019bjj}. For energies $\sqrt{s_{\rm NN}}$ $\geq$ 19.6 GeV, the measured data follow closely with each other and also with the previous measurement from 200 GeV, which is consistent with a picture of coalescence and recombination dynamics over a broad $p_{\rm T}$ range of 1–4 GeV/c~\cite{STAR:2015vvs}. The ratios at $\sqrt{s_{\rm NN}}$ $\leq$ 11.5 GeV seem to deviate from the trend observed at higher beam energies. In particular, the ratios at 11.5 GeV appear to turn down around $p_{\rm T}$ of 2 GeV/c while those at higher beam energies such as 39 GeV peak at $p_{\rm T}$ of 3 GeV/c or above. Since the $\Omega$ and $\phi$ particles have small hadronic rescattering cross sections, the change in these $\Omega/\phi$ ratios may indicate a significant change in the hadron formation dynamics and/or on strange quark $p_{\rm T}$ distribution at the lower energies.

%--==========================================================
%---===============================================
\subsection{Collectivity}

Collective observables including radial flow and anisotropic flow are powerful tools for extracting parameters of the EOS and understanding the property of the medium created in high-energy nuclear collisions~\cite{Bzdak:2019pkr,Ma23flow,Ma24flow}.  In this session, the energy dependence of $v_1, v_2$, its scaling and EOS parameters will be discussed.

\begin{figure}[htb]
\centering  
\includegraphics[width=0.5\textwidth]{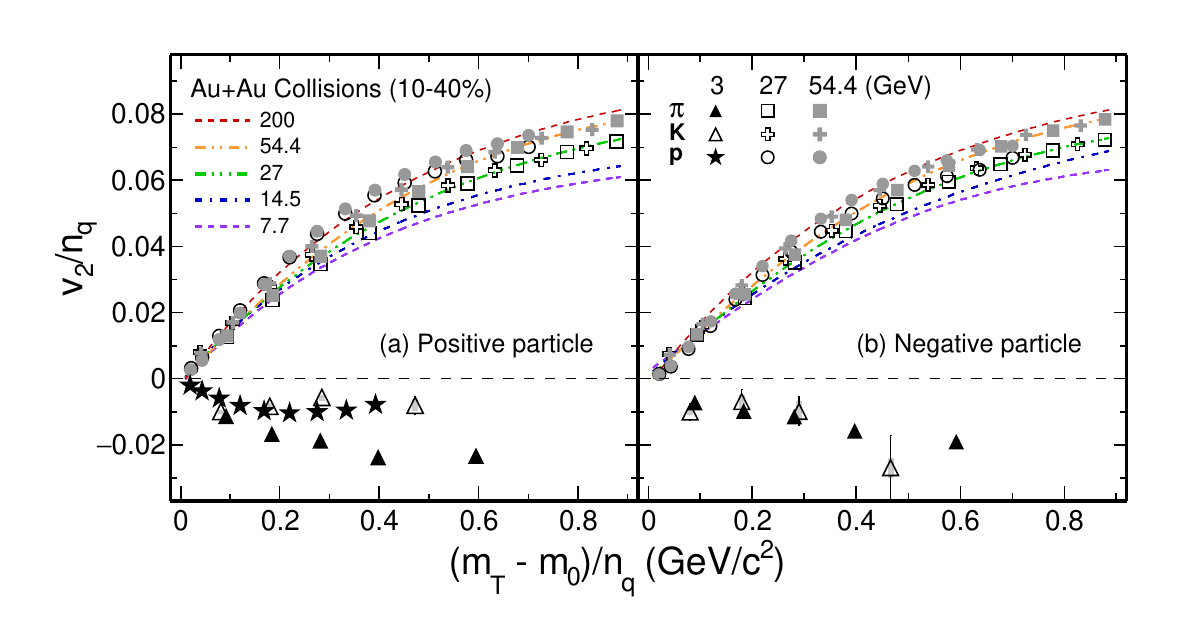}
\vspace{-0.65cm}
\caption{Number of constituent quarks ($n_q$) scaled elliptic flow $v_2/n_q$ is shown as a function of scaled transverse kinetic energy $(m_T-m_0)/n_q$ for pions, kaons, and protons from Au+Au collisions in 10-40\% centrality at $\sqrt{s_{\rm NN}}$ = 3, 27, and 54.4 GeV for positively charged particles (left panel) and negatively charged particles (right panel). Colored dashed lines represent the scaling fit to data from Au+Au collisions at 7.7, 14.5, 27, 54.4, and 200 GeV from the STAR experiment at RHIC ~\cite{Adamczyk:2013gv,Adamczyk:2015fum,Abelev:2008ae}. Statistical and systematic uncertainties are shown as bars and gray bands, respectively.}
\label{3GeV_ncq}
\end{figure}

The elliptic flow scaled by the number of constituent quarks (NCQ), $v_2/n_q$, for the copiously produced hadrons $\pi^{\pm}$ (squares), $K^{\pm}$ (crosses), $p$ and $\bar{p}$ (circles), is shown as a function of the scaled transverse kinetic energy $(m_T-m_0)/n_q$ in Fig.\ref{3GeV_ncq}. The data are from 10-40\% mid-central Au+Au collisions at RHIC. Data points from collisions at 27 and 54.4 GeV are shown as open and closed symbols, respectively. The colored dashed lines, also displayed in the figure, represent the scaling fit to data for pions, kaons, and protons in Au+Au collisions at 7.7, 14.5, 27, 54.4, and 200 GeV for both positively and negatively charged particles \cite{Dong:2004ve,Adamczyk:2013gw}.
Although the overall quark number scaling is evident, it has been observed that the best scaling is reached in the RHIC top energy $\sqrt{s_{NN}}$ = 200 GeV collisions~\cite{Adamczyk:2015ukd}. As the collision energy decreases, the scaling deteriorates. Particles and antiparticles are no longer consistent with the single-particle NCQ scaling due to the mixture of the transported and produced quarks~\cite{Adamczyk:2013gw}. More detailed discussions on the effects of transported quarks on collectivity can be found in Refs.~\cite{Adamczyk:2017nxg,Dunlop:2011cf}. As one of the important evidence for the QGP formation in high energy collisions at RHIC, the observed NCQ scaling originates from partonic collectivity~\cite{Molnar:2003ff,Adamczyk:2015ukd,Adamczyk:2017xur}.
Interestingly, in the analysis of the elliptic flow of light nuclei in low- and intermediate-energy nuclear reactions, a similar scaling law has been found, i.e., the elliptic flow of light nuclei is scaled according to the number of component nucleons\cite{Yan006}. Inspired by literature~\cite{Yan006}, the STAR experiment of relativistic heavy ion collisions \cite{STAR:2016ydv} also confirms the nucleon number scaling rate of the elliptic flow of light nuclei, i.e., it fulfills the theoretical prediction of Ref.~\cite{Yan006}. The similarity between the NCQ scaling  of elliptic flows and the nucleon-number scaling law of light nuclei lies in the merger mechanism of hadron formation or nucleosynthesis, while the difference lies in the difference of whether the merger is at the quark level or the nucleon level.  
On the other hand, the LHC-ALICE Collaboration reported measurements of higher-order anisotropic flows \cite{ALICE:2011ab}, which for the first time experimentally gave measurements of the triangular flow $v_3$. Theoretically, the NCQ scaling of the higher order collective flow were generalized in Ref. \cite{HanLX} and confirmed in experimental measurement \cite{STAR:2022ncy}, which can also be regarded as a further probe of the quark gluon plasma. 

At low energy, in $\sqrt{s_{\rm NN}}$ = 3.0 GeV Au+Au collisions, a total different scaling behavior is evident, as shown in Fig. \ref{3GeV_ncq}. Opposite to that observed in high-energy collisions, all $v_2$ values are negative, which is a characteristic of nuclear shadowing in such non-central collisions. There is no sign of NCQ scaling at this low energy \cite{Tian09}. These results clearly indicate different properties for the matter produced. With baryonic mean field, hadronic transport model calculations from JAM~\cite{Nara:1999dz} and UrQMD~\cite{Bass:1998ca,Bleicher:1999xi} reproduce the observed negative values of $v_2$ for protons as well as $\Lambda$s. In other words, in the Au+Au collisions at $\sqrt{s_{\rm NN}}$ = 3 GeV, partonic interactions no longer dominate, and baryonic scatterings take over, indicating that predominantly hadronic matter is created in such low-energy collisions.

%--==========================================================
\begin{figure}[htb]
\centering  
\includegraphics[width=0.5\textwidth]{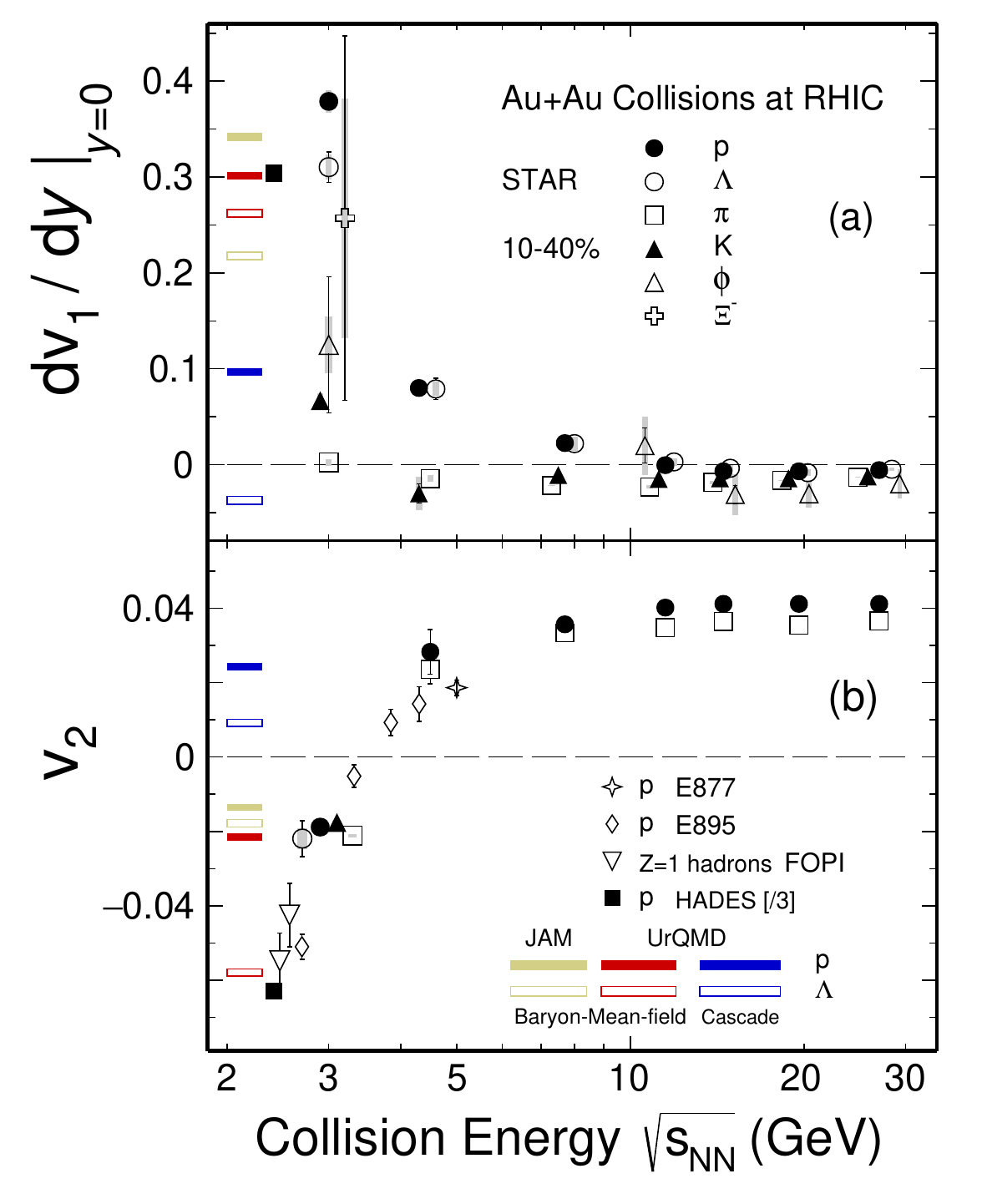}
\vspace{-0.65cm}
\caption{Collision energy dependence of directed flow slope $dv_1/dy|_{y=0}$ for $p$, $\Lambda$, charged $\pi$s, and kaons (including $K^{\pm}$ and $K^0_S$), $\phi$, and $\Xi^{-}$ (top panel). The bottom panel shows elliptic flow $v_2$ for protons and $\pi$s from heavy ion collisions. Statistical and systematic uncertainties are shown as bars and gray bands, respectively. The JAM and UrQMD results are displayed as colored bands: golden, red, and blue bands represent JAM mean-field, UrQMD mean-field, and cascade mode, respectively.}
\label{3GeV_eng}
\end{figure}
Now we turn to the $p_T$-integrated results and examine $v_1$ and $v_2$ together. The collision energy dependence of directed and elliptic flow is summarized in Fig.~\ref{3GeV_eng}, where panel (a) shows the slope of the $p_{\rm T}$-integrated directed flow at midrapidity, $dv_{1}/dy|{y = 0}$, for $\pi$, $K$, $p$, $\Lambda$, and multi-strange hadrons $\phi$ and $\Xi^{-}$ from Au+Au collisions in the 10-40\% centrality interval. Here, $K$ and $\pi$ represent the combined results of $K^{\pm}$ and $K_{S}^{0}$, and $\pi^{\pm}$, respectively. Panel (b) displays the $p_{\rm T}$-integrated $v_2$ at midrapidity for $\pi$, $K$, $p$, and $\Lambda$ as open squares, filled triangles, filled circles, and open circles, respectively.
Due to partonic collectivity in Au+Au collisions at high energy~\cite{Snellings:1999bt}, all observed $v_1$ slopes and $v_2$ at midrapidity are found to be negative and positive, respectively, which is opposite to what is observed at low energy. This can be seen in the results from the 3.0 GeV Au+Au collisions shown in Fig.~\ref{3GeV_eng}. The early strong partonic expansion leads to positive $v_2$ with NCQ scaling in high-energy collisions, whereas at 3 GeV, both weaker pressure gradients and the shadowing of the spectators result in negative $v_2$ values, where the scaling is absent.
Results from calculations using the hadronic transport models JAM and UrQMD are also shown as colored bands in the figure. By including the baryonic mean-field, both the JAM and UrQMD models reproduced the trends for both $dv_{1}/dy|_{y = 0}$ and $v_2$ for baryons, including protons and $\Lambda$. The consistency of the transport models (JAM and UrQMD) with the baryonic mean-field for all measured baryons implies that the dominant degrees of freedom at a collision energy of 3 GeV are the interacting baryons. The signatures for the transition from partonic dominance to hadronic and then to baryonic dominance regions have also been discussed in Refs.~\cite{Bzdak:2019pkr,Adamczyk:2014ipa,Adamczyk:2017iwn,Adamczyk:2017nxg} for the ratios of $K^+/\pi^+$ and net-particle $v_1$ slopes, respectively. The data from 3 GeV Au+Au collisions clearly reveals that baryonic interactions dictate the collision dynamics.

%--==========================================================
\begin{figure}[htb]
\centering  
\includegraphics[width=0.475\textwidth]{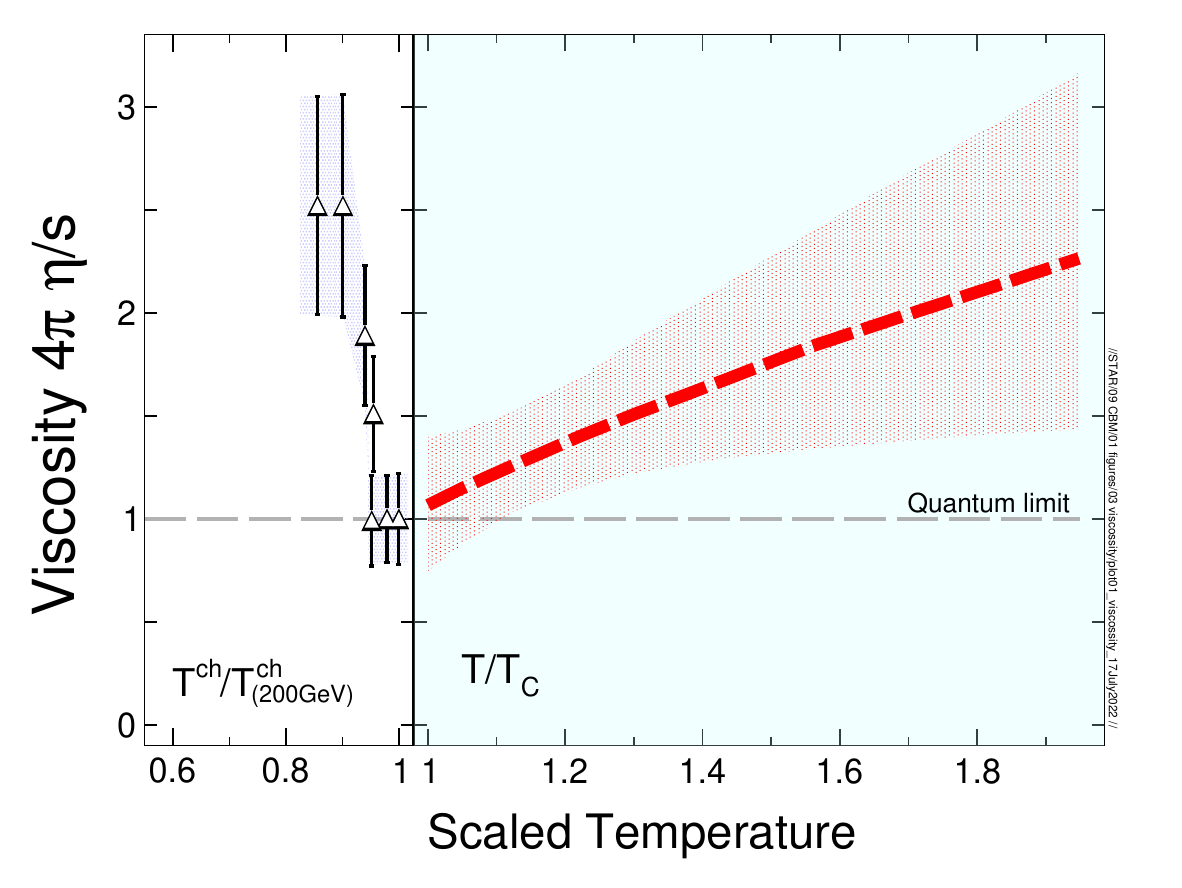}
\vspace{-0.65cm}
\caption{The effective values of shear viscosity-to-entropy ratio, $4\pi \eta/s$, shown as a function of the scaled temperature. The horizontal dashed line indicates the quantum lower limit. Left panel: the extracted $4\pi \eta/s$ from the energy dependence of the measured $v_2$~\cite{STAR:2012och} and $v_3$~\cite{STAR:2016vqt}, shown as the scaled chemical freeze-out temperature ${\rm T_{ch}/T_{ch}}$(200 GeV). Right panel: temperature evolution of $4\pi \eta/s$, extracted from Bayesian analyses~\cite{Bernhard:2019bmu,Xu:2017obm}. }
\label{eta_o_s}
\end{figure}

The results of collectivity, the EOS, and phase structure are closely connected. By comparing measurements with calculations, the parameters of the EOS for each collision can be readily extracted~\cite{Bernhard:2019bmu,Xu:2017obm}. As an example, Fig~\ref{eta_o_s} shows the ratio of shear viscosity to entropy as a function of scaled temperature~\cite{Huang:2023ibh}. 
In the left panel, chemical freeze-out temperature from each energy~\cite{STAR:2017sal} is used and normalized to that from Au+Au collisions at $\sqrt{s_{\rm NN}}=$ 200 GeV. As one can see, in the high energy limit, $\sqrt{s_{\rm NN}}=$ 39 - 200 GeV, the ratio reaches unity, namely the quantum limit, implying that the medium created in such collisions is dominated by partonic interactions with a minimum value of $4\pi \eta/s$. At lower collision energies, on the other hand, hadronic interactions become dominant, and the medium shows a rapid increase in the viscosity-to-entropy ratio. 
The right panel shows the temperature evolution of the shear viscosity-to-entropy ratio as a function of the scaled temperature $T/T_{\rm C}$. Here, $T_{\rm C}$ represents the critical temperature in the calculation~\cite{Bernhard:2019bmu,Xu:2017obm}. 
The entire curve is extracted from the experimental results of $R_{\rm AA}$ and $v_2$ at Au+Au collisions at $\sqrt{s_{\rm NN}}=$ 200 GeV. The observed $V$-shaped feature is quite similar to what is described in Ref.~\cite{Csernai:2006zz} for a system dominated by electromagnetic interactions. The phase transition is universal and independent on the degrees of freedom of the medium under study. The unique feature is a clear evidence of the crossover transition in strong interaction. We here refer to a recent review for a comprehensive discussion for the shear viscosity and phase transition in nucleon and quark levels~\cite{Deng2024}. 

%--==========================================================
%---===============================================
\subsection{Chirality}
Quark interactions with topological gluon configurations can induce chirality imbalance and local parity violation in QCD~\cite{Lee:1974ma,Kharzeev:1998kz,Kharzeev:1999cz}. In relativistic heavy ion collisions, this can lead to observable electric charge separation along the direction of the strong magnetic field produced primarily by spectator protons~\cite{Fukushima:2008xe,Muller:2010jd,Kharzeev:2015znc}. This is called the chiral magnetic effect (CME).
An observation of the CME-induced charge separation would confirm a fundamental property of QCD. Measurements of the electric charge separations can provide a means to study the non-trivial QCD topological structures and are therefore of paramount importance. Extensive theoretical and experimental efforts have been devoted to the search for CME~\cite{Kharzeev:2015znc,Zhao:2019hta,Li:2020dwr}. 

\begin{figure*}
	\centering
\includegraphics[width=0.8\textwidth]{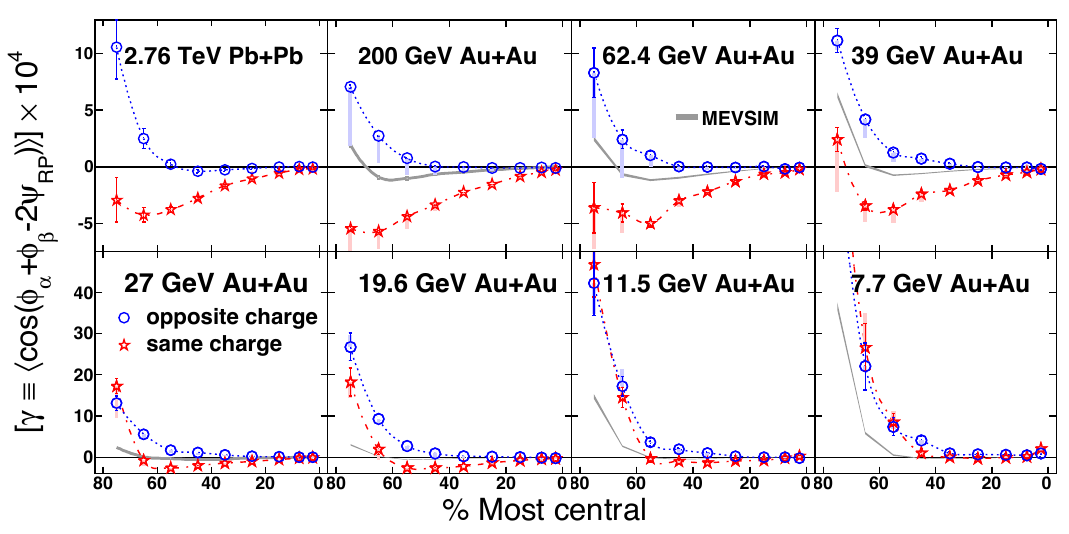}
\caption[]{(Color online) 
	The $\gamma$ correlators as functions of centrality in Au+Au collisions at $\sqrt{s_{\rm NN}}=$ 7.7-200~GeV from STAR~\cite{STAR:2009wot,STAR:2014uiw} and for Pb+Pb collisions at $\sqrt{s_{\rm NN}}=$ 2.76~TeV from ALICE~\cite{ALICE:2012nhw}. 
	}
\label{CMEBES}
\end{figure*}

The commonly used observable to measure charge separation is the three-point correlator difference~\cite{Voloshin:2004vk}, $\Delta\gamma\equiv\gamma_{\rm OS}-\gamma_{\rm SS}$. 
Here $\gamma=\langle\cos(\alpha+\beta-2\psi_2)\rangle$, $\alpha$ and $\beta$ are the azimuthal angles of two charged particles and $\psi_2$ 
is that of the second-order harmonic plane; $\gamma_{\rm OS}$ stands for the $\gamma$ of opposite electric charge sign  (OS) 
and $\gamma_{\rm SS}$ for that of same-sign pairs (SS). The first $\gamma$ measurements were made by the STAR collaboration in Au+Au collisions at top RHIC energy in 2009~\cite{STAR:2009wot}.
Significant $\Delta\gamma$ has indeed been observed.
Further measurements were made at lower RHIC energies by STAR~\cite{STAR:2014uiw} and at higher LHC energy by ALICE~\cite{ALICE:2012nhw}.
Figure~\ref{CMEBES} shows the $\gamma_{\rm OS}$ and $\gamma_{\rm SS}$ correlators as a function of the collision centrality in Au+Au collisions at $\sqrt{s_{\rm NN}}$ = 7.7-200 GeV at RHIC and in Pb+Pb collisions at 2.76 TeV at the LHC.
At high collision energies, charge-dependent signals are observed; $\gamma_{\rm OS}$ is larger than $\gamma_{\rm SS}$. 
The difference between $\gamma_{\rm OS}$ and $\gamma_{\rm SS}$, i.e.~$\Delta\gamma$, decreases with increasing centrality, 
which would be consistent with the expectation of the magnetic field strength to decrease with increasing centrality. 
At the low collision energy of $\sqrt{s_{\rm NN}}$ =7.7 GeV, the difference between the $\gamma_{\rm OS}$ and $\gamma_{\rm SS}$ disappears, 
which could be consistent with the disappearance of the CME in the presumably hadronic dominant interactions at this energy. 
Thus, these results are qualitatively consistent with the CME expectation. 

One of the difficulties to interpret the positive $\Delta\gamma$ to be from the CME is the major charge-dependent background contributions to the observable~\cite{Wang:2009kd,Bzdak:2009fc,Schlichting:2010qia}, such as those from resonance decays and jets. 
The $\Delta\gamma$ variable is ambiguous between an OS pair from the CME back-to-back 
perpendicular to $\psi_2$ and an OS pair from a resonance decay along $\psi_2$. 
More resonances are produced along the $\psi_2$ than perpendicular to it, 
the relative difference of which is quantified by the elliptical anisotropy parameter $v_2$ of the resonances. (Jet correlations also exhibit azimuthal anisotropy because of jet quenching effects in heavy ion collisions~\cite{Wang:2000bf}.)
The CME background arises from the coupling of this elliptical anisotropy and genuine particle correlations from resonance decays and jets, among others.
Calculations using the blast wave parameterizations of the measured particle production data can indeed reproduce  essentially the entirety of the measured $\gamma$ correlations~\cite{Schlichting:2010qia}.

\begin{figure*}
\centering
\includegraphics[width=0.36\textwidth]{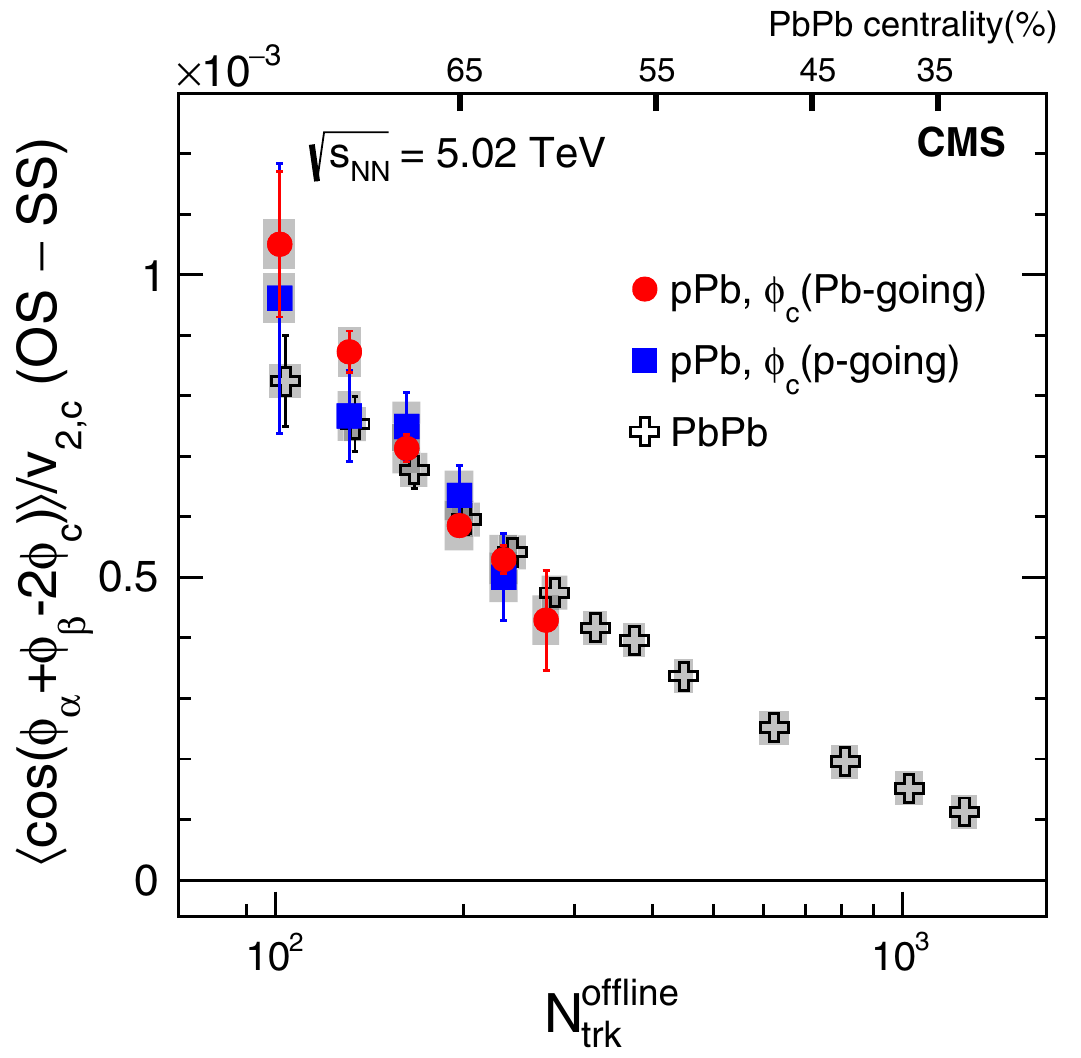}
\includegraphics[width=0.46\textwidth]{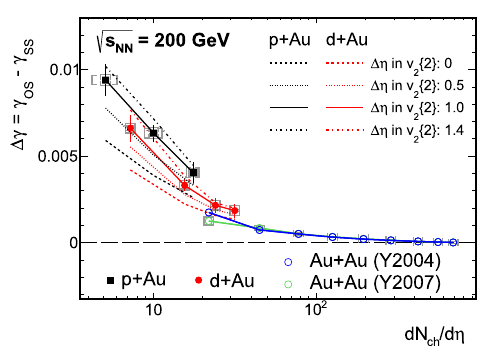}
\caption[]{(Color online) 
	The $\Delta\gamma$ correlators as functions of multiplicity in p+Pb and Pb+Pb collisions from LHC-CMS~\cite{CMS:2016wfo} (left), and in p/d+Au and Au+Au collisions from RHIC-STAR~\cite{STAR:2019xzd} (right). 
	}
\label{CME-small}
\end{figure*}
The CME and the $v_{2}$-related background are driven by different physics: 
the CME is sensitive to the magnetic field which is mostly perpendicular to the spectator plane,
while the $v_{2}$-related background is connected to the participant plane.
In non-central heavy ion collisions, the participant plane is generally 
aligned with the reaction plane, the $\Delta\gamma$ measurement is thus entangled by the two contributions: the possible CME and the $v_2$-induced background.
In small-system p+A or d+A collisions, however, the participant plane is determined purely by geometry fluctuations, 
uncorrelated to the magnetic field direction~\cite{CMS:2016wfo}. 
As a result any CME signal would average to zero in small-system collisions. Background sources, on the other hand, 
contribute to small-system collisions similarly as to heavy ion collisions. 
Figure~\ref{CME-small} (left) show the first $\Delta\gamma$ measurements in small system p+A collisions from CMS~\cite{CMS:2016wfo}. Within uncertainties, the results in p+Pb and Pb+Pb collisions exhibit the same magnitude and trend as a function of multiplicity.
Figure~\ref{CME-small} (right) show the $\Delta\gamma$ measurements in small system p/d+A collisions from STAR~\cite{STAR:2019xzd}. The trends of the magnitudes are similar, decreasing with increasing multiplicity. These results indicate that there are strong correlations in small systems contributing to the $\gamma$ correlators. These correlations may be of genuine three-particle correlation nature, and thus can explain the peripheral heavy ion data but insufficient for mid-central heavy ion data as they are strongly diluted by event multiplicity. Some of the correlations, on the other hand, may be of flow nature as there are indications of collective flow in those small systems~\cite{Adamczyk:2015xjc,STAR:2022pfn}, especially at the LHC energies~\cite{Khachatryan:2010gv}. Nevertheless, the small system results suggest the complicate nature of the backgrounds which must be rigorously removed before addressing the important physics of the CME. 

Since the major background is induced by $v_2$, it is interesting to examine the $\Delta\gamma$ observable with varying $v_2$ while holding constant the expected CME signal.
The Event Shape Engineering (ESE) method is performed based on the magnitude of the flow vector to possibly access to the initial participant geometry.
By restricting to a given narrow centrality, the ESE selecting of events is not expected to affect the magnetic filed. 
The different dependence of the CME signal and background on $v_{2}$ ($q_{2}$) could possibly be used to disentangle the CME signal from background. Using the ESE method, the ALICE experiment showed that the CME fraction in the measured $\Delta\gamma$ is consistent with zero~\cite{ALICE:2017sss}. 

\begin{figure*}
	\centering 
	\includegraphics[width=0.95\textwidth]{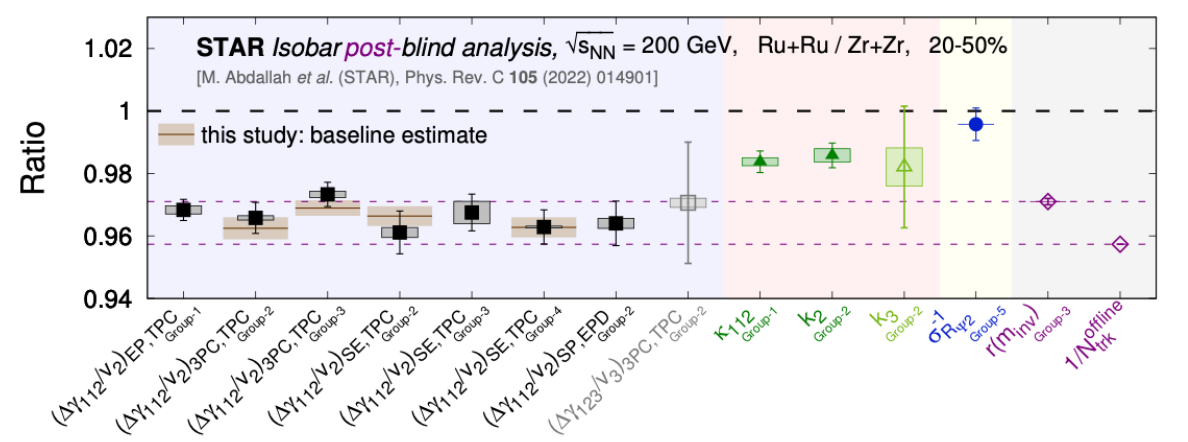} 
	\caption{(Color online)
         Compilation of results from the isobar analysis. Only results contrasting between the two isobar systems are shown. Results are shown in terms of the ratio of measures in Ru+Ru collisions over Zr+Zr collisions. Solid dark symbols show CME-sensitive measures whereas open light symbols show counterpart measures that are supposed to be insensitive to CME. The vertical lines indicate statistical uncertainties whereas boxes indicate systematic uncertainties. The colors in the background are intended to separate different types of measures. The fact that CME-sensitive observable ratios lie below unity leads to the conclusion that no predefined CME signatures are observed in this blind analysis~\cite{STAR:2021mii}. The estimated background baselines from non-flow contamination for the four cumulant measurements of the isobar $\Delta\gamma/v_{2}$ ratios are shown by the horizontal bars (central values) and the shaded areas (total uncertainties, the quadratic sum of the statistical and systematic uncertainties on the background baseline estimates)~\cite{STAR:2023gzg,STAR:2023ioo}.
         }
	\label{CMEisobar}
\end{figure*}

It is also interesting to examine the $\Delta\gamma$ observable with varying magnetic field but keeping the $v_2$ relatively constant. 
To gauge differently the magnetic field relative to the $v_{2}$,
isobaric collisions and Uranium+Uranium collisions have been proposed~\cite{Voloshin:2010ut}.
The isobaric collisions are proposed to study the two systems with similar $v_{2}$ but different magnetic field strength~\cite{Voloshin:2010ut},
such as $^{96}_{44}\rm{Ru}$ and $^{96}_{40}\rm{Zr}$, which have the same mass number but differ by charge (proton) number. 
One would thus expect very similar $v_{2}$ at mid-rapidity in $^{96}_{44}\rm{Ru} + ^{96}_{44}\rm{Ru}$ and $^{96}_{40}\rm{Zr} + ^{96}_{40}\rm{Zr}$ collisions, 
but the magnetic field, proportional to the nuclei electric charge, could vary by 10\%.
The variation of the magnetic field strength between $^{96}_{44}\rm{Ru} + ^{96}_{44}\rm{Ru}$ and $^{96}_{40}\rm{Zr} + ^{96}_{40}\rm{Zr}$ collisions provides an ideal way to disentangle the signal of the chiral magnetic effect from $v_{2}$ related background, 
as the $v_{2}$ related backgrounds are expected to be very similar between these two systems. 

Figure~\ref{CMEisobar} shows the ratio of $\Delta\gamma/v_2$ in Ru+Ru over Zr+Zr collisions, among other observables,  from the isobar analysis~\cite{STAR:2021mii,STAR:2023gzg,STAR:2023ioo}. The CME-sensitive observable ratios lie below unity leading to the conclusion that no predefined CME signatures--one of which is a larger-than-unity Ru+Ru over Zr+Zr ratio of $\Delta\gamma/v_2$--are observed in this blind analysis. 
This is rather counter intuitive at the first glance, but well understood from nuclear structure considerations. In fact, it was predicted that the $^{96}$Zr nucleus is larger than $^{96}$Ru because of its thicker neutron skin, resulting in a slightly smaller energy density and fewer produced particles in Zr+Zr than Ru+Ru collisions~\cite{Xu:2017zcn,Li:2018oec,Li:2019kkh}. The larger $^{96}$Zr nucleus also gives smaller eccentricity at a given centrality and thus smaller $v_2$~\cite{Xu:2017zcn,Li:2018oec}. While the non-identical $v_2$ is properly taken into account in the blind analysis observable $\Delta\gamma/v_2$, the non-identical event multiplicities are not. 
After properly factoring in the multiplicity, the isobar ratios of $N\Delta\gamma/v_2$ from various analyses shown in Fig.~\ref{CMEisobar} indicate a positive signal of a few standard deviations~\cite{STAR:2021mii,Kharzeev:2022hqz}.
However, non-flow contamination exists in the $\Delta\gamma/v_2$ ratio variable~\cite{Feng:2021pgf}. One such contamination is the aforementioned genuine three-particle correlations because the $\Delta\gamma$ is measured by the three-particle correlator in the STAR TPC. The other is due to the fact that two-particle $v_2$ cumulant measurements are contaminated by non-flow correlations and such $v_2$ values are used to compute the $\Delta\gamma$ from the three-particle correlator measurement.
Rigorous studies of non-flow contamination have been carried out in the post-blind analysis, and improved background baselines are derived~\cite{STAR:2023gzg,STAR:2023ioo}. Figure~\ref{CMEisobar} shows the measured isobar ratios of $\Delta\gamma/v_2$ from the blind analysis together with the estimated background baselines from the post-blind analysis. The results show that the isobar ratios are consistent with the baselines, indicating that no statistically significant CME signals have been observed in the isobar data.

The STAR isobar data, without any clear evidence for a possible CME-related signal difference possibly arising from the charge difference (44 in Ru versus 40 in Zr), have provided important lessons for experimental searches for the CME. Firstly, the difference in nuclear shape and/or neutron skin between isobaric nuclei could induce percent-level background variations, which are not easily estimated with theoretical calculations or controlled with experimental constraints. Thus, searches for small differences in the CME signal due to the magnetic field variation in isobar collisions will be extremely challenging. Secondly, the strength of the magnetic field plays a critical role in the CME signal, so larger nuclei would be preferable in search of a possible CME-induced signal in $\Delta\gamma$ correlations. Thirdly, we need a better understanding of the background sources in the $\Delta\gamma$ correlator and how to suppress background from elliptic flow and non-flow correlations.

The major background source in the CME observable $\Delta\gamma$ is induced by elliptic flow ($v_{2}$). The original event shape engineering approach~\cite{Schukraft:2012ah,ALICE:2017sss,CMS:2017lrw} used particles from separate rapidity or pseudorapidity regions to define event classes. This approach is able to select event shapes sensitive to the eccentricity of the initial overlapping participants and the corresponding geometrical fluctuations. However, for particles of interest used for measuring the CME-sensitive observable $\Delta\gamma$ in a different rapidity region, the event-by-event $v_{2}$ background has contributions from both eccentricity and particle emission pattern fluctuations. Petersen and Muller~\cite{Petersen:2013vca} pointed out that the emission pattern fluctuations dominate the event-by-event $v_{2}$ fluctuations. Recently, Z. Xu {\it et al.} proposed a novel event shape selection (ESS) approach to suppress the background in the CME $\Delta\gamma$ measurement~\cite{Xu:2023elq}.  They found that to suppress the apparent flow-induced background in $\Delta\gamma$, the combined event-by-event information from eccentricity and the emission pattern fluctuations from particles of interest should be used to select azimuthally round shape events for correlator measurements. With this ESS approach, it is possible to suppress the flow-related background. Using AMPT and AVFD model simulations, Z. Xu {\it et al.}~\cite{Xu:2023elq} showed that the most effective ESS approach is to use particle pairs to construct the event shape variable to form event shape classes so that the CME sensitive correlator can be calculated at the limit of zero elliptic flow for particles of interest. This is consistent with the expectation that the background in $\Delta\gamma$ has significant contributions from particle pair emission coupled with elliptic flow.

The RHIC BES-II also provides a unique venue for the CME searches, covering the center of mass energies from 7.7 to 27 GeV. At these beam energies, the STAR Event Plane Detector (EPD), added during the BES-II program, can register spectator protons from the colliding beams. This capability allows for an accurate estimation of the reaction plane, enhancing the sensitivity to the magnetic field direction and suppressing non-flow contributions to the background. For Au+Au collisions at the top RHIC energy, spectator neutrons may be detected by the zero-degree calorimeter (ZDC), though the corresponding event plane resolution is not as good as that in the BES-II data. Theoretical calculations expect that the initial magnetic field would be smaller in Au+Au collisions from BES-II than that from the top RHIC energy. However, the dynamics of the QGP formation and the time evolution of the magnetic field in the QGP as a function of collision energy have not been fully understood. Recent STAR measurements of the deflection of charged particles by the magnetic field in heavy ion collisions indicate significant imprints of magnetic-field effects at these BES-II energies~\cite{STAR:2023jdd}. The STAR collaboration reported preliminary results on the CME search from the RHIC BES-II data at the Quark Matter 2023 conference, demonstrating a promising approach to focus on Au+Au collisions with the innovative experimental technique for background suppression~\cite{Xu:2023wcy}.

As aforementioned, the $\Delta\gamma$ measurement in heavy ion collisions is entangled by two contributions, one from the CME and the other from the $v_2$ induced background. They are sensitive to difference planes, with which $\Delta\gamma$ can be measured. 
The background is related to $v_{2}$, determined by the participant geometry, therefore is the largest with respect to the participant plane ($\psi_{\rm PP}$).
The CME-driven charge separation is along the magnetic field direction ($\psi_{B}$), 
different from $\psi_{\rm PP}$.
The $\psi_{B}$ and $\psi_{\rm PP}$ are in general correlated to the $\psi_{\rm RP}$, the impact parameter direction, therefore are correlated to each other. 
While the magnetic field is mainly produced by spectator protons, their positions fluctuate, thus $\psi_{B}$ is not always perpendicular to the $\psi_{\rm RP}$. 
The position fluctuations of participant nucleons and spectator protons are independent, thus $\psi_{\rm PP}$ and $\psi_{B}$ fluctuate independently about $\psi_{\rm RP}$. 
A new approach has been proposed to measure $\Delta\gamma$ with respect to $\psi_{\rm SP}$ and $\psi_{\rm PP}$ to disentangle the CME signal from the $v_2$ background~\cite{Xu:2017qfs,Voloshin:2018qsm}. This is exploited by STAR by measuring $\Delta\gamma$ with respect to the first-order harmonic plane from the ZDC and the second-order harmonic plane from the TPC. Because the former aligns better with the spectator plane and the latter aligns better with the participant plane, these measurements contain different amounts of the harmonic plane sensitive flow backgrounds and the magnetic field-sensitive CME signal, and can thus be used to extract the possible CME.

\begin{figure}
	\centering 
	\includegraphics[width=0.5\textwidth]{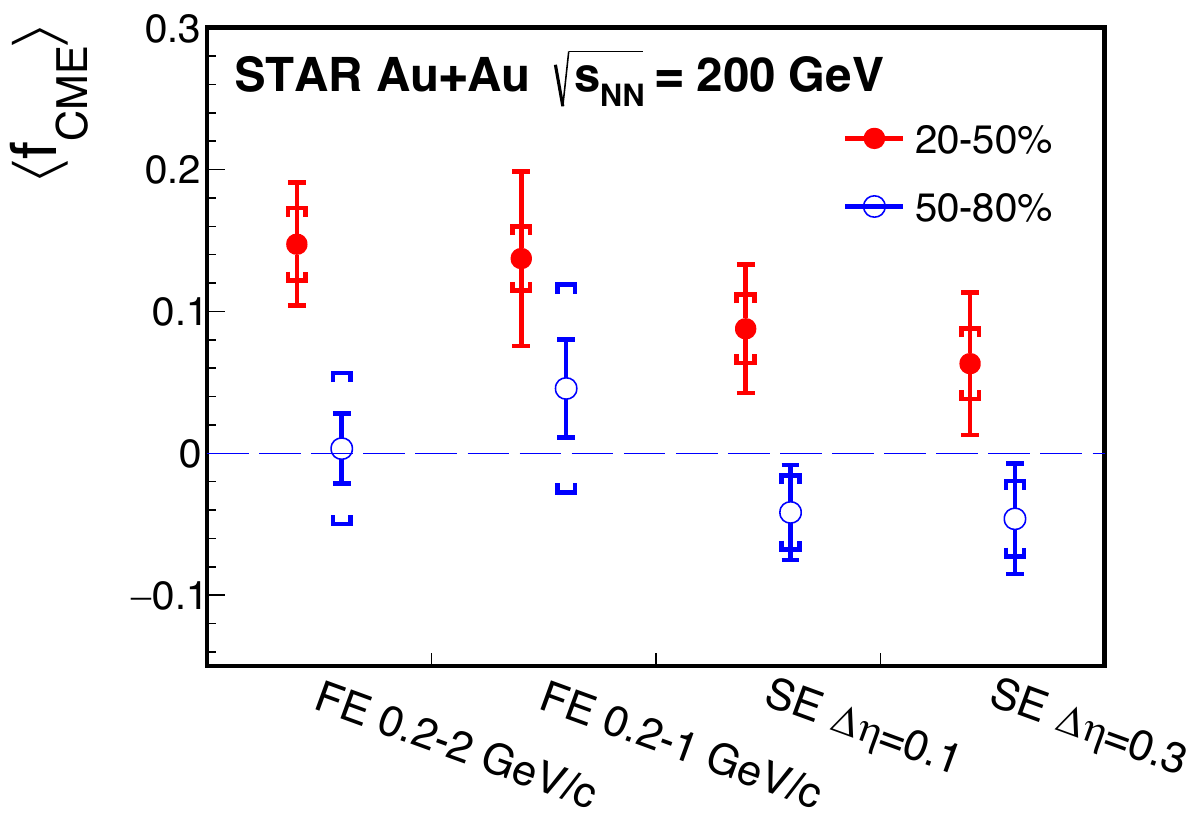} 
	\caption{(Color online)
	The flow-background removed CME signal fraction $\langle{f_{\rm CME}}\rangle$ in 50--80\% (open markers) and 20--50\% (solid markers) centrality Au+Au collisions at $\sqrt{s_{\rm NN}}$ = 200 GeV~\cite{STAR:2021pwb}. Results are shown for full-event (FE) analysis method with two $p_\perp$ ranges and for sub-event (SE) analysis method with two $\Delta\eta$ gaps. Error bars show statistical uncertainties; the caps indicate the systematic uncertainties.
	}
	\label{CMEPPRP}
\end{figure}
STAR reported such measurements in Au+Au collisions at $\sqrt{s_{\rm NN}}$ = 200 GeV~\cite{STAR:2021pwb}, as shown in Fig.~\ref{CMEPPRP}. It is found that the charge separation, with the flow background removed, is consistent with zero in peripheral collisions. In mid-central collisions, on the other hand, intriguing  indication of finite CME signals is seen on the order of 1-3$\sigma$ standard deviations.   

In RHIC 2023-2025, STAR is expected to collect about 20 B events, which is about a factor of ~10 more compared to the data used for Fig.~\ref{CMEPPRP}. More precise results are expected in the near future.  Besides, new analyses utilizing Event-Shape-Engineering with particle pair anisotropy and invariant mass that are ongoing, and results are expected soon.

%--==========================================================
%---===============================================
\subsection{Criticality}
In high energy nuclear collisions where the baryon density is vanishingly small, the transition from QGP to hadronic matter is smooth crossover~\cite{Aoki:2006we}. At finite density and lower temperature, however, the transition is speculated to be first-order, with an associated phase boundary. The point that connects the smooth crossover and the first-order phase boundary is the QCD critical point. Since 2010, RHIC has conducted two rounds of beam energy scan (BES) campaigns primarily aimed at investigating the QCD critical point. The BES programs cover an energy range  from $\sqrt{s_{NN}}$  = 200 GeV to 3.0 GeV, corresponding to a baryonic chemical potential of 20 $\leq \mu_B \leq$ 750 MeV. As of the summer of 2022, both BES-I and BES-II have been completed.

\begin{figure}[htpb]
    \centering
    \includegraphics[width=0.475\textwidth]{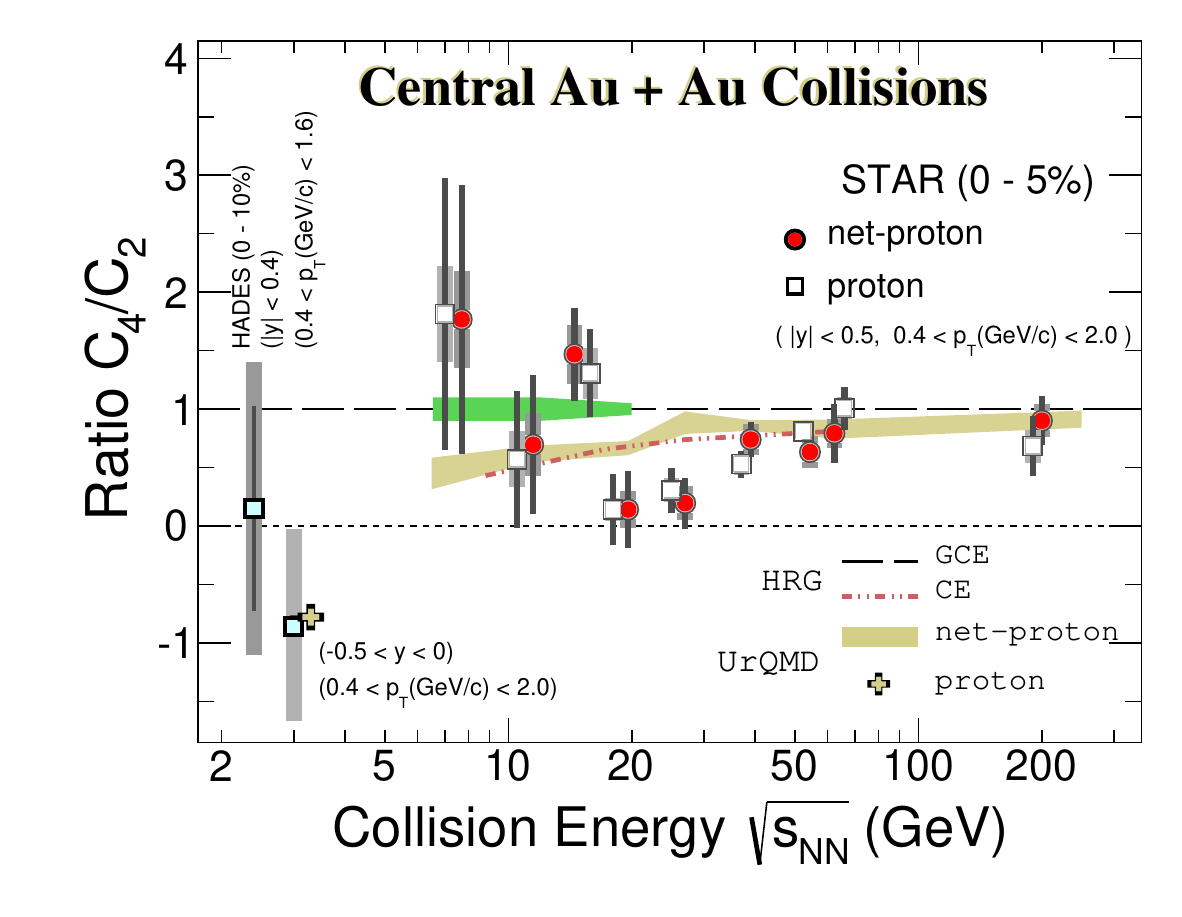}
    \caption{Collision energy dependence of the ratios of cumulants, $C_4/C_2$, for proton (squares) and net-proton (red circles) from top 0–5\% Au+Au collisions at RHIC~\cite{STAR:2020tga}. The points for protons are shifted horizontally for clarity. The new result for proton from $\sqrt{s_{\rm NN}}$ = 3.0 GeV collisions is shown as a filled square. HADES data of $\sqrt{s_{\rm NN}}$ = 2.4 GeV 0–10\% collisions is also shown. Results from the HRG model and transport model UrQMD~\cite{Bass:1998ca,Bleicher:1999xi} are shown.}
    \label{BESII_netp}
    \end{figure}

Due to their high sensitivity on correlation length, high-order cumulants of protons and net-protons (event-by-event number: net-p =  $p - \bar{p}$) distributions are used in the search for the QCD critical point~\cite{Luo:2017faz}. The experimental results shown as a function of the collision energy are depicted in Fig.~\ref{BESII_netp}. Overall, the ratios of $C_4/C_2$ for net-protons from collider mode ($\sqrt{s_{\rm NN}} \geq$ 7.7 GeV)~\cite{STAR:2021iop,STAR:2020tga} and protons from the fixed-target mode are decrease as collision energy decreases due to the baryon number conservation. Both Hadronic Resonance Gas model and hadronic transport model UrQMD~\cite{Bass:1998ca,Bleicher:1999xi} calculations reproduce the trend. As a function of collision energy, a rise and then fall of the net-proton $C_4/C_2$ (or $\kappa\sigma^2$) has been predicted to indicate the critical behavior expected near the critical point in the QCD phase diagram. While results of $C_4/C_2$ ratios from BES-I had shown dip like energy dependence around 20 GeV, the statistics at lower collision energy are too poor to draw any conclusion.
Note that at low energy, or in another words, in high baryon density region, both HADES ($\sqrt{s_{\rm NN}}$  = 2.4 GeV) and STAR ($\sqrt{s_{\rm NN}}$  = 3.0 GeV~\cite{STAR:2022etb,STAR:2021fge}) high moment of proton results are below Poisson baseline and the non-critical hadronic transport model calculations reproduced the data at the high baryon density region. This implies that in this energy regime is dominated by hadronic interactions. In order to look for the oscillation pattern in the energy dependence of the ratio of the $C_4/C_2$, RHIC had the second beam energy scan (BES-II). The analysis of the BES-II data are under way.

\begin{figure}[htpb]
     \centering
     \includegraphics[scale=0.42]{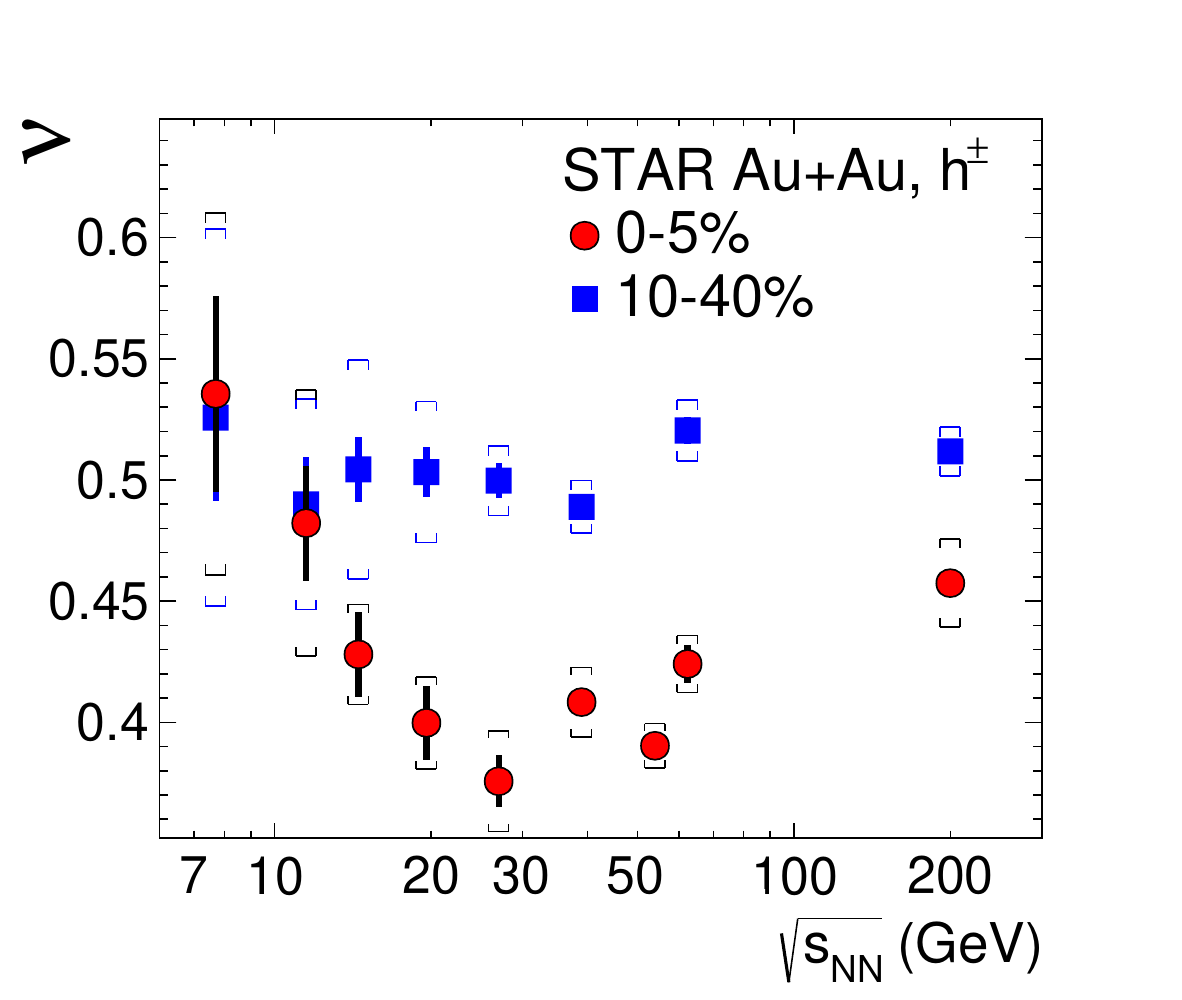}
     \caption{Energy dependence of the scaling exponent ($\nu$) for identified charged hadrons ($h^{\pm}$) in Au$+$Au collisions at $\sqrt{s_\mathrm{_{NN}}}$ = 7.7-200 GeV~\cite{STAR:2023jpm}. Red circles and blue squares represent $\nu$ in the most central collisions (0-5\%) and the mid-central collisions (10-40\%), respectively. The statistical and systematic errors are shown in bars and brackets, respectively.}
     \label{Fig:nuenergy}
\end{figure}

%%%%%%%%%%%%%%%%%%%%%%%%%%%%%%%%%%%%%%
On the other hand, it is predicted that density fluctuations near the QCD critical point can be probed via an intermittency analysis in relativistic heavy ion collisions~\cite{Hwa:1992uq,Antoniou:2006zb}. Figure~\ref{Fig:nuenergy} shows the energy dependence of the scaling exponent ($\nu$) for identified charged hadrons in Au$+$Au collisions for two different collision centralities (0-5\% and 10-40\%)~\cite{STAR:2023jpm}. In the most central collisions, $\nu$ exhibits a non-monotonic behavior as a function of collision energy, reaching a minimum around $\sqrt{s_\mathrm{_{NN}}}$ =  20-30 GeV. In contrast, for 10-40\% central collisions, $\nu$ remains approximately constant with increasing $\sqrt{s_\mathrm{_{NN}}}$. The observed non-monotonic energy dependence of $\nu$ in the most central collisions could indicate density fluctuations induced by the QCD critical point. However, at $\sqrt{s_\mathrm{_{NN}}}\leq$ 11.5 GeV, there are large systematic and statistical uncertainties for $\nu$. Higher statistics data from the BES-II program are needed to confirm this energy dependence. The measured value of $\nu$ is significantly smaller than the theoretical predictions of $\nu$= 1.30 from GL theory and 1.0 from the 2D Ising model. These theoretical values are derived from calculations over the entire phase space without constraints on acceptance, whereas the experimental measurements are limited to the available transverse momentum space. It is anticipated that $\nu$ would increase if measured over the entire phase space, particularly including higher $p_{T}$ regions.
Therefore, theoretical calculations that consider a reduced transverse momentum phase space and equivalent experimental acceptance, are required to understand the measured scaling exponent. The transport-based UrQMD model is unable to calculate $\nu$ due to the absence of the power-law scaling of $\Delta F_{q}(M)\propto \Delta F_{2}(M)^{\beta_{q}}$. Consequently, models that exhibits such power-law scaling is required to produce a non-critical baseline for comparison with experimental data.

Recently, a study of the information entropy \cite{Ma1999}
of the net-proton multiplicity distribution using the ultra-relativistic quantum molecular dynamics model \cite{Deng2024b}.
The ratios of the net-proton information entropies of the UrQMD result with the EoS:CH
(chiral+hadronic gas EoS with first-order transition and critical endpoint), the UrQMD result with EoS:BM (bag model EoS with strong first-order phase transition between QGP and hadronic phase), 
and the STAR experimental data to the UrQMD result w.o. hydrodynamic EOS, 
are compared. 
The results show that the STAR experimental data, extracted from Ref.~\cite{STAR:2020tga}
     display an enhancement of about 20 GeV with respect to the baseline entropy without hydrodynamics, which is consistent with the minimum $\kappa \sigma^{2}$ value reported in Ref.~\cite{STAR:2021iop}. On the other hand, the UrQMD simulations with the EoS:BM and EoS:CH equations of state also show slightly pronounced enhancements, but at energies that occur at a higher value, about 30 GeV, consistent with recent observations of $N_{t}N_{p}/N_{d}^{2}$ \cite{STAR:2022hbp}
     as well as the analysis of the intermittency scaling exponent \cite{STAR:2023jpm}, which gives a peak or dip around $\sqrt{s_{\rm NN}}$=20 -- 30 GeV, which could indicate the CEP. Thus, the information entropy could also be seen as an indication of an alternative observable at the QGP phase transition. 

%--==========================================================
\subsection{Global Polarization of QCD Matter}

In non-central relativistic heavy ion collisions, huge orbital angular momenta (OAM) and vorticity fields are produced in the QGP~\cite{Shuryak:1980tp}. They can lead to the hadron polarization and spin alignment along the direction of the system OAM through spin-orbit couplings~\cite{Liang:2004ph,Liang:2004xn,Gao:2007bc} or spin-vorticity couplings~\cite{Betz:2007kg,Becattini:2007sr}, a phenomenon called global polarization. Such polarization phenomenon in relativistic heavy ion collisions possesses some unique features which are different from the conventional observations. For example, its measurement is not mediated by a magnetic field, like in the well-known Barnett effect~\cite{Barnett:1915uqc}. The global spin polarization of particles is directly observed in relativistic heavy ion collisions, which is not possible in ordinary matter. Second, the QGP at very high energy is almost neutral by charge conjugation. If it was precisely neutral, the observation of polarization by magnetization would be impossible because particles and antiparticles have opposite magnetic moments. In fact, $\Lambda$ and $\bar{\Lambda}$ in relativistic heavy ion collisions at high energy have almost the same mean polarization, which supports the polarization is a strong interaction driven phenomenon. If the electromagnetic field was responsible for this effect, the sign of the mean spin vector components would be opposite. Hence, while for non-relativistic matter it is impossible to resolve polarization by rotation and by magnetization, which lie at the very heart of the Barnett effect~\cite{Barnett:1915uqc} and Einstein-de Hass effect~\cite{Einstein:1915}, in relativistic matter, because of the existence of antiparticles, the rotation and magnetization effects can be distinguished. And QGP is the first relativistic system through which the distinction has been observed~\cite{STAR:2017ckg}.

{\it Global Polarization of Hyperon}

The global polarization of hyperons can be determined from the angular distribution of hyperon decay products in hyperon's rest frame with respect to the system OAM:
\begin{equation}
    \centering
    \frac{dN}{d\cos{\theta^*}} \propto 1 + \alpha_{\rm H} P_{\rm H} \cos{\theta^*},
    \label{eq:HypPol}
\end{equation}
where $\alpha_H$ is the hyperon decay parameter, $P_H$ is the hyperon polarization, and $\theta^*$ is the angle between the polarization vector and the direction of the daughter baryon momentum in the hyperon rest frame. Since the system OAM is perpendicular to the reaction plane, the global polarization can be measured via the distribution of the azimuthal angle of the hyperon decay baryon in the hyperon rest frame with respect to the reaction plane. The reaction plane is defined by the direction of the incoming nuclei (beam direction) and the impact parameter vector ($\hat{b}$)\cite{Poskanzer:1998yz}. We refer to Refs.~\cite{STAR:2018gyt,Xu:2023lnd} for the analysis detains and focus on the results here. 

\begin{figure}[h]       
\centering
\includegraphics[width=0.47\textwidth]{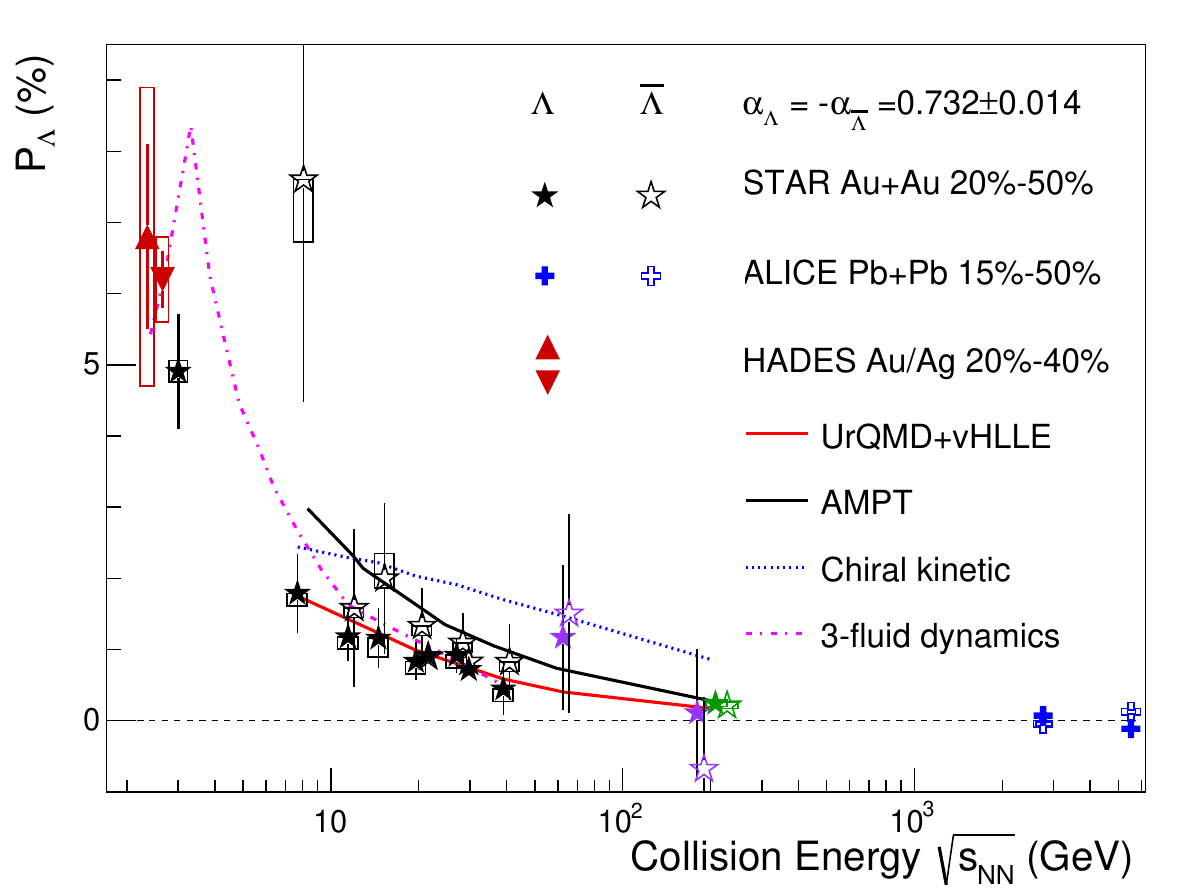}
\caption{Global $\Lambda$ and $\bar{\Lambda}$ polarization 
as a function of $\sqrt{s_{\rm NN}}$ in mid-central heavy ion collisions~\cite{STAR:2007ccu,STAR:2017ckg,STAR:2018gyt,ALICE:2019onw,STAR:2021beb,HADES:2022enx,STAR:2023nvo}. For clarify, data points of the same collision energy from updated measurement are shifted a little bit in the x-axis. 
Calculations with a hybrid model (UrQMD+vHLLE)~\cite{Karpenko:2016jyx}, chiral-kinetic transport (Chiral kinetic)~\cite{Sun:2017xhx} and a multi-phase transport model (AMPT)~\cite{Li:2017slc}
are compared to the higher $\sqrt{s_{\rm NN}}$ data only, while the hydrodynamics 3-fluid model with different equation of state predicts a sharp rising 
$P_{\Lambda}$ at lower $\sqrt{s_{\rm NN}}$~\cite{Ivanov:2020udj}.
}
\label{fig:globHypPol}
\end{figure}

Figure~\ref{fig:globHypPol} shows the first measurement of $P_H$ at $\sqrt{s_{\rm NN}}=$62.4 and 200~GeV at STAR experiment that were consistent with zero~\cite{STAR:2007ccu}. 
The later STAR measurements at $\sqrt{s_{\rm NN}}=$3, 7.7--39~GeV~\cite{STAR:2017ckg,STAR:2021beb} and with higher statistics at $\sqrt{s_{\rm NN}}=$200~GeV~\cite{STAR:2018gyt} indicate 
statistically significant global polarization $P_H>$0, while high-statistics ALICE measurements at $\sqrt{s_{\rm NN}}=$2.76 and 5.02~TeV demonstrate $P_H$ is consistent with zero at the LHC energies~\cite{ALICE:2019onw}.
$P_H$ is observed to increase with collision centrality, which is in agreement with larger system OAM from central collisions to peripheral collisions.
Fig.~\ref{fig:globHypPol} also shows a measurement of $\Lambda$ polarization going to lower energies of Au+Au collisions at $\sqrt{s_{\rm NN}}=$2.4 GeV and Ag+Ag collisions at $\sqrt{s_{\rm NN}}=$2.55 GeV by HADES experiment~\cite{HADES:2022enx}. An increasing trend of $P_H$ as the decreasing of $\sqrt{s_{\rm NN}}$ is observed. The collision energy dependence of experimental data can be reasonably describes by theoretical calculations, as displayed also in the figure, including hydrodynamic~\cite{Karpenko:2016jyx,Ivanov:2020udj} and transport simulations~\cite{Sun:2017xhx,Li:2017slc}. 
Some models also predict that $P_H$ would vanish at $\sqrt{s_{\rm NN}}$=2$m_N$, and thus $P_H$ may peak around 3 GeV~\cite{Ivanov:2020udj,Guo:2021udq}.
It is also interesting to investigate the $P_H$ dependencies versus hyperon transverse momentum $p_{\rm T}$ and its rapidity $y$, as different models gave even opposite trend for high rapidity region~\cite{Liang:2019pst,Sun:2017xhx}.
The available measurements mostly cover mid-rapidity and observed $P_H$ is constant versus $p_{\rm T}$ and $y$ within uncertainties. 
Future measurements at large rapidity region remain with special interest in particular after the STAR forward detector upgrade. 
There are also discussions of collision system dependence of $P_H$, for example in smaller colliding systems~\cite{Shen2022}. 
Recently, STAR experiment measured the $\Lambda$ global polarization in isobar Ru+Ru and Zr+Zr collisions at $\sqrt{s_{\rm NN}}=$200 GeV, and also observed $\Lambda$ polarization along the beam direction relative to the second and third harmonic event planes originating from local vorticity~\cite{STAR:2023eck}.

All particles and antiparticles of the same spin should have the same global polarization assuming OAM is the only driven source of polarization. A difference could arise from effects of the initial magnetic field, from the fact that particles and their antiparticles have opposite magnetic moments. 
In addition, different particles could be produced at different times or regions as the system freezes out, or through meson-baryon interactions.
The measurement of $\Lambda$ and $\bar{\Lambda}$ polarization in the $\sqrt{s_{\rm NN}}=$ 7.7--39~GeV demonstrate no difference within current uncertainties. Therefore, to establish the global nature
of the polarization, it is very important to measure the polarization of different particles, and if possible, particles of different spins.

{\it Global Spin Alignment of Vector Meson}

The global polarization has its imprint on vector mesons such as $\phi(1020)$ and $K^{*0}(892)$. Unlike $\Lambda$ ($\bar{\Lambda}$) hyperons that can undergo weak decay with parity violation, the polarization of vector mesons cannot be directly measured since they mainly decay through the strong interaction, in which parity is conserved. Nevertheless the spin state of a spin-1 vector meson can be described by a $3\times 3$ spin density matrix with unit trace~\cite{Schilling:1969um}.
The diagonal elements of this matrix, namely, $\rho_{11}, \rho_{00}$ and $\rho_{\rm{-1} \rm{-1}}$, are probabilities for the spin component to take the values of 1, 0, and -1 respectively along a quantization axis, which is a chosen axis onto which the projection of OAM has well-determined values. When the three spin states have equal probability to be occupied, all three elements are $1/3$ and there is no spin alignment. If $\rho_{00} \neq 1/3$, the spin of the vector meson is aligned with the spin quantization direction.
For a vector meson decaying into two spin-0 daughters, the angular distribution of one of its decay products in the vector meson rest frame can be written as 
\begin{eqnarray}\label{eq:extract_rho00}
\frac{dN}{d\mathrm{cos}\theta^*} \propto (1-\rho_{00}) + (3 \rho_{00} -1) \mathrm{cos}^2\theta^* ,
\label{eq:Vectormeson}
\end{eqnarray}
where $\theta^*$ is same to the definition of Eq.(\ref{eq:HypPol}), the polar angle between the quantization axis and the momentum direction of one of the decay products. For our study of global spin alignment, the quantization axis is chosen to be the direction of the system OAM, which is perpendicular to the reaction plane. By fitting the angular distribution of decay products with Eq.(\ref{eq:Vectormeson}), one can infer the $\rho_{00}$ value.

The search for global spin alignment of $\phi(1020)$ and $K^{*0}(892)$ mesons for Au+Au collisions at $\sqrt{s_{\rm NN}} = 200$ GeV was started in parallel with the $\Lambda$ polarization. 
Due to limited statistics at the beginning, no significant result was reported at that time~\cite{STAR:2008lcm}. 

\begin{figure}[htb]       
\centering
\includegraphics[width=0.47\textwidth]{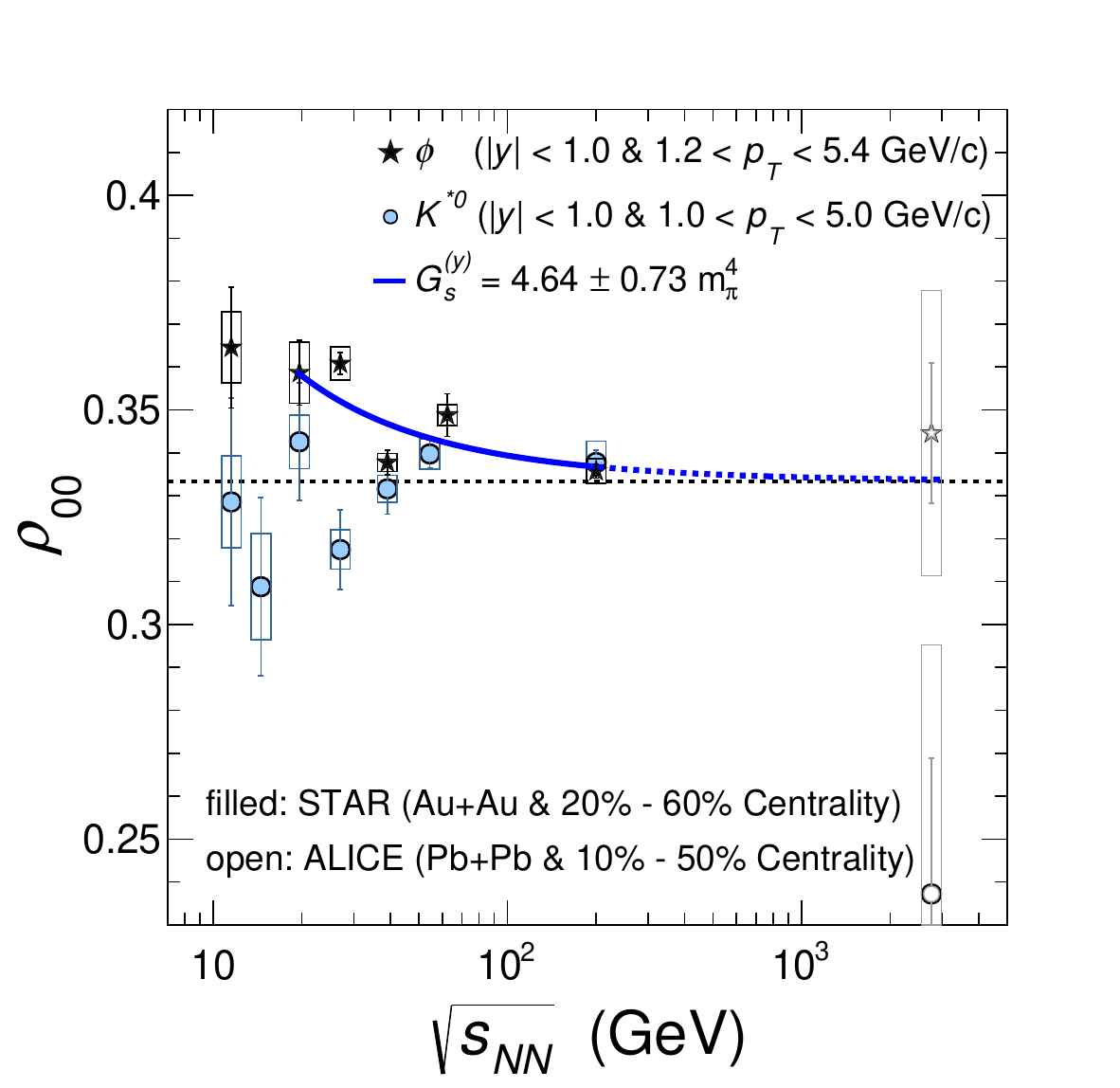}\vspace{-.5cm}
\caption{Global spin alignment of $\phi$ and $K^{*0}$ vector mesons in heavy ion collisions. The measured matrix element $\rho_{00}$ as a function of beam energy for the $\phi$ and $K^{*0}$ vector mesons within the indicated windows of centrality, transverse momentum ($p_{\rm T}$) and rapidity ($y$). The open symbols indicate ALICE results \cite{Acharya:2019vpe} for Pb+Pb collisions at 2.76 TeV at $p_{\rm T}$ values of 2.0 and 1.4 GeV/c for the $\phi$ and $K^{*0}$ mesons, respectively. The blue solid curve is a fit to data in the range of $\sqrt{s_{\rm NN}} = 19.6$ to 200 GeV, based on a theoretical calculation with a $\phi$-meson field~\cite{Sheng:2019kmk}. Parameter sensitivity of $\rho_{00}$ to the $\phi$-meson field is shown in Ref.~\cite{Sheng:2022wsy}. The blue dashed line is an extension of the solid curve with the fitted parameter $G_s^{(y)}$. The black dashed line represents $\rho_{00}=$ 1/3.}
\label{alignment_BES}
\end{figure}

Figure~\ref{alignment_BES} presents the $\phi(1020)$ mesons spin alignment in Au+Au collisions at beam energies between $\sqrt{s_{\rm NN}} = 11.5$ and 200 GeV~\cite{STAR:2022fan}. The STAR measurements presented in Fig.~\ref{alignment_BES} are for centralities between 20\% and 60\% where a maximum OAM of the collision system is expected. The quantization axis is the normal to the 2nd-order event plane~\cite{Poskanzer:1998yz} (a proxy for the reaction plane), determined using STAR charged particle information. The $\phi$-meson results are presented for transverse momentum $1.2 < p_{\rm T} < 5.4$ GeV/$c$, and $\rho_{00}$ for this species is significantly above 1/3 for collision energies of 62 GeV and below, indicating finite global spin alignment. The $\rho_{00}$ for $\phi$ mesons, averaged over beam energies of 62 GeV and below is 0.3512 $\pm$ 0.0017 (stat.) $\pm$ 0.0017 (syst.). Taking the total uncertainty as the sum in quadrature of statistical and systematical uncertainties, our results indicate that the $\phi$-meson $\rho_{00}$ is above 1/3 with a significance of 7.4~$\sigma$~\cite{STAR:2022fan}.

Figure~\ref{alignment_BES} also presents the beam-energy dependence of $\rho_{00}$ for $K^{*0}$ within $1.0~<~p_{\rm T}~<~5.0$~GeV/$c$. We observe that $\rho_{00}$ for $K^{*0}$ is largely consistent with 1/3, in marked contrast to the case for $\phi$. The $\rho_{00}$ for $K^{*0}$, averaged over beam energies of 54.4 GeV and below is 0.3356 $\pm$ 0.0034 (stat.) $\pm$ 0.0043 (syst.),
and the deviation from 1/3 has a $\sim 0.42 \sigma$ significance~\cite{STAR:2022fan}. Measurements from the ALICE collaboration for Pb+Pb collisions at $\sqrt{s_{\rm NN}}$ = 2.76~TeV~\cite{Acharya:2019vpe}, taken from the closest data points~\cite{Acharya:2019vpe} to the mean $p_{\rm T}$ for the range of $1.0~<~p_{\rm T}~<~5.0$~GeV/$c$, are also shown for comparison in Fig. ~\ref{alignment_BES}. They are consistent with 1/3 considering the large statistical uncertainties.

According to the quark coalescence for hadron production in heavy ion collisions, $\Lambda$ polarization depends linearly on quark polarization where vector meson polarization depends quadratically on it~\cite{Liang:2004ph,Liang:2004xn}. One would therefore expect the polarization for $\phi$ to be smaller than the one measured for the $\Lambda$. However, the measured $\rho_{00}$ of $\phi$ is orders of magnitude larger than what one would expect from the same vorticity that causes the measured $\Lambda$ and $\bar{\Lambda}$ polarization in the same collisions. Contributions from electromagnetic fields and other possible conventional mechanisms are also orders of magnitude smaller compared to data~\cite{Sheng:2019kmk,Xia:2020tyd,Gao:2021rom,Muller:2021hpe}. A new mechanism of vector meson spin alignment is interpreted as evidence for a strong force field being capable of describing both the $\rho_{00}$ of $\phi$ and $K^{*0}$~\cite{Sheng:2022wsy}. It is also pointed out that the difference of $\Lambda$ polarization and vector meson spin alignment can be understood as followings: the $\Lambda$ polarization gives information on the mean values of quark polarization, while the $\rho_{00}$ gives information on the correlation of quark polarization and antiquark polarization inside the vector meson~\cite{Lv:2024uev}. Thus measurements of vector meson spin alignment provide an novel way to probe the quark spin correlations, which information may be also accessible via the measurements of hyperon-hyperon and hyperon-antihyperon spin correlations~\cite{Lv:2024uev}. These different scenarios open exciting discovery potential for the spin polarization measurements. For example, one may expect that the strong force correlation will provide a set of new information about the short distance structure of QGP and the nature of QCD phase diagram~\cite{Wang:2023fvy,Chen:2023hnb}.

%--==========================================================
%---===============================================
\subsection{Light Cluster Formation}
Light nuclei and hypernuclei are loosely bound objects of nucleons and hyperons with binding energies of several MeV. Their formation in heavy ion collisions provides important information about the properties of nuclear matter at high densities and temperatures, such as the nucleon-nucleon/hyperon interactions, the equation of state which may offer insights into the inner structure of compact stars.

The production of light nuclei in relativistic nucleus-nucleus collisions is studied since early 1960~\cite{Butler:1961} and their production mechanisms are still under debate~\cite{Andronic:2010qu,Braun-Munzinger:2018hat,Chen:2018tnh}. The thermal/statistical and nucleon coalescence models are  two widely recognized and effective for explaining the production of light nuclei in high-energy heavy ion collisions. In the thermal model, the formation of light nuclei is similar to that of hadrons, with yields calculated based on particle masses and the thermodynamic properties near the chemical freeze-out of the collision system~\cite{Andronic:2010qu,Braun-Munzinger:2018hat}. The coalescence model assumes that light nuclei are emerge through the combination of nucleons when they come into close to each other near the time of kinetic freeze-out~\cite{sato1981coalescence,oliinychenko2021overview,Shao:2022eyd}. 

\begin{figure}[htb]       
\centering
\includegraphics[width=0.47\textwidth]{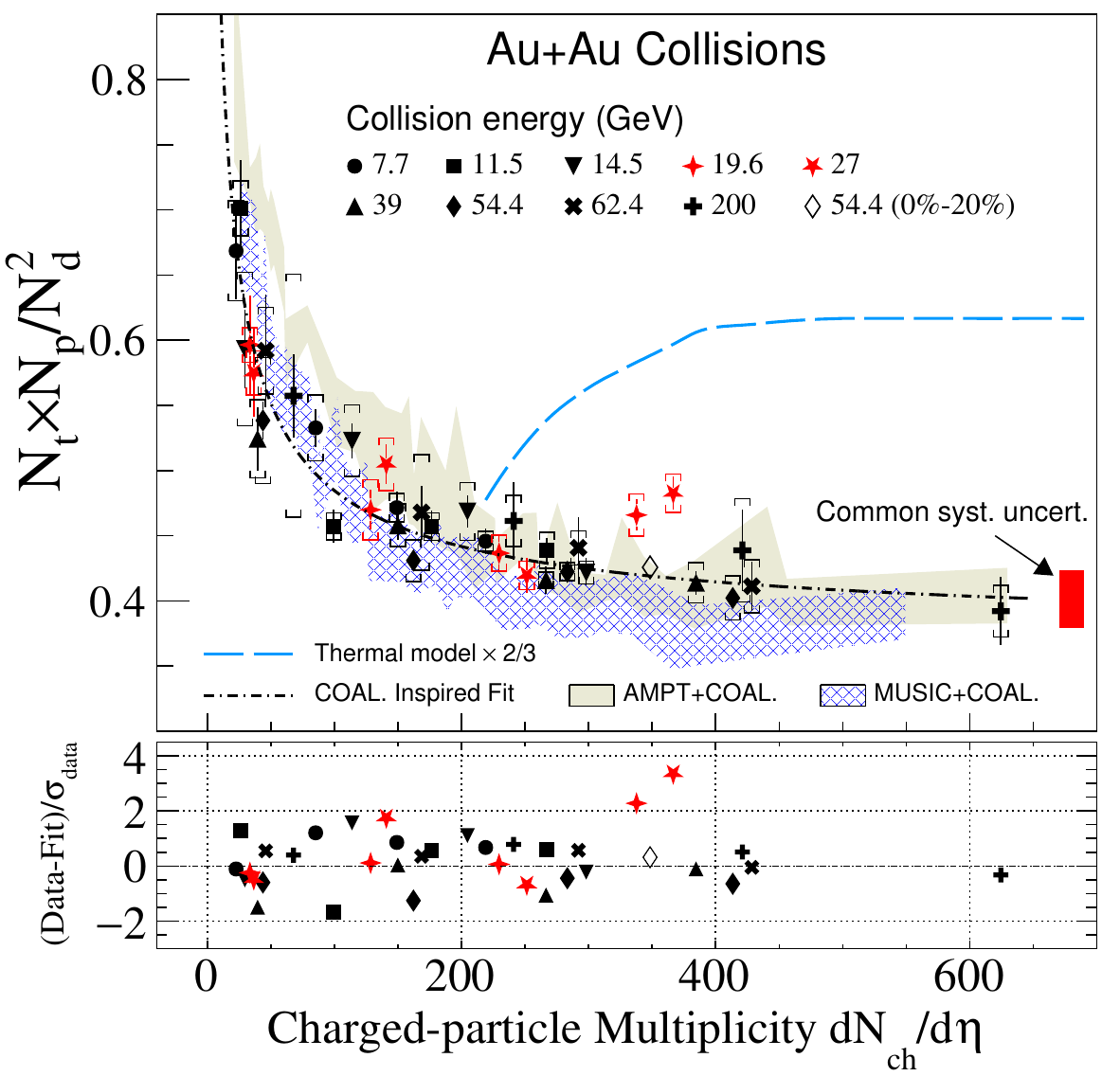}\vspace{-.5cm}
\caption{The yield ratio $N_t\times N_p/N^2_d$
as a function of charged-particle
multiplicity $dN_{\rm ch}/d\eta$ ( $|\eta|<$\,0.5) in Au+Au collisions at $\sqrt{s_{\rm NN}}$ = 7.7 – 200\,GeV for various collision centralities. The black dot-dashed line denotes
the coalescence-inspired fit.
The significance of the deviation relative to the fit is shown in the lower panel. The results calculated from thermal model are shown as the blue long-dashed line. Calculations from AMPT and MUSIC+UrQMD hybrid models are shown as shaded bands~\cite{Zhao:2021dka}.}
\label{fig_III_H_lightNuclearYieldRatio}
\end{figure}

Based on coalescence model, it was predicted that the compound yield ratio $N_t \times N_p/N^2_d$ of tritons ($N_t$), deuterons ($N_d$), and protons ($N_p$), is sensitive to the neutron density fluctuations, making it a promising observable to search for the signature of the critical point (CP) and/or a first-order phase transition in heavy ion collisions~\cite{Zhao:2022xkz,Sun:2020pjz,Shao:2020lbq,Shuryak:2018lgd}. The expected signature of CP is the non-monotonic variation as a function of collision energy.

Figure~\ref{fig_III_H_lightNuclearYieldRatio} shows the charged-particle multiplicity $dN_{\rm ch}/d\eta$ ($|\eta| < $ 0.5) dependence of the yield ratio $N_t \times N_p/N^2_d$ in Au+Au collisions at $\sqrt{s_{_{\rm NN}}}$ = 7.7 -- 200\,GeV combining all centrality bins~\cite{STAR:2022hbp}.
It is observed that the yield ratio exhibits scaling, regardless of collision energy and centrality. The shaded bands are the corresponding results from the calculations of hadronic transport AMPT and MUSIC+UrQMD hybrid models, in which neither critical point nor first-order phase transition is included. These two models are employed to generate the nucleon phase space at kinetic freeze-out, when light nuclei are formed via nucleon coalescence. It is found that the overall trend of the experimental data is well described by the model calculations. The
light blue dashed line is the result calculated from the thermal
model at chemical freeze-out ($T_{\rm ch}$ = 157\,MeV at 200 GeV) for central Au+Au collisions, which overestimates the experimental data by more than a factor of two at $dN_{\rm ch}/d\eta$ $\sim$ 600 which could be due to the effects of
hadronic rescatterings during hadronic expansion. The black dot-dashed line is a fit to the data based on the coalescence model. The lower panel of the Fig.~\ref{fig_III_H_lightNuclearYieldRatio} shows that most of the measurements are within significance of 2$\sigma$ from the coalescence baseline, except there are enhancements observed for the yield ratios in the 0-10\% most central Au+Au collisions at 19.6 and 27 GeV with significance of 2.3$\sigma$ and 3.4$\sigma$, respectively. It is worthwhile to point out that in the net-proton higher moments and charged particle intermittency measurements, non-monotonic behaviors were observed at around collision energy of $\sqrt{s_{_{\rm NN}}}$ = 20 GeV. Further studies from dynamical modeling of heavy ion collisions with a realistic equation of state are required to confirm if the enhancements are due to large baryon density fluctuations near the critical point. These systematic measurements of triton yields and yield ratios over a broad energy range provide important insights into the production dynamics of light nuclei and our understanding of the QCD phase diagram.

\begin{figure}[htb]       
\centering
\includegraphics[width=0.47\textwidth]{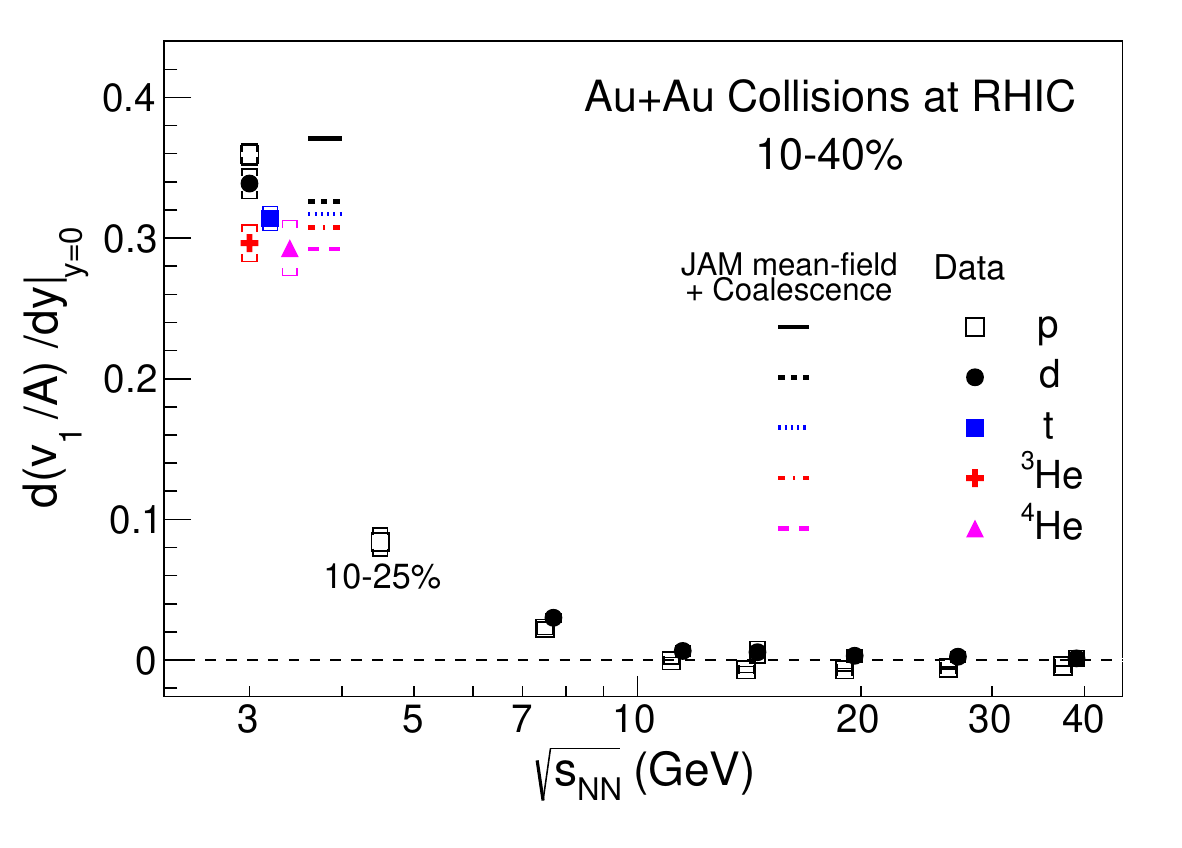}\vspace{-.5cm}
\caption{Light nucleus $v_{1}$ slopes $dv_1/dy|_{y=0}$ scaled by the atomic mass number
as a function of collision energy in 10-40\% mid-central Au+Au collisions~\cite{STAR:2021ozh,adam2020beam,abdallah2021flow}
For clarity, the data points are shifted horizontally. Results of the JAM model in the mean-field mode plus coalescence calculations are shown as color bars.}
\label{fig_III_H_lightNucleaFlow}
\end{figure}

Similar to the number of constituent quark scaling of hadron flow, the light nuclei flow is expected to exhibit an approximate scaling with the mass number $A$ scaling under the coalescence assumption~\cite{Yan006}
\begin{equation}
\centering
v_{n}^{A}(p_{\rm T},y)/A \approx v_{n}^{p}(p_{\rm T}/A,y).
\end{equation}
However, unlike quarks, whose flow cannot be directly measured, both proton and light nuclei flow can be directly measured in experiments to validate the coalescence model. Fig.~\ref{fig_III_H_lightNucleaFlow} shows the light nucleus $v_{1}$ slopes $dv_1/dy|_{y=0}$, which are utilized to characterize the strength of $v_{1}$, scaled by the atomic mass number as a function of collision energy from $\sqrt{s_{\rm NN}}=$ 3 -- 40 GeV at STAR experiment~\cite{STAR:2021ozh,adam2020beam,abdallah2021flow}. Overall, the $dv_1/dy|_{y=0}$ decreases decrease monotonically with increasing collision energy for both protons and light nuclei. At $\sqrt{s_{\rm NN}}=$ 3 GeV, the $dv_1/dy|_{y=0}$  follow an approximate scaling with the atomic mass number $A$. The transport model calculation with a baryon mean-field and an afterburner coalescence qualitatively reproduce the measurements for both protons and light nuclei, as indicated by the short lines near the data points. The results indicate that the light nuclei are likely formed via the coalescence of nucleons at $\sqrt{s_{\rm NN}}=$ 3 GeV Au+Au collisions, where baryonic interactions dominate the collision dynamics~\cite{STAR:2021ozh}.
At $\sqrt{s_{\rm NN}} = 7.7$ GeV, the $A$ scaling still holds for $v_{1}$ of the deuteron. However, as we move to higher energies, the $dv_1/dy|_{y=0}$ values for protons become negative, while the corresponding value for deuterons keeps to be positive but with larger uncertainties~\cite{adam2020beam}. This discrepancy in the scaling behavior of light nuclei $dv_1/dy|_{y=0}$ at energies below 7.7 GeV and above 11.5 GeV may indicate a different production mechanism or system evolution, as it is expected that the QGP is formed at higher energies and the interactions occur at the partonic level~\cite{Adamczyk:2015ukd,Adamczyk:2013gw}.

Hypernuclei are nuclei containing at least one hyperon. As such, they are excellent experimental probes to study the hyperon-nucleon ($Y$–$N$) interaction~\cite{STAR:2010gyg}, an important ingredient in the EOS of dense nuclear matter~\cite{Steinheimer:2012tb,Chen:2023mel}. Similar to light nuclei production in heavy ion collisions, statistical thermal hadronization~\cite{Andronic:2010qu} and coalescence~\cite{Steinheimer:2012tb} have been proposed to describe hypernuclei formation. While thermal model calculations primarily depend only on the freeze-out temperature and the baryon-chemical potential, the $Y$–$N$ interaction plays an important role in the coalescence approach, through its influence on the dynamics of hyperon transportation in nuclear medium, as well as its connection to the coalescence criterion for hypernuclei formation from hyperons and nucleons~\cite{Steinheimer:2012tb}.

\begin{figure}[htb]       
\centering
\includegraphics[width=0.45\textwidth]{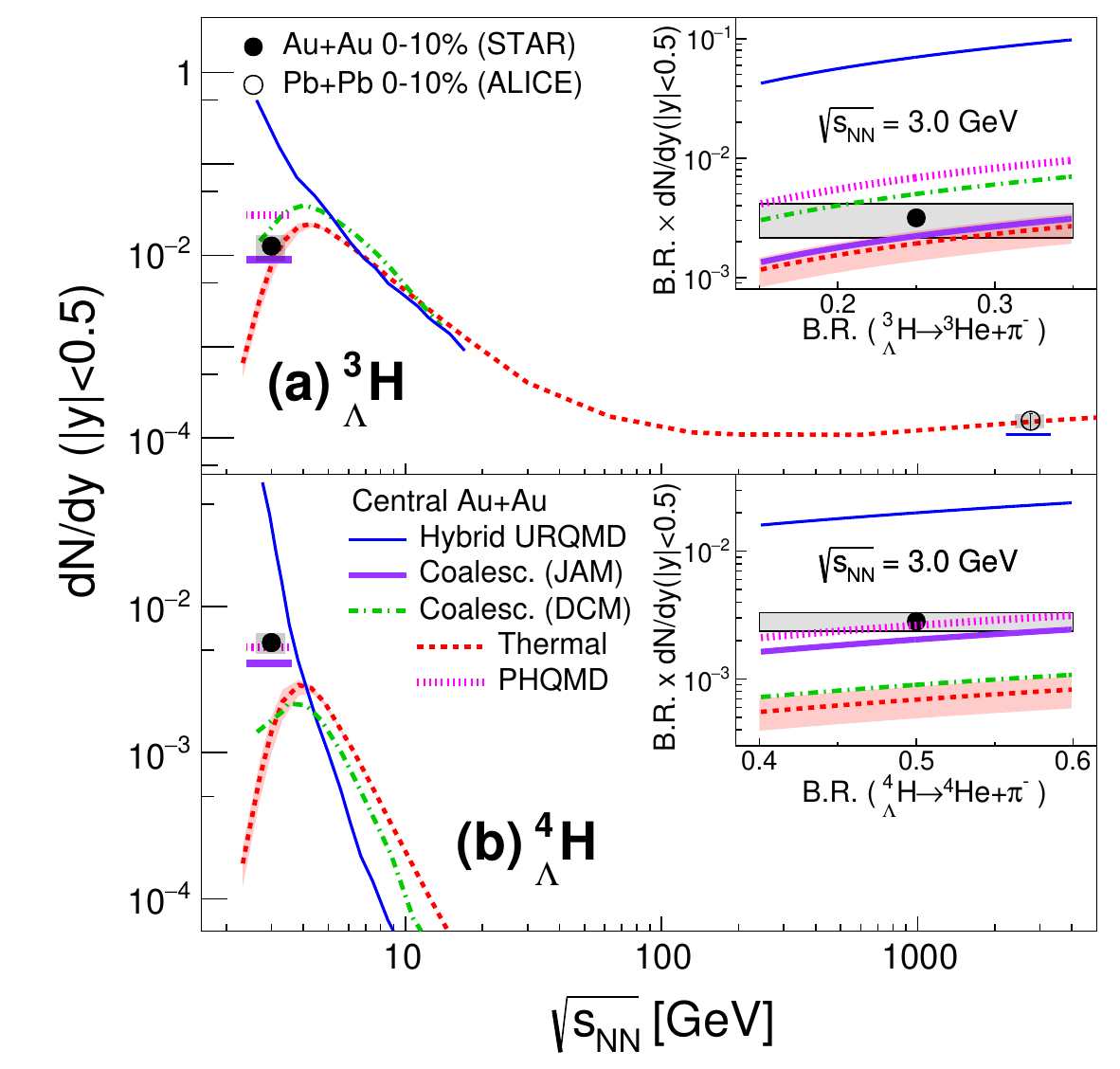}\vspace{-.5cm}
\caption{Beam energy dependent ${}^3_{\Lambda}$H (a) and ${}^4_{\Lambda}$H (b) yields at $|y|$$<$\,0.5 in central heavy ion collisions compared to theoretical model calculations. The data points assume a B.R. of 25(50)\% for ${}^3_{\Lambda}$H (${}^4_{\Lambda}$H) $\rightarrow$ ${}^3$He (${}^4$He) + $\pi^-$. The insets show their yields at $|y| <$ 0.5 times the B.R. as a function of the B.R.~\cite{STAR:2021orx}.}
\label{fig_III_H_hypernuclearYield}
\end{figure}

Figure~\ref{fig_III_H_hypernuclearYield} shows the ${}^3_{\Lambda}$H and ${}^4_{\Lambda}$H mid-rapidity yields for central Au+Au collisions of $\sqrt{s_{\rm NN}}=$ 3 GeV in comparison with the measurement at LHC. Instead, the insets show the $dN/dy \times B.R.$ as a function of B.R.. We observe that the ${}^3_{\Lambda}$H yield in Au+Au collisions at $\sqrt{s_{\rm NN}}=$ 3.0 GeV is significantly enhanced compared to the yield at LHC, likely driven by the increase in baryon density at low energies. Calculations from the thermal model~\cite{Andronic:2010qu}, which adopts the canonical ensemble for strangeness that is mandatory at low beam energies are compared to data. Interestingly, while the ${}^3_{\Lambda}$H yields at 3.0 GeV and 2.76 TeV are well described by the model, the ${}^4_{\Lambda}$H yield is underestimated by approximately a factor of 4. Coalescence calculations using DCM, an intra-nuclear cascade model to describe the dynamical stage of the reaction~\cite{Steinheimer:2012tb}, are consistent with the ${}^3_{\Lambda}$H yield while underestimating the ${}^4_{\Lambda}$H yield, whereas the coalescence (JAM) calculations are consistent with both. We note that in the DCM model, the same coalescence parameters are assumed for two hypernuclei, while in the JAM model, parameters are tuned separately for ${}^3_{\Lambda}$H and ${}^4_{\Lambda}$H to fit the data. It is expected that the calculated hypernuclei yields depend on the choice of the coalescence parameters~\cite{Steinheimer:2012tb}. Recent calculations from PHQMD~\cite{Glassel:2021rod}, a microscopic transport model which utilizes a dynamical description of hypernuclei formation, is consistent with the measured yields within uncertainties. Compared to the JAM model which adopts a baryonic mean-field approach, baryonic interactions in PHQMD are modelled by density dependent 2-body baryonic potentials. Meanwhile, the UrQMD-hydro hybrid model overestimates the yields at 3.0 GeV by an order of magnitude. The STAR measurements possess distinguishing power between different production models, and provide new baselines for the strangeness canonical volume in thermal models and coalescence parameters in transport coalescence models. Such constraints can be utilized to improve model estimations on the production of exotic strange matter in
the high baryon density region.

\begin{figure}[htb]       
\centering
\includegraphics[width=0.47\textwidth]{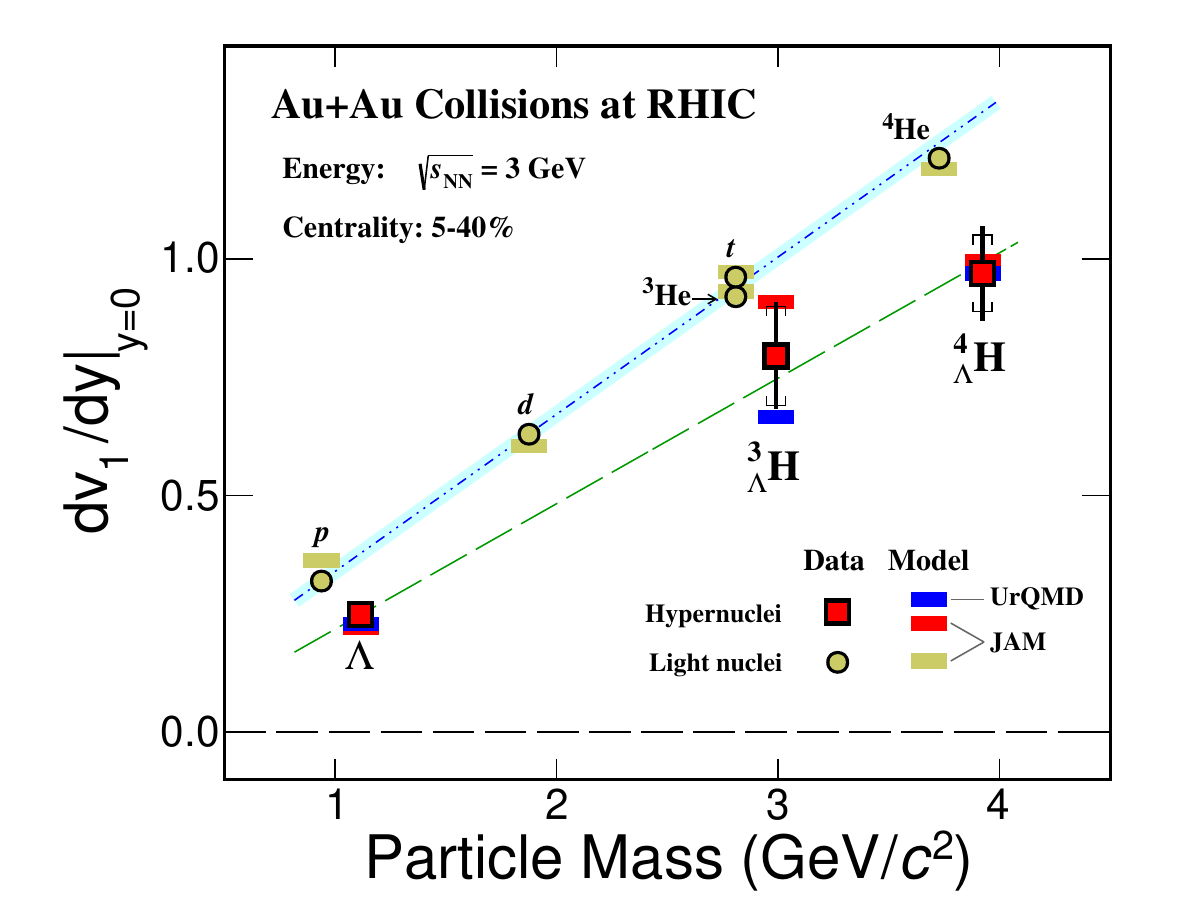}\vspace{-.5cm}
\caption{Mass dependence of light nuclei and hypernuclei $v_{1}$ slopes $dv_1/dy|_{y=0}$  from the $\sqrt{s_{\rm NN}}=3$ GeV 5\%-40\% centrality Au + Au collisions~\cite{STAR:2022fnj}.The dashed lines are the results of a linear fit to the measured light nuclei and hypernuclei $dv_1/dy|_{y=0}$, respectively.The calculations of transport models plus coalescence afterburner are shown as gold and red bars from the JAM model, and blue bars from the UrQMD model.}
\label{fig_III_H_hyperNucleadv1dy}
\end{figure}

The STAR experiment reported the first observation of the $v_1$ of hypernuclei ${}^3_{\Lambda}$H and ${}^4_{\Lambda}$H in 3 GeV Au+Au collisions~\cite{STAR:2022fnj}, as shown in Fig.~\ref{fig_III_H_hyperNucleadv1dy}.
The mass dependence of $dv_1/dy|_{y=0}$ for $\Lambda$ and hypernuclei is similar to that of light nuclei, increasing linearly with the particle mass, i.e., following a baryon mass number scaling.  While it is noteworthy that the $dv_1/dy|_{y=0}$ values for hypernuclei is systematically lower compared to those for nuclei of equivalent mass numbers. This discrepancy may be attributed to the fact that the $dv_1/dy|_{y=0}$ for $\Lambda$ is lower than that for protons.
The calculations using transport model plus an afterburner qualitatively reproduce the data within uncertainties, suggesting that the hypernuclei are produced via coalescence of hyperon and light nuclei core
in such heavy ion collisions.
If hypernuclei are formed through the coalescence process, both their $v_1$ and yield could be affected the interactions involving hyperons and nucleons ($Y$-$N$), which is essential for understanding the inner structure of compact stellar objects.
The linear fits to the extracted $dv_1/dy|_{y=0}$ in Fig.~\ref{fig_III_H_hyperNucleadv1dy} shows comparable slopes considering uncertainties for both light nuclei and hypernuclei, but their central value are slightly different. This difference may originate from the differences in nucleon-nucleon and $Y$-$N$ interactions.
Thus, more precise measurements with increased statistics, especially at high baryon density, will be crucial in elucidating the production mechanisms of hypernuclei and hyperon-nucleon interactions in the future.
%--==========================================================
%---===============================================
\subsection{Heavy Flavor Hadron Production} 

Heavy flavor hadrons are hardrons with at least one constituent heavy flavor quark. They are penetrating probes of QGP. Heavy flavor quarks are predominantly produced through initial hard scattering processes in heavy ion collisions thanks to their large masses. These initial hard processes happen before the formation of QGP. Consequently, heavy flavor quarks experience the whole evolution of QGP created in heavy ion collisions. Heavy flavor quarks interact with the deconfined quarks, mainly light flavor quarks, and gluons when they transit QGP and approach thermalization. Their thermal relaxation time is expected to be comparable to or longer than the lifetime of the QGP created in heavy ion collisions. Heavy flavor quarks may gain collectivity from the collectively expanding hot medium. The collectivity of heavy flavor quarks is sensitive to the hot medium transport properties, especially the parameter called the heavy flavor diffusion coefficient $\mathcal{D}_s$~\cite{Akiba:2015jwa}.

Significant elliptic flow ($v_2$) for charmed meson $D^0$ is observed in Au+Au collisions at $\sqrt{s_{\rm NN}}$ = 200 GeV by the STAR collaboration~\cite{STAR:2017kkh}. The $D^0$s are fully reconstructed via the two body decay to charged pions and kaons with branching ratio of $(3.95 \pm 0.03)\%$. The random combinatorial background of pions and kaons originating from primary vertices are significantly suppressed by precise measurements of the distance of closest approach (DCA) between tracks and primary vertex thanks to the relative large $c\tau$ of $D^0$ mesons ($\approx 123~\mu$m). The precise measurements of DCA are provided by the Heavy Flavor Tracker installed in STAR during 2014 and 2016. The $v_2$ results for $D^0$ mesons in 10-40\% central Au+Au collisions at $\sqrt{s_\textrm{NN}}=200$ GeV is compared to those for light flavor hadrons ($K_S$, $\Lambda$, $\Xi^-$)~\cite{STAR:2008ftz}. At $p_{\rm T}<2$ GeV/$c$, the $v_2$ for $D^0$ mesons is found to be smaller than those of light flavor hadrons, exhibiting mass-ordering behavior expected from hydrodynamics. At $p_{\rm T}>2$ GeV/$c$, $D^0$ $v_2$ is consistent with that of light flavor mesons such as $K_S$. The comparison of $v_2/n_q$ as a function of $(m_{\rm T}-m_{\rm 0})/n_q$, where $n_q$ is the number of constituent quarks (NCQ) in a hadron, among these hadrons show that $D^0$ elliptic flow follows the universal trend as the light hadrons. These comparisons indicate that charm quarks gain significant flow through interaction with the strongly coupled QGP created in 10-40\% Au+Au collisions at RHIC top energy. Recent phenomenological models constrained by the $D^0$ $v_2$ measurement as well as measurements of heavy flavor quarks $v_2$ using single electrons from heavy flavor hadron decays (HFE) suggest that the dimensionless charm quark spatial diffusion coefficient $2\pi T \mathcal{D}_s$ is 2-5 in the vicinity of the critical temperature (Ref.~\cite{Dong:2019byy} and references therein). The value is consistent with theoretical calculations from quenched lattice QCD within large uncertainties. The dependence of the heavy flavor quarks diffusion coefficient $D_s$ on heavy flavor quark momentum, as well as temperature and baryon chemical potential of QGP is yet to be determined. The measurements of heavy flavor quarks collectivity in Au+Au collisions at energy below the RHIC top energy enabled by the RHIC BES program can shed new light on the temperature and baryon chemical potential dependence of the QGP transport parameter $\mathcal{D}_s$.

The elliptic flow of heavy flavor hadrons from RHIC BES program was measured in Au+Au collisions at $\sqrt{s_\textrm{NN}}=$27, 39, 54.4 and 62.4 GeV~\cite{STAR:2023eui,STAR:2014yia}. The 39 GeV and 62.4 GeV data were taken in 2010 during the first phase of the RHIC BES program. The number of events used for the analyses are 87 and 38 million, respectively. The 27 GeV and 54.4 GeV data were taken in 2018 and 2017 between the first phase and the second phase of the RHIC BES program. The number of events passed the event-level criterion were 240 and 570 million, respectively. Due to an order of magnitude difference of the number of events, combined with significant energy dependence of heavy flavor hadron production cross section, the precision of the measurements at different energies varies a lot. The results from the 54.4 GeV collisions have the best precision. While the fully reconstruction of heavy flavor hadrons in these data is not possible due to lack of silicon vertex devices, the electrons from heavy flavor hadron decays are used as proxy of heavy flavor hadrons. Electrons are identified using the inverse velocity calculated from the path length measured by the STAR TPC, and time of flight measured by the Vertex Position Detector providing start time measurement and the TOF detector providing stop time measurement. The electron candidates are further selected by the ionization energy loss in the gas of the TPC. The number of electrons are corrected for purity. The dominate source of background for heavy flavor decay electrons are photonic electrons which are produced via Dalitz decay of light mesons such as $\pi^0$ and $\eta$, and photon conversion in the detector material. The yield of non-photonic electrons (NPE) is calculated as:
\begin{equation}
N^\textrm{NPE} = N^\textrm{INC} - N^\textrm{PE},
\end{equation}
where $N^\textrm{INC}$ and $N^\textrm{PE}$ represent the yield of inclusive and photonic electrons (PE), respectively. Photonic electron candidates are selected via the invariant mass distribution of inclusive electron candidates and partner electrons from the same event. The yield of photonic electron can be expressed as:
\begin{equation}
   N^\textrm{PE} = (N^\textrm{UL} - N^{LS})/\varepsilon,  
\end{equation}
where $N^\textrm{UL}$ and $N^{LS}$ are respectively the raw yield of unlike-sign and like-sign pairs and $\varepsilon$ is the partner electron finding efficiency. The $v_2$ of inclusive electron and photonic electrons is extracted by the event-plane $\eta$-sub method. The $v_2$ of NPE is calculated by:
\begin{equation}
    N^\textrm{NPE} v_2^\textrm{NPE} = N^\textrm{INC} v_2^\textrm{INC} - N^\textrm{PE} v_2^\textrm{PE} -\sum{f_h \times N^\textrm{INC} v_2^h},
\end{equation}
where $v_2^\textrm{INC}$, $v_2^\textrm{PE}$ and $v_2^h$ are $v_2$ of inclusive electrons, photonic electrons and hadrons contaminated in inclusive electron candidates, $f_h$ is the hadron contamination fraction.

In addition to photonic electrons, other major background sources for heavy flavor decay electrons are from decay of kaons ($K_{e3}$) and vector mesons ($\rho$, $\omega$ and $\phi$). They are subtracted by:
\begin{equation}
v_2^\textrm{HFE} = v_2^\textrm{NPE}(1+f_{K_{e3}}+f_\textrm{VM})-f_{K_{e3}} \times v_2^{K_{e3}} + f_\textrm{VM}\times v_2^\textrm{VM},
\end{equation}
where $f_{K_{e3}}$ and $f_\textrm{VM}$ are the estimated yield ratio of electrons decays of kaons and vector mesons, respectively, to HFE. The residual non-flow contribution is estimated according to the HFE-hadron correlation in p+p collisions and the hadron multiplicity in Au+Au collisions.

\begin{figure}[th]       
\centering
\includegraphics[width=0.48\textwidth]{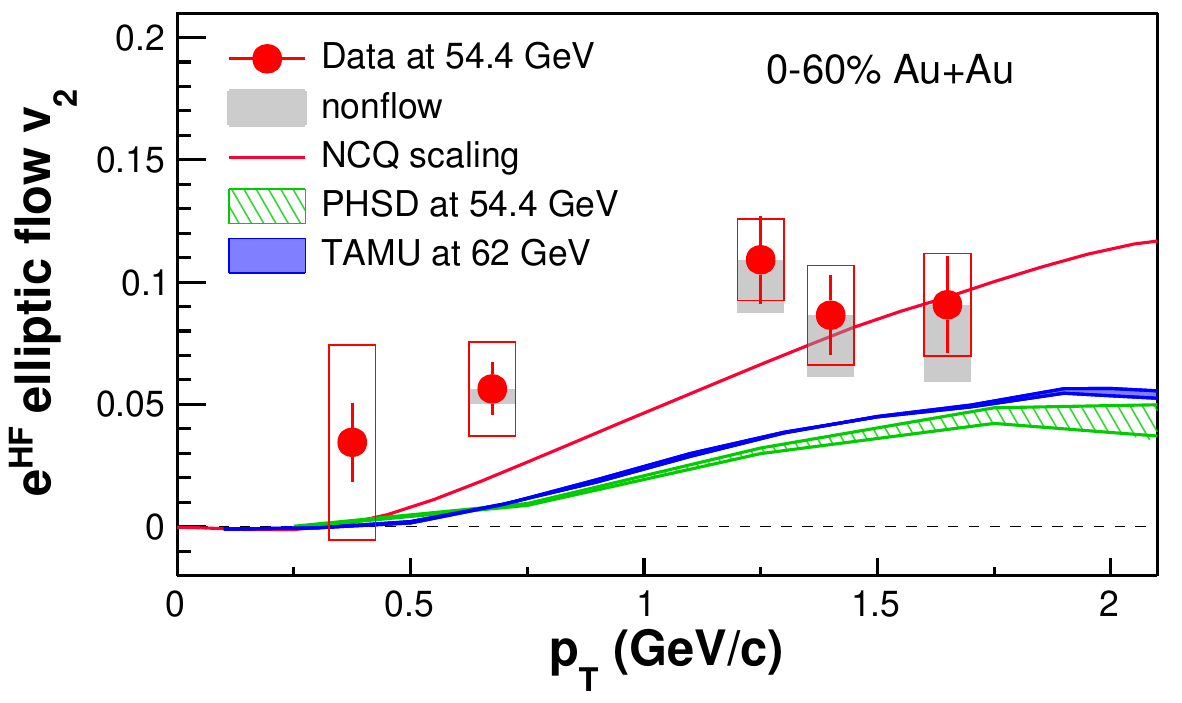}
\vspace{-.5cm}
\caption{$v_2$ as a function of $p_{\rm T}$ for heavy flavor decay electrons at midrapidity in Au+Au collisions at $\sqrt{s_{\rm NN}}$ = 54.4 GeV~\cite{STAR:2023eui}, compared to TAMU~\cite{He:2014epa} and PHSD~\cite{Song:2015sfa,Song:2016rzw} model calculations. }
\label{fig_HFe_v2_Energy}
\end{figure}

Fig.~\ref{fig_HFe_v2_Energy} shows elliptic flow coefficient $v_2$ of heavy flavor decay electrons as a function of $p_T$ at mid-rapidity ($|y|<0.8$) in Au+Au collisions at $\sqrt{s_\textrm{NN}}$ = 54.4 GeV. The error bars and boxes depict statistical and systematic uncertainties, respectively. The hatched areas indicate the estimated non-flow contributions. Significant $v_2$ is observed at $0.5 < p_T < 2$ GeV/$c$, the average $v_2$ in the $p_T$ range from 1.2 to 2.0 GeV/$c$ is $0.094 \pm 0.008~\textrm{(stat.)} \pm 0.014~\textrm{(syst.)}$, while the estimated upper limit of non-flow is only 0.02. The dashed curve represents the projected charm quark decay electron $v_2$ assuming open charm hadron $v_2$ follows NCQ scaling with other light hadron. Because charm quark is the dominate contributor of HFE in this $p_T$ range, the significant $v_2$ and the consistency between the data and the dashed curve indicate that charm quarks interact with the hot medium and may reach local thermal equilibrium in Au+Au collisions, even though the center-of-mass energy is nearly a factor of 4 lower than the RHIC top energy.

The two bands shown in Fig.~\ref{fig_HFe_v2_Energy} are calculations from two phenomenological models, TAMU~\cite{He:2014epa} and PHSD~\cite{Song:2015sfa,Song:2016rzw}. Both models assume that heavy flavor quarks interact with the QGP medium elastically. The assumption is generally accepted in the low $p_{\rm T}$ region. The elastic scattering are implemented in different way in the two models. In the TAMU model, the microscopic elastic interaction between heavy flavor quarks and quarks/gluons in the hot, dense medium are evaluated using non-perturbative T-Matrix calculations. The heavy flavor quark transport coefficient calculated is then fed into macroscopic Langevin simulation of heavy quark diffusion through the background medium, which are modeled by ideal 2+1D hydrodynamics. In the PHSD model, heavy flavor quarks interact with the off-shell massive partons in the QGP medium. The masses and width of the partons in the QGP medium and the scattering probability are given by the dynamical quasi-particle model. In both models, the heavy flavor quarks hadronized through both coalescence and fragmentation. In the PHSD model, the hadronized heavy flavor hadrons subsequently interact with other hadrons in the hadronic phase. Although the calculations from both TAMU and PHSD models are systematical lower than the measurements, the deviation is only 1-2$\sigma$ at $p_{\rm T}>0.5$ GeV/$c$ if taken the estimated upper limit of non-flow contribution into account. Furthermore, both models do not consider the contribution from charm baryons, whose yield is measured to be even enhance in heavy ion collisions relative to that of mesons~\cite{STAR:2019ank}. This contribution will slightly increase HFE $v_2$ at $p_{\rm T} > 1$ GeV/$c$.

\begin{figure*}[ht]       
\centering
\includegraphics[width=0.9\textwidth]{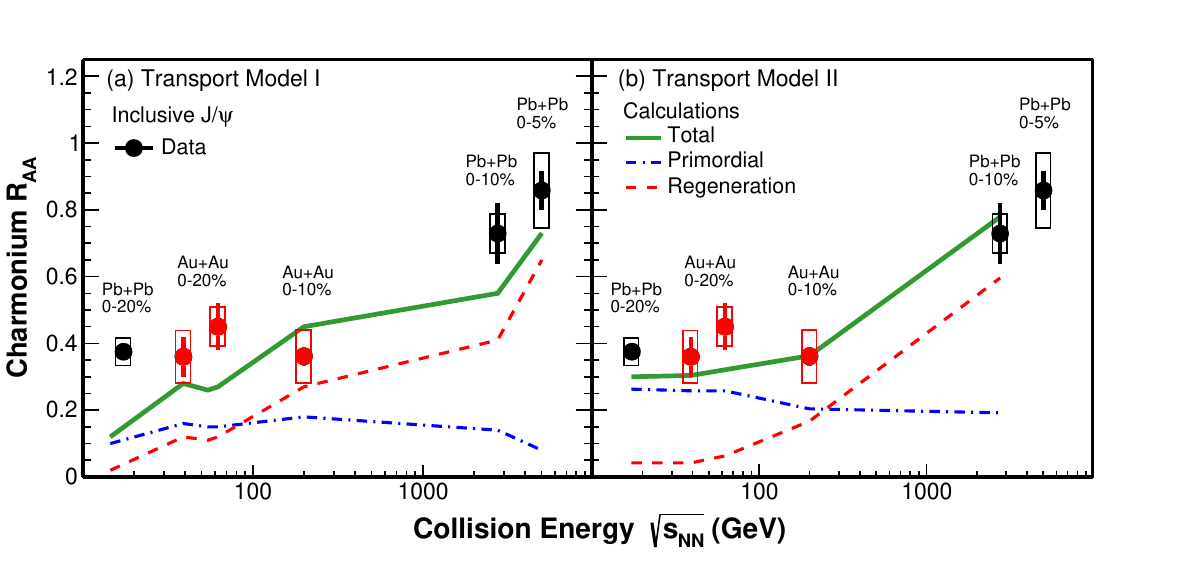}\vspace{-.5cm}
\caption{$J/\psi$ nuclear modification factors ($R_{\textrm{AA}}$) as a function of center-of-mass energy in central heavy ion collisions. The solid circles represent the measurements at SPS, RHIC and the LHC~\cite{NA50:2000brc,STAR:2016utm,STAR:2019fge,ALICE:2013osk,ALICE:2023gco}, and the curves in the left (right) panel depict calculations from transport model I~\cite{Zhao:2022ggw} (II~\cite{Zhao:2010nk}).}
\label{fig_Jpsi_RAA}
\end{figure*}

Heavy quarkonium is a bound state of heavy flavor quark and its antiquark. The pairs of heavy flavor quark and its antiquark are produced predominantly in the initial scattering in heavy ion collisions and are tightly bound together thus less sensitive to the interaction with other particles. However, it is believed that the color potential of the bound states is subject to be modified when QGP is formed, resulting in dissociation of heavy quarkonium~\cite{Matsui:1986dk,Xu:1995eb,Yao:2018sgn,Sharma:2012dy}. The suppression of quarkonium yield in heavy ion collisions arising from the modification of the potential is considered as the `smoking-gun' signature of deconfinement in QGP. The suppression is sensitive to the temperature profile of QGP because the modification of potential between a heavy quark and its antiquark in QGP is sensitive the temperature of the medium.

The suppression of $J/\psi$ in heavy ion collisions was extensively studied in experiments at CERN SPS~\cite{Kluberg:2005yh}. It was found that the production yield of quarkonium in heavy ion collisions is also affected by cold nuclear matter (CNM) effects. Suppression of $J/\psi$ yield, beyond the expected CNM effects based on the results from proton and nucleus collisions, was observed in central Pb+Pb collisions at 17.3 GeV and was considered as evidence of deconfinement in QGP~\cite{NA50:2000brc}. 

However, the first quarkonium measurement in heavy ion collisions at RHIC was very puzzling. The $J/\psi$ suppression, quantified by the nuclear modification factors, and its centrality dependence measured in Au+Au collisions at $\sqrt{s_\textrm{NN}}=200$ GeV~\cite{PHENIX:2006gsi} is found to be consistent with that observed in Pb+Pb collisions at $\sqrt{s_\textrm{NN}}=17.3$ GeV. $J/\psi$ suppression is expected to be stronger at higher collision energy where the initial temperature is believed to be systematically higher. Furthermore, the suppression was observed to significantly stronger at forward rapidity that at mid-rapidity, where the energy density is believed to be higher compared to the part at forward rapidity. An additional production mechanism, (re)combination of charm quark and anti-charm quark in the QGP medium, was proposed to solve the problem~\cite{Braun-Munzinger:2000csl,Grandchamp:2003uw}. Unlike the static and dynamic screening of the potential between heavy quark and its antiquark, the (re)combination mechanism could enhance quarkonium yield in heavy ion collisions. Because the yield of quarkonium from the (re)combination mechanism is approximately proportional to the square of total heavy quark cross section, which exhibits significant increase trend as collisions energy increases, the contribution from (re)combination mechanism should have clear center-of-mass energy dependence. The RHIC BES program provides an unique opportunity to vary the initial temperature and number of charm quark pairs in the same event and to shed new light on the production mechanism of quarkonium in heavy ion collisions.

The data for the $J/\psi$ production study during the RHIC BES is taken in 2010 by the STAR experiment at $\sqrt{s_\textrm{NN}}=39$ and 62.4 GeV~\cite{STAR:2016utm}. The total number of minimum bias triggered events are 182 million and 94 million respectively. The $J/\psi$s are reconstructed through their decays into electron-positron pairs. The electron daughters are identified by combining the information from the STAR TPC and the TOF. The random combinatorial background is reconstructed using the mixed-event technique. The invariant mass spectrum for unlike-sign pairs from mixed-events are normalized to that for like-sign pairs from same events, and subtracted from that for unlike-sign pairs from same events. The combinatorial background subtracted invariant mass spectrum is fit to $J/\psi$ template from Monte Carlo simulations plus a linear function for residual background to extract the $J/\psi$ yields. The nuclear modification factors ($R_\textrm{AA}$) is calculated as:
\begin{equation}
R_\textrm{AA} = \frac{1}{T_\textrm{AA}} \frac{d^2N_\textrm{AA}/dp_Tdy}{d^2\sigma_\textrm{pp}/dp_Tdy},
\end{equation}
where $d^2N_\textrm{AA}/dp_Tdy$ is the efficiency and acceptance corrected $J/\psi$ $p_T$ spectrum measured in A+A collisions, $T_\textrm{AA}$ is the nuclear overlap function from Glauber Monte Carlo simulations, and $d^2\sigma_\textrm{pp}/dp_Tdy$ is the $J/\psi$ production cross section in $p$+$p$ collisions at the same energy as that of A+A collisions. The production cross section of $J/\psi$ in $p$+$p$ collisions at $\sqrt{s}=39$ and 62.4 GeV are derived from an interpolation based on world wide $J/\psi$ data because there are no measurement available for $p$+$p$ collisions at $\sqrt{s}=39$ and 62.4 GeV and previous measurements near these two energies from the Intersecting Storage Ring collider experiments show discrepancies among different measurements~\cite{Zha:2015eca}. 

Figure~\ref{fig_Jpsi_RAA} shows the center-of-mass energy dependence of $J/\psi$ nuclear modification factors measured at midrapidity in central heavy ion collisions from SPS, RHIC to LHC energies. The data from RHIC are measured in Au+Au collisions, while the data from SPS and LHC are measured in Pb+Pb collisions~\cite{NA50:2000brc,STAR:2016utm,STAR:2019fge,ALICE:2013osk,ALICE:2023gco}. The error bars and boxes represent statistical and systematic uncertainties. It shows that the $J/\psi$ $R_\textrm{AA}$ remains a constant from $\sqrt{s_\textrm{NN}}=17.3$ to 200 GeV and substantially increases from RHIC top energy to LHC energies. 

The curves in the left and right panels of Fig.~\ref{fig_Jpsi_RAA} depict the calculations from two transport models~\cite{Zhao:2022ggw,Zhao:2010nk}. The dot-dashed lines represent the contribution from primordial $J/\psi$s which are affected by the static/dynamic color-screening of the potential in QGP medium and CNM effects, while the dashed lines represent the contribution from (re)combination. The solid lines represent the sum of the two components. Although both transport models can describe the data, except for the transport model I at SPS energy, the decomposed contributions from primordial and (re)combination are quite different, indicating further constraint on the understanding of $J/\psi$ production mechanisms in heavy ion collisions is urgent before using it to extract the properties of QGP. STAR has taken much larger data samples at different energies during the second phase of RHIC BES program. Preliminary results show that $J/\psi$ suppression can be measured precisely in Au+Au collisions at 54.4 GeV, and the suppression measurements can be extended to energy down to 14.6 GeV, an energy below SPS top energy. These new data will shed new lights on the production mechanism of $J/\psi$ in heavy ion collisions.

%--==========================================================
%---===============================================
\subsection{Di-lepton Production}

Photons and dileptons ($e^+e^-$ or $\mu^+\mu^-$) emerge at various stages throughout the space-time evolution of the nuclear medium formed in ultra-relativistic heavy ion collisions. As penetrating electromagnetic probes, dileptons remain unaffected by strong interactions, preserving un-distorted information about their sources. These sources manifest differently in various lepton-pair invariant mass ($M_{ll}$) regions, typically categorized into three classes. In the low-mass region (LMR), below the $\phi$ mass ($M_{ll} < $1.1 GeV/$c^2$), predominant contributions arise from decays of light mesons ($\pi^0$, $\eta$, $\rho^{0}$, $\omega$, $\phi$). Investigation of $\rho^{0}$ spectra modifications allows for probing the in-medium hadronic properties, which are particularly sensitive to mechanisms of chiral symmetry restoration in QCD matter~\cite{Hohler:2013eba}. The expected modifications in dilepton yields within the LMR provide insights into the medium's lifetime and the transition from hadronic to partonic degrees of freedom~\cite{Rapp:1999us}. In the intermediate-mass region (IMR), which lies between the $\phi$ and $J/\psi$ masses ($M_{ll} \simeq 1.2 - 3 \text{ GeV}/c^2$), the invariant mass spectrum appears as a continuum arising from both heavy flavor decays and QGP thermal radiation. This provides an opportunity to directly measure the average temperature of the QGP~\cite{Rapp:2014hha} by extracting the inverse slope of the mass spectra, which remain unaffected by the blue shift of the expanding system. In the high-mass region (HMR), defined as $M_{ll}\geq$3GeV/$c^2$, primary sources contributing to the dielectron spectrum are heavy flavor/quarkonium decays and the Drell-Yan process. Detailed discussions on the associated physics for HMR have been provided in the preceding section and will not be reiterated here.

\begin{figure}[htb]
\centering  
\includegraphics[width=0.475\textwidth]{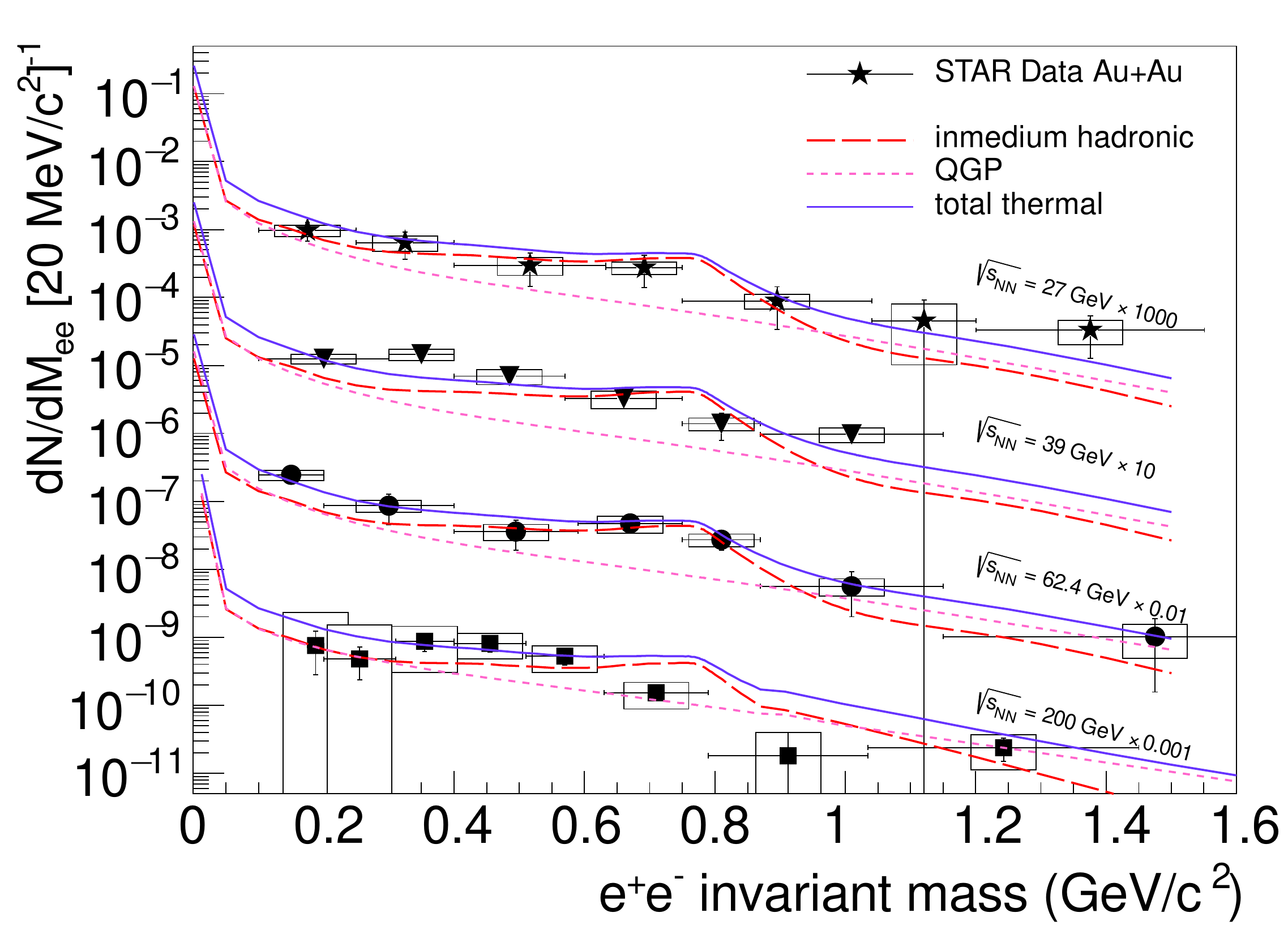}
\caption{The dielectron invariant mass spectra in Au+Au collisions at $\sqrt{s_{\rm{NN}}} =$ 27, 39, 62.4, and 200 GeV measured by the STAR Collaboration~\cite{STAR:2018xaj,STAR:2015zal}. The spectra are shown after the subtraction of hadronic background contributions (cocktail), compared with theoretical model calculations~\cite{Rapp:2014hha,Rapp:1999us,vanHees:2007th}. The theoretical predictions represent the total thermal radiation (blue solid lines), including contributions from both in-medium hadronic processes (red dashed lines) and QGP thermal radiation (red dotted lines). The figure is sourced from Ref.~\cite{STAR:2018xaj}.}
\label{dielectron-mass-spectrum}
\end{figure}

To achieve precise measurements of the aforementioned dileptons, it is crucial to utilize detectors with large, uniform acceptance and excellent lepton identification capabilities. The integration of the TOF detector has paved the way for dilepton measurements at STAR. Specifically, by combining timing measurements from the TOF detector with momentum and ionization energy loss ($\langle dE/dx \rangle$) measurements from the TPC, robust identification of electrons over a wide $p_{\rm T}$ range is achieved. This identification is characterized by high efficiency and purity, facilitating comprehensive dielectron analysis. The identified electron and positron candidates are paired by opposite and same sign charges, called unlike-sign and like-sign pairs, respectively. The like-sign pairs are used to statistically model both the combinatorial and correlated backgrounds. Moreover, the subtraction of decay products from light mesons, known as the "cocktail", is achieved through simulations. These dielectron spectra are of particular interest, as they are anticipated to carry radiation signatures from various stages of heavy ion collisions prior to freeze-out.

The acceptance-corrected excess dielectron mass spectra, following the careful removal of background contributions, have been thoroughly measured by the STAR Collaboration across various collision energies~\cite{STAR:2018xaj,STAR:2015zal}, as illustrated in Fig.~\ref{dielectron-mass-spectrum}. Accompanying these measurements are model calculations~\cite{Rapp:2014hha,Rapp:1999us,vanHees:2007th} depicting the total thermal radiation (solid lines), which consider contributions from both in-medium hadronic processes (dashed lines) and the QGP phase (dotted lines). Remarkably, the model predictions provide a coherent framework for interpreting the measured dielectron spectra across a wide energy range and invariant mass regions. In the low-mass region, the predominant hadronic radiation is primarily attributed to the in-medium $\rho$ broadening, stemming from its interactions with the hadronic medium, particularly baryons. Notably, this model also yields a consistent description of the invariant mass spectrum of dimuon pairs measured by the NA60 experiment at the SPS~\cite{NA60:2008ctj}. The observed in-medium $\rho$ broadening serves as a compelling indicator of the partial restoration of chiral symmetry within the hot QCD medium~\cite{HotQCD:2012vvd}. However, in the intermediate-mass region, the contribution from QGP radiation is anticipated, although current measurements still lack precision in this regime. Consequently, the search for and exploration of QGP thermal radiation remain pivotal future endeavors in the dilepton programs at both RHIC and LHC experiments.

\begin{figure}[htb]
\centering  
\includegraphics[width=0.475\textwidth]{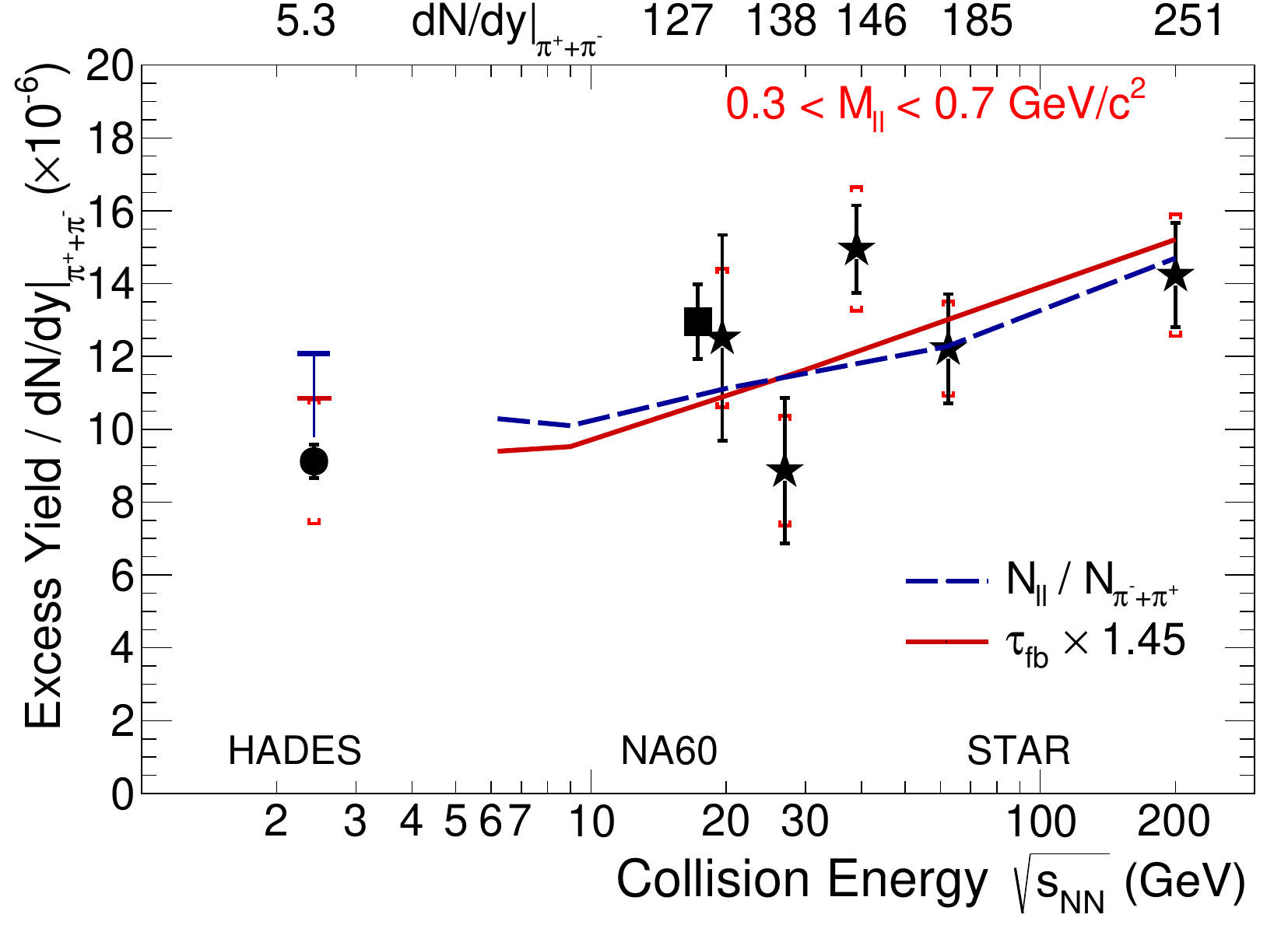}
\caption{The collision energy dependence of the integrated dielectron yield, normalized by the charged pion yield $dN/dy$, in the mass region $0.3 < M_{ll} < 0.7$ GeV/$c^2$, as measured by HADES~\cite{Agakishiev:2011vf}, NA60~\cite{NA60:2008ctj}, and STAR Collaborations~\cite{STAR:2015tnn,STAR:2015zal,STAR:2018xaj}. This comparative analysis is supplemented by model calculations of dielectron yields (dashed blue lines) and fireball lifetime (solid red lines)~\cite{vanHees:2007th}. The figure is sourced from Ref.~\cite{Luo:2022mtp}.}
\label{dielectron-mass-low}
\end{figure}

To quantitatively compare the excess in the LMR, the integrated excess yield of dielectrons in the mass region $0.3 < M_{ll} < 0.7$ GeV/$c^2$ is normalized to the charged pion yield $dN/dy$ to cancel out the volume effect. Figure~\ref{dielectron-mass-low} shows the collision energy dependence of the integrated dielectron yield, as measured by HADES~\cite{Agakishiev:2011vf}, NA60~\cite{NA60:2008ctj}, and STAR~\cite{STAR:2015tnn,STAR:2015zal,STAR:2018xaj} Collaborations. The figure also includes theoretical model calculations depicting the dielectron yields (dashed blue lines) and the fireball lifetime (solid red lines)~\cite{vanHees:2007th}. Impressively, the model provides a commendable description of the energy dependence, illustrating a modest increase from the SPS to the top RHIC energy. This observed increase effectively tracks the fireball lifetime over a broad spectrum of collision energies. Notably, the STAR measurements presented here pertain to the BES phase I. However, the subsequent analysis of BES-II data promises to extend these measurements from 19.6 GeV down to 7.7 GeV, providing fresh insights into the properties of the hot medium within the high baryon density regime.

\begin{figure}[htb]
\centering  
\includegraphics[width=0.475\textwidth]{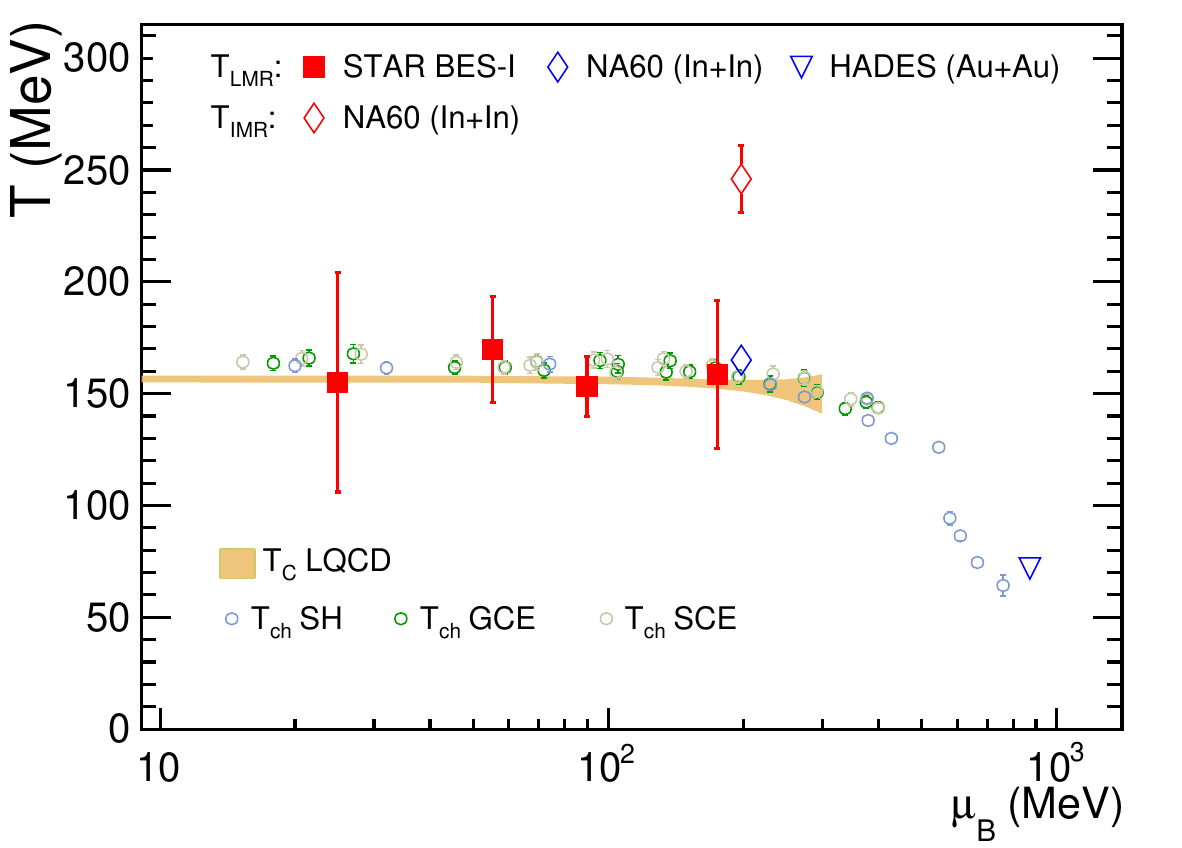}
\caption{Temperatures versus baryon chemical potential, with temperatures extracted from the in-medium $\rho$ mass spectra at LMR. The earlier QGP stage region from NA60 data~\cite{NA60:2008dcb} (diamonds) and LMR from HADES data~\cite{HADES:2019auv} (inverted triangle) are also depicted. Chemical freeze-out temperatures extracted from statistical thermal models (SH, GCE, SCE) \cite{Andronic:2017pug,STAR:2017sal} are represented as open circles. The QCD critical temperature $T_{\rm{C}}$ at finite $\mu_{\rm B}$, predicted by LQCD calculations \cite{HotQCD:2018pds}, is shown as a yellow band. The figure is adapted from STAR~\cite{STAR:2024bpc}.}
\label{dielectron-T-muB}
\end{figure}

Traditionally, LMR dileptons have been used to explore the in-medium broadening of the $\rho$ meson and its association with chiral symmetry restoration. The impact on the invariant mass distribution of dileptons is often overlooked, as it is considered a trivial thermal factor incorporated into models for data comparison~\cite{Rapp:1999us,Rapp:2014hha}. Recent observations by STAR~\cite{STAR:2024bpc} indicate that the broadening of the $\rho$ meson significantly determines the temperature of the thermal source responsible for LMR radiation. To extract this temperature, a fitting function that combines the in-medium resonance structure with the continuum thermal distribution is applied to the measured mass spectrum. In a vacuum, the mass line shape of the $\rho$ decaying into dileptons is represented by a relativistic Breit-Wigner function, $f^{\rm{BW}}(M)$. Within a hot QCD medium, this line shape is modified by multiplying $f^{\rm{BW}}(M)$ with a Boltzmann factor, $e^{-M/T}$, to account for phase space effects~\cite{STAR:2024bpc}. Furthermore, if the $\rho$ is completely dissolved in the medium, its mass spectral structure spreads out and approaches a smooth distribution similar to the dielectron continuum from QGP thermal radiation, described by $M^{3/2}e^{-M/T}$\cite{Rapp:2014hha}. Fig.~\ref{dielectron-T-muB} illustrates temperatures derived from BES-I dielectron data as a function of the baryon chemical potential $\mu_{\rm{B}}$. The chemical freeze-out temperature $T_{\rm{ch}}$ and $\mu_{\rm{B}}$ are determined by applying statistical thermal models to the yields of hadron production. $T_{\rm{ch}}$ from several statistical thermal models and the QCD critical temperature $T_{\rm{C}}$ from lattice QCD~\cite{Andronic:2017pug,STAR:2017sal,HotQCD:2018pds} are shown in Fig. \ref{dielectron-T-muB} as open circles and a shaded band, respectively. Similarly, temperatures extracted from previously published low-mass thermal dielectron spectra \cite{NA60:2008dcb,HADES:2019auv} are presented. Notably, temperatures extracted from BES-I and SPS LMR closely align with $T_{\rm{ch}}$ from statistical thermal models and $T_{\rm{C}}$ from lattice QCD. This alignment suggests that dielectron emission at LMR is mainly influenced by $\rho$ broadening during the phase transition (or mixed phase), and the chemical freeze-out temperature at RHIC BES energies lies at the phase transition boundary. Recent analyses by the STAR Collaboration have extracted temperatures from IMR thermal dileptons for Au+Au collisions at $\sqrt{s_{NN}} = 27$ and 54 GeV, as reported in Ref.~\cite{STAR:2024bpc}. The extracted temperatures are significantly higher than those from statistical thermal models and lattice QCD calculations, indicating that IMR dileptons primarily originate from the earlier partonic stage of the collisions. Further details can be found in Ref.~\cite{STAR:2024bpc}.

In relativistic heavy ion collisions, dileptons emerge not only from hadronic processes but also through the interaction of the intense electromagnetic fields accompanying the colliding ions, known as the Breit-Wheeler process~\cite{Bertulani:2005ru,Baur:2007zz}. These fields can be treated as a spectrum of equivalent photons, with the photon flux being proportional to the square of the particle's charge ($Z^2$), resulting in dilepton production scaling with $Z^4$. Initially, dilepton production from the two-photon process was studied in ultra-peripheral collisions, where the impact parameter is large enough to avoid hadronic interactions. However, recent observations have shown that such photo-production also occurs in hadronic heavy ion collisions~\cite{STAR:2018ldd,ATLAS:2018pfw}, prompting theoretical advancements to describe these processes~\cite{Zha:2018ytv,Klusek-Gawenda:2015hja,Klein:2018cjh}. In events with hadronic overlap, dilepton photo-production occurs alongside hadronic interactions, offering a new method to probe the QGP, especially its electromagnetic properties. Data from peripheral collisions show a discrepancy in the $p_{\rm T}^{2}$ distribution between experimental results and theoretical models without considering the impact parameter dependence of photon kinematics, suggesting potential alternative origins of $p_{\rm T}^{2}$ broadening, possibly linked to a postulated, trapped magnetic field or Coulomb scattering in the hot and dense medium~\cite{STAR:2018ldd,ATLAS:2018pfw}. However, theoretical calculations that account for impact parameter dependence can explain the observed broadening~\cite{Zha:2018tlq,Klein:2020jom,Klusek-Gawenda:2020eja,Wang:2021kxm,Li:2019sin}, indicating the significant influence of the initial electromagnetic field strength, which were later confirmed by the CMS measurements~\cite{CMS:2020skx}. Future precision measurements at STAR, CMS, and ATLAS will further explore these effects, potentially revealing medium induced modifications in dilepton kinematics.

%--==========================================================
%--==========================================================
\section{Summary and Outlook}\label{Sec:summary}

%--==========================================================
%\subsection{Summary}

\begin{figure}[htb]
    \includegraphics[width=0.5\textwidth]{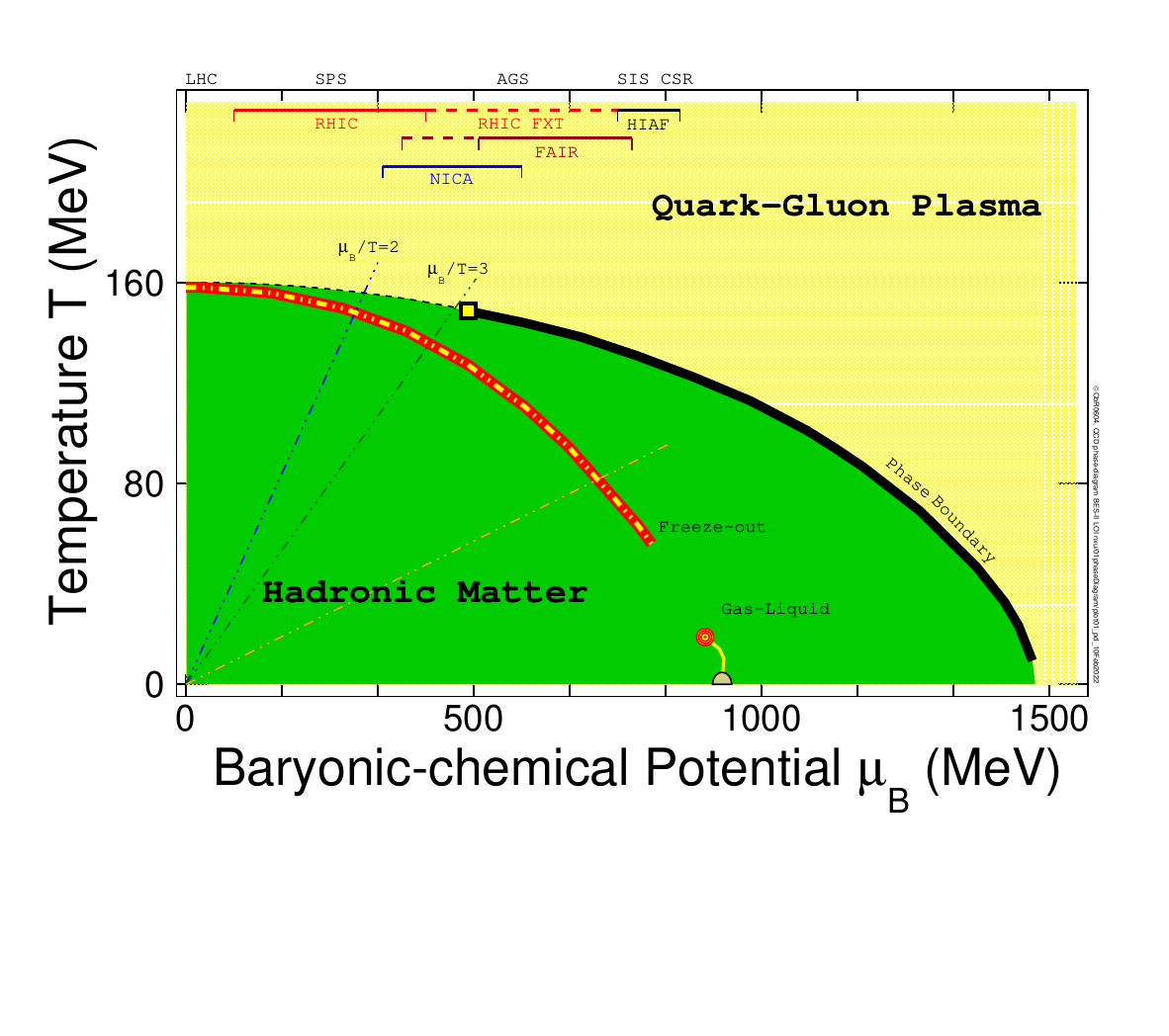}
\vspace{-1.5cm}
\caption{Sketch of the QCD phase diagram. The dashed line represents the smooth crossover region up to $\mu_{\rm B}/T \leq 3$. Black solid line represents the speculated first-order phase boundary. The empirical thermal freeze-out results from global hadron yield data are shown as the red-yellow line~\cite{Cleymans:2005xv}.   The liquid-gas transition region that features a second order critical point is shown by the red-circle, and a first-order transition line is shown by the yellow line, which connects the critical point to the ground state of nuclear matter. The coverage of the RHIC BES program, STAR fixed target program (FXT), future HIAF and FAIR facilities are indicated at the top of the figure. }
\label{fig1:phasestructure} 
\end{figure}

Since the discovery of the new form of matter, the strongly coupled Quark-Gluon Plasma (QGP)~\cite{STAR:2005gfr}, created in high-energy nuclear collisions in the early 2000, scientists have been asking: ``What is the structure of the QCD phase diagram in the high baryon density region?" and ``Is there a QCD critical point?". Model studies have shown that a first-order phase boundary is expected at the finite baryon density while at vanishing $\mu_{\rm B}$ the transition between the QGP and the hadronic matter is a smooth crossover. In such a scenario, the first-order phase transition line must end at a critical point and in a finite system such as nuclear collisions the critical point may turn into a critical region, see Fig.~\ref{fig1:phasestructure}.  More discussions on experimental results and Lattice calculations can be found in Refs.~\cite{Bzdak:2019pkr,STAR:2021rls}. The energy scan program at RHIC offers unprecedented high statistics data of nuclear collisions from the center of mass energy of $\sqrt{s_{\rm NN}}$ = 3\footnote{The STAR fixed-target (FXT) program became viable scientific endeavor due to the endcap Time-of-Flight detector constructed by Chinese colleagues for the CBM experiment at FAIR.} to 200 GeV, corresponding to the baryonic-chemical potential of $\mu_{\rm B}$ = 750 to 20 MeV. Measured data of Net-proton high moments from 200 - 39 GeV, {\it i.e.} $\mu_{\rm B}/T \le 2$, are consistent with smooth crossover transition~\cite{STAR:2020tga} as predicted by the first principal LGT calculations, see Fig.~\ref{fig1:phasestructure}. In lower energy region, or at larger net baryon densities, the collected data will allow us to probe the possible QCD critical region.

%{\it Outlook}: 
Thanks to the growth of the scientific community of high energy heavy ion physics and the development of state-of-art detector technologies boosted by the joint RHIC STAR-China research program,  the Chinese scientific program on high baryon density physics will continue to flourish based on a number of domestic facilities, from the Heavy Ion Research Facility in Lanzhou-Cooling Storage Ring (HIRFL-CSR) \cite{Yuan2020} to the Heavy Ion Accelerator Facility in Huizhou (HIAF) \cite{ZHW2020}, for example. The RHIC BES program revealed exciting physical dynamics and scientific opportunities at the high baryon density regime. Future investigations of properties of nuclear matter at  
moderate  $T$ and $\mu_{\rm B}$, created in the heavy ion collisions from sub-GeV/u (at HIRFL) to a few GeV/u (at HIAF) beam energies, are expected to shed new insight on QCD at extreme conditions.

The HIRFL-CSR external-target experiment (CEE) is a spectrometer covering a wide range of solid-angle in the center of mass reference frame, currently under construction with the supports from NSFC and CAS. With promising performance in tracking and particle identification for charged particles, CEE foresees plenty of opportunities in the studies of collision dynamics and nuclear matter properties at densities ranging from $\rho_0$ to $2.5~\rho_0$, where $\rho_0$ is the nuclear saturation density \cite{DGuo2024,LMLyu2017}.  For instance, one can carry out systematic measurements with CEE including productions of pions, kaons, strangeness baryons and the collective flow to probe the nuclear matter EOS. In parallel, the study the medium effect of baryon-baryon interactions in the cold nuclear matter near saturation density can be done using proton-induced collisions at CEE as well. In addition, measurements of the quark effect by the short range correlation of nucleons in nuclei can be extended at CEE in the near future~\cite{Ye:2024mls}.  

The Chinese team is also well positioned in the international community of heavy ion physics. At LHC, we are playing important roles in all experiments including ALICE~\cite{ALICE:2011ab}, ATLAS~\cite{ATLAS:2018pfw}, CMS~\cite{CMS:2012qk} and LHCb~\cite{Hadjidakis:2018ifr} exploring the properties of the QCD matter at vanishing net-baryon density. At high baryon density, on the other hand, the Chinese team has made substantial investments in both experiment CBM at FAIR~\cite{CBM:2016kpk} and MPD at NICA~\cite{Kisiel:2020spj}. Part of the TOF detector successfully employed in STAR BES-II program was made in China, supported in part by NSFC, for the CBM experiment~\cite{cbm2012} at FAIR. It is important to note that understanding nuclear matter at high baryon densities offers unique opportunities to study dynamics related to the inner structure of compact stars~\cite{Luo:2022mtp}.    

\section*{Acknowledgments}
We are grateful for the STAR Collaboration at RHIC.
This work is supported in part by the National Key Research and Development Program of China under Contract No. 2022YFA1604900, by the National Natural Science Foundation of China under Grant No. 12025501 and the Strategic Priority Research Program of Chinese Academy of Sciences under Grant No. XDB34000000.
%\vspace{10cm}

%%%%%%%%%%%%%%%%%%  Main article  %%%%%%%%%%%%%%%%%%
%\appendix
\vspace{1.cm}
{\it Postscript:}

This review is dedicated to Professor Wenqing Shen on the occasion of his 80th Birthday. Elected as an academician of the Chinese Academy of Sciences in 1999, Prof. Shen served in many academic leadership roles including President of the Shanghai Branch of the Chinese Academy of Sciences, Deputy Director of the National Natural Science Foundation of China, and Chairman of the Shanghai Association of Science and Technology. His research areas range from low-energy nuclear physics to heavy ion physics. In 2000, he led six Chinese institutions to join the RHIC-STAR collaboration. Since then, Chinese research teams have participated in a number of RHIC detector upgrades and have lead many outstanding physics analyses \cite{Shen07}. Thanks to his leadership, the Chinese nuclear physics community has expanded from traditional low-energy nuclear physics to high-energy nuclear physics. In the two decades after 2000, the Chinese high-energy nuclear physics community has achieved remarkable success. This review article reflects partially topics of the Chinese high-energy nuclear physics team's exploration of QCD matter.
As we celebrate Professor Shen's 80th birthday for his leadership and scientific achievements, we also reflect on how international collaborations can bring people with diverse culture background together and how collaborative teams can achieve greater scientific goals than individuals can. In retrospect, the key elements for the successful RHIC STAR-China project were all present in 2000, as the Chinese wisdom states, ``at the right time, in the right place and with the right people". We note that Dr. Timothy Hallman, then deputy spokesperson of the STAR collaboration and Professor Wenqing Shen played critical roles in initiating and fostering the STAR-China project.  Over the past twenty-five years, the project has grown, in hardware contributions to STAR from a small time-of-flight patch to a full scale time-of-flight detector, and to time-project-chamber upgrade, in physics analyses from the study of the light hadron spectra to dileptons, to quarkonia, to hypernuclei, to proton spin, and to global alignments. The Chinese heavy ion physics community has became one of the most vigorous international communities on studies of the strong interaction and the QCD at high temperature and densities. We are grateful to Professor Shen's leadership and vision. %þWe sincerely wish him many happy returns and 寿比南山、福如东海！

%--==========================================================
%--==========================================================
\bibliographystyle{woc.bst}
\bibliography{example}

\end{document}